\documentclass[longauth]{aa}  
 
\usepackage{natbib}
\usepackage{colortbl}
\definecolor{Gray}{gray}{0.9}
\usepackage{graphicx}
\usepackage{txfonts}
\usepackage{hyperref} \hypersetup{ colorlinks=true, citecolor={blue} ,linkcolor=blue, filecolor=blue, urlcolor=blue, pdftitle={Overleaf Example}}

\newcommand{\IREX}[0]{ \frac{\FD+\FH}{\FS}}
\newcommand{\VTP}[0]{ V_{tot}^{\prime}}
\newcommand{\VT}[0]{ V_{tot}}
\newcommand{\VC}[0]{ V_{Cont}}
\newcommand{\FC}[0]{F_{Cont}}

\newcommand{\FLC}[0]{F_{L/C}}
\newcommand{\FLCP}[0]{F^{ \prime}_{L/C}}

\newcommand{\FS}[0]{F_{*}}
\newcommand{\FH}[0]{F_{Halo}}
\newcommand{\FD}[0]{F_{Disk}}
\newcommand{\FL}[0]{F_{Line}}
\newcommand{\VS}[0]{V_{*}}
\newcommand{\VD}[0]{V_{Disk}}
\newcommand{\VL}[0]{V_{Line}}
\newcommand{\VH}[0]{V_{Halo}}

\newcommand{\OL}[0]{ O_{Line}}

\newcommand{\OT}[0]{ O_{tot}}
\newcommand{\OC}[0]{O_{Cont}}

\newcommand{\dpht}[0]{ e^{i\Phi_{tot}}}

\newcommand{\dphl}[0]{ e^{i\Phi_{Line}}}
\newcommand{\bg}[0]{Br\textrm{$\gamma$} }
\newcommand{\mm}[0]{\textrm{$\mathrm{\mu}$m} }

\begin{document} 

   \title{The GRAVITY young stellar object survey}
   \subtitle{IX. Spatially resolved kinematics of hot hydrogen gas in the star-disk interaction region of T~Tauri stars}
   


   \author{GRAVITY Collaboration: J. A. Wojtczak \inst{1}
        \and    L. Labadie\inst{1}
        \and    K. Perraut\inst{2}
        \and    B. Tessore\inst{2}
        \and    A. Soulain\inst{2}
        \and    V. Ganci\inst{1,5 }
        \and    J. Bouvier\inst{2}
        \and    C. Dougados\inst{2}
        \and    E. Alécian\inst{2}
        \and    H. Nowacki\inst{2}
        \and    G. Cozzo\inst{2,15}
        \and    W. Brandner\inst{3}
        \and    A. Caratti o Garatti\inst{3,6,7}
        \and    P. Garcia\inst{9,10}
        \and    R. Garcia Lopez\inst{3,6,11}
        \and    J. Sanchez-Bermudez\inst{3,8}
        \and    A. Amorim\inst{9,12}
        \and    M. Benisty\inst{2}
        \and    J.-P. Berger\inst{2}
        \and    G. Bourdarot\inst{4}
        \and    P. Caselli\inst{4}
        \and    Y. Clénet\inst{13}
        \and    P. T. de Zeeuw\inst{4,14}
        \and    R. Davies\inst{4}
        \and    A. Drescher\inst{4}
        \and    G. Duvert\inst{2}
        \and    A. Eckart\inst{1,5}
        \and    F. Eisenhauer\inst{4}
        \and    F. Eupen\inst{1}
        \and    N. M. Förster-Schreiber\inst{4}
        \and    E. Gendron\inst{13}
        \and    S. Gillessen\inst{4}
        \and    S. Grant\inst{4}
        \and    R. Grellmann\inst{1}
        \and    G. Heißel\inst{13}
        \and    Th. Henning\inst{3}
        \and    S. Hippler\inst{3}
        \and    M. Horrobin\inst{1}
        \and    Z. Hubert\inst{2}
        \and    L. Jocou\inst{2}
        \and    P. Kervella\inst{13}
        \and    S. Lacour\inst{13}
        \and    V. Lapeyrère\inst{13}
        \and    J.-B. Le Bouquin\inst{2}
        \and    P. Léna\inst{13}
        \and    D. Lutz\inst{4}
        \and    F. Mang\inst{4}
        \and    T. Ott\inst{4}
        \and    T. Paumard\inst{13}
        \and    G. Perrin\inst{13}
        \and    S. Scheithauer\inst{3}
        \and    J. Shangguan\inst{4}
        \and    T. Shimizu\inst{4}
        \and    S. Spezzano\inst{4}
        \and    O. Straub\inst{4}
        \and    C. Straubmeier\inst{1}
        \and    E. Sturm \inst{4}
        \and    E. van\,Dishoeck \inst{4,14}
        \and    F. Vincent\inst{13}
        \and    F. Widmann\inst{4}
        }

\institute{I. Physikalisches Institut, Universität zu Köln, Zülpicher Str. 77, 50937, Köln, Germany \\
\email{wojtczak@ph1.uni-koeln.de}
\and
Univ. Grenoble Alpes, CNRS, IPAG, 38000 Grenoble, France 
\and
Max Planck Institute for Astronomy, Königstuhl 17, 69117 Heidelberg, Germany 
\and
Max Planck Institute for Extraterrestrial Physics, Giessenbachstrasse,
85741 Garching bei München, Germany
\and
Max-Planck-Institute for Radio Astronomy, Auf dem Hügel 69,
53121 Bonn, Germany
\and
Dublin Institute for Advanced Studies, 31 Fitzwilliam Place,
D02,XF86 Dublin, Ireland 
\and 
INAF – Osservatorio Astronomico di Capodimonte, via Moiariello
16, 80131 Napoli, Italy 
\and
Instituto de Astronomía, Universidad Nacional Autónoma de México, Apdo. Postal 70264, Ciudad de México 04510, Mexico 
\and 
CENTRA, Centro de Astrofísica e Gravitação, Instituto Superior Técnico, Avenida Rovisco Pais 1, 1049 Lisboa, Portugal 
\and
Universidade do Porto, Faculdade de Engenharia, Rua Dr. Roberto
Frias, 4200-465 Porto, Portugal 
\and
School of Physics, University College Dublin, Belfield, Dublin 4,
Ireland
\and
 Universidade de Lisboa, Faculdade de Ciências, Campo Grande,
1749-016 Lisboa, Portugal
\and
LESIA, Observatoire de Paris, PSL Research University, CNRS,
Sorbonne Universités, UPMC Univ. Paris 06, Univ. Paris Diderot,
Sorbonne Paris Cité, France
\and
Sterrewacht Leiden, Leiden University, Postbus 9513, 2300 RA Leiden, The Netherlands
\and
Dipartimento di Fisica \& Chimica, Università di Palermo, Piazza del Parlamento 1, I-90134 Palermo, Italy
} 
   \date{Received -; accepted -}

 
  \abstract
  {Hot atomic hydrogen emission lines in pre-main sequence stars serve as tracers for physical processes in the innermost regions of circumstellar accretion disks, where the interaction between a star and disk is the dominant influence on the formation of infalls and outflows. In the highly magnetically active T~Tauri stars, this interaction region is particularly shaped by the stellar magnetic field and the associated magnetosphere, covering the inner five stellar radii around the central star. Even for the closest T~Tauri stars, a region as compact as this is only observed on the sky plane at sub-mas scales. To resolve it spatially, the capabilities of optical long baseline interferometry are required.}
  {We aim to spatially and spectrally resolve the \bg hydrogen emission line with the methods of interferometry in order to examine the kinematics of the hydrogen gas emission region in the inner accretion disk of a sample of solar-like young stellar objects. The goal is to identify trends and categories among the sources of our sample and to discuss whether or not they can be tied to different origin mechanisms associated with \bg emission in T~Tauri stars, chiefly and most prominently magnetospheric accretion.}
  {We observed a sample of seven T~Tauri stars for the first time with VLTI GRAVITY, recording spectra and spectrally dispersed interferometric quantities across the \bg line at 2.16 $\mu$m in the near-infrared K-band. We used
  the visibilities and differential phases to extract the size of the \bg emission region and the photocentre shifts on a channel-by-channel basis, probing the variation of spatial extent at different radial velocities. To assist in the interpretation, we also made use of radiative transfer models of magnetospheric accretion to establish a baseline of expected interferometric signatures if accretion is the primary driver of \bg emission.}
  {From among our sample, we find that five of the seven T~Tauri stars show an emission region with a half-flux radius in the four to seven stellar radii range that is broadly expected for magnetospheric truncation. Two of the five objects also show \bg emission primarily originating from within the co-rotation radius, which is an important criterion for magnetospheric accretion. Two objects exhibit extended emission on a scale beyond 10 R$_*$, one of them is even beyond the K-band continuum half-flux radius of 11.3 R$_*$. The observed photocentre shifts across the line can be either similar to what is expected for disks in rotation or show patterns of higher complexity. }
  {Based on the observational findings and the comparison with the radiative transfer models, we find strong evidence to suggest that for the two weakest accretors in the sample, magnetospheric accretion is the primary driver of \bg radiation. The results for the remaining sources imply either partial or strong contributions coming from additional, spatially extended emission components in the form of outflows, such as stellar or disk winds. We expect that in actively accreting T~Tauri stars, these phenomena typically occur simultaneously on different spatial scales. Through more advanced modelling, interferometry will be a key factor in disentangling their distinct contributions to the total \bg flux arising from the innermost disk regions.}

   
 \authorrunning {J.~A.~Wojtczak et al.}
    \titlerunning{Spatially resolved kinematics of hot hydrogen gas}
   \maketitle
%

\section{Introduction}

\begin{figure*}[!t]
    \centering
    \includegraphics[width=\linewidth]{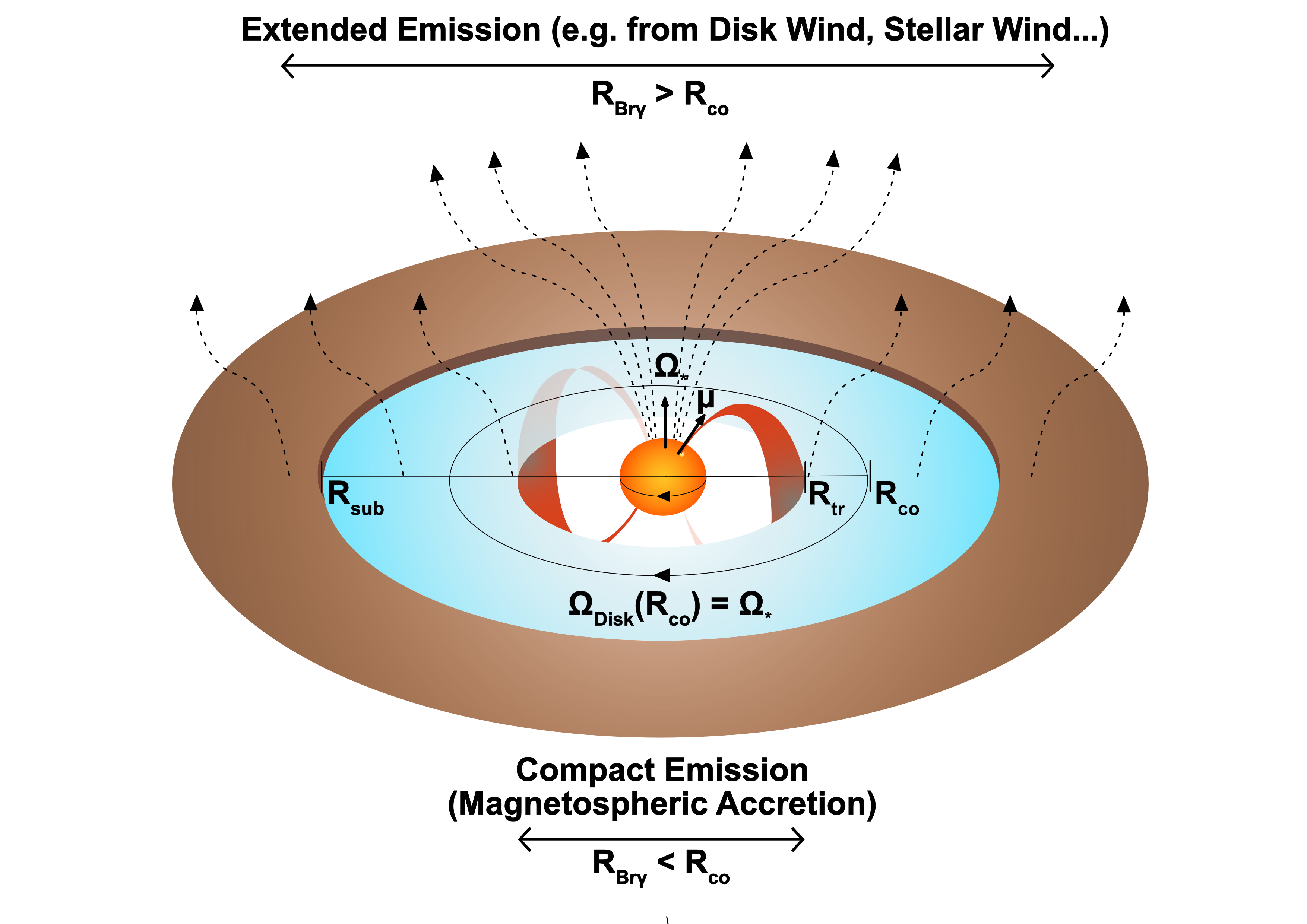}
    \caption{Schematic depiction of the innermost parts of the circumstellar disk of a T~Tauri star. \bg emission in T~Tauri stars is associated with the concept of magnetospheric accretion, but outflows caused by certain types of disk or stellar winds (dashed lines) can heat the hydrogen gas to the necessary temperatures to produce \bg emission as well. Different origin mechanisms can be tied to different spatial scales. Magnetospheric accretion flows (depicted in red) can only stably form in a compact region within the co-rotation radius R$_{\textrm{co}}$, where the gaseous disk rotates at the same angular velocity $\Omega$ as the star. Spatially extended \bg emission from outside R$_{\textrm{co}}$ cannot be attributed to magnetospheric accretion and must be caused by some other emission component. \\
    This figure specifically shows the common case of a non-axisymmetric magnetosphere, featuring a tilt between the stellar rotational axis $\Omega_*$ and the magnetic dipole $\mu$. The obliquity of the magnetosphere leads to the formation of two accretion columns, which funnel gas from a wide arc at the truncation radius onto one primary hot spot per hemisphere, as first suggested in \citet{Bertout1988} and \citet{Calvet1992}.
    The equations and physical quantities relevant for the determination of the co-rotation radius $R_{\rm co}$ and the magnetic truncation radius $R_{\rm tr}$ are discussed in Section \ref{sec:6}. For the equations used to estimate the sublimation radius $R_{\rm sub}$, which indicates the area at which the disk becomes sufficiently hot to destroy dust grains, we refer to \cite{Perraut2021}.
    }
    \label{fig:magneto}
\end{figure*}
Since the interplay between a young stellar object (YSO) and its circumstellar disk has a major influence on the different phases of its development, the study of such a star-disk interaction is of great importance in providing insight into many aspects of star formation. Active research into this topic is connected to a variety of astrophysical concepts, ranging from high level models of stellar evolution down to the environmental conditions in the cradles of planet formation and the mechanics that govern mass accretion during the pre-main sequence stage of a YSO.
%
%
A number of ideas regarding the processes behind accretion and ejection of matter in the inner disk have been extensively explored in theoretical models and are relatively well established on a conceptual level. However, the exact nature of the physical interaction in the innermost circumstellar region inside 0.1\,au from the central star is still debated because the observational evidence for many of these concepts remains mostly indirect. 
In this context, a distinction arises between intermediate-mass Herbig stars and low-mass T~Tauri stars. The latter has a lower effective temperature and weaker mass-accretion rates, but also exhibits significant magnetic field strengths of 1-2\,kG due their convective nature (e.g. \citealt{JohnsKrull2007}). This substantial difference in the level of magnetic activity has given rise to the paradigm of magnetospheric accretion. As a theory concerning accretion in low-mass YSOs, it has attained preeminence due to its ability to reconcile accretion properties with the  presence of stellar hot spots and the observations of hydrogen line profiles \citep{Camenzind1990, Calvet1992}. 
It has superseded the earlier theory of boundary layer accretion, which is still thought to govern the processes of mass accretion in magnetically weak intermediate mass Herbig Ae/Be type YSOs \citep{Mendigutia2020}.\\
In the model of magnetospheric accretion, it is proposed that the inner gaseous disk is truncated by the strong stellar magnetic field because of the magnetic pressure exerted perpendicular to the magnetic field lines threading the disk. Thus, a region of a few stellar radii surrounding the star is formed where the disk plane is almost completely evacuated from gas particles \citep{Bouvier2007}. Instead of further spiralling inwards, the gas particles can only move freely along the field lines, which effectively serve as funnels that channel the gas onto the star. As the gas impacts the stellar surface at high latitudes, it creates heated shock regions close to the poles \citep{Hartmann2016}.

\noindent Hydrogen emission lines serve as tracers for the kinematics of the hot gas in the inner star-disk system. In particular the near-infrared (NIR) \bg line at 2.16\,$\mu$m, which is directly relevant for this work, requires the gas to be at temperatures of 8000 to 10000\,K  (\cite{Muzerolle98a}, \cite{Tambovtseva2016})  in order to produce strong emission. A heated gaseous disk in Keplerian rotation can, in principle, be the cause for the \bg emission in YSOs, in particular for intermediate-mass Herbig Ae/Be stars \citep{Kraus2008}. However, even in those cases, theoretical considerations indicate that the total line-flux contribution coming from the hot disk layers is faint compared to the contribution from outflows and infalls (see again \citet{Tambovtseva2016}).
In T Tauri stars, with their much lower effective temperatures and
comparatively lower mass-accretion rates, the radiation field strength alone is
insufficient to heat the gas disk up to such temperatures at scales of ~0.1 au 
 \citep{Natta1993}. It is then reasonable to assume that only specific heating mechanisms acting much closer to the star in the accretion funnel flows or in the stellar-disk wind-launching regions result in the formation of \bg emission lines. 
Consequently, \bg emission in T\,Tauri stars is an excellent tracer of the processes associated with such gas-heating mechanisms, in particular for the magnetospheric accretion model which is believed to play a central role in the evolution of low-mass pre-main sequence stars.\\
Fig.~\ref{fig:magneto} illustrates how different possible mechanisms producing \bg emission are tied to different spatial scales.
\bg emission coming from extended parts of the gaseous disk closer to the sublimation radius is likely to trace outflows in the form of a disk or even a stellar wind, launched from  either the circumstellar disk or the polar region of the star, respectively \citep{Ferreira2013}. 
For Herbig stars, a wind contribution in the \bg emission has been proposed in previous interferometric works \citep{GarciaLopez2015,Kurosawa2016}. 
On the contrary, \bg emission arising from a more compact region close to the star is considered, in particular in magnetically active T\,Tauri stars, to originate in the accretion columns located inside the magnetospheric truncation radius $R_{tr}$, at which point the ram pressure of the infalling material is compensated for by the magnetic pressure exerted perpendicular to the field lines \citep{Bessolaz2008}. A useful parameter to differentiate compact versus extended emission in the context of magnetospheric accretion can be the co-rotation radius $R_{\rm co}$, which is the radius at which the angular velocities of the gaseous disk and of the star equate \citep{Bouvier2013}.  
Since the formation of stable funnels is only possible if the disk is truncated within the co-rotation radius \citep{Romanova2016}, \bg emission detected inside $R_{\rm co}$ indicates the probable presence of magnetospheric accretion flows. Following Fig.~\ref{fig:magneto}, it is clear that the mechanisms involved in the accretion process of matter in low-mass pre-main sequence stars are acting on a scale of a few (for the magnetic accretion) to a few tens (in the case of winds) of stellar radii.\\
Prior to the advent of NIR long baseline interferometry, observational evidence for the magnetospheric accretion scenario has mainly been discussed in the context of high spectral resolution spectroscopy studies by \citet{Muzerolle1998}, who have shown that the magnetospheric accretion model could predict the \bg profiles. The process of magnetospheric accretion would influence the shape and strength of hydrogen emission lines, with a characteristic inverse P~Cygni profile caused by infalling optically thick gas (e.g. \citealt{Calvet1992}, \citealt{Edwards1994}, \citealt{Petrov2014}), typically observed at high inclination of the system. Through theoretical simulations,
\citet{Kurosawa2011} investigated how the combination of the magnetospheric model with a (stellar-disk) wind model can qualitatively reproduce the observed profiles. 
However, without spatially resolving the star-disk interacting regions, such a picture remains incomplete since such spectroscopic features with their overall line shapes are not unique to magnetospheric accretion.\\ 
With the advent of long baseline interferometry delivering sub-milliarcsecond resolution and the concomitant improvement in the sensitivity of the technique, a larger reservoir of \bg emitting T~Tauri stars at a few hundred parsecs distance has become accessible. This has been exploited in the past in the works of \citet{Eisner2010} and \citet{Eisner2014}, who examined \bg emission in a number of T~Tauri stars with the Keck Interferometer, as well as in recent pioneering studies with the Very Large Telescope Interferometer (VLTI) by \citet{GarciaLopez2017} for S CrA, \cite{GarciaLopez2020} for TW\,Hya, and \citet{Bouvier2020a} for DoAr 44. \\
In this paper, we report on observations made with GRAVITY \citep{GravityCollaboration2017}, the second generation beam combiner instrument of the VLTI, of a small sample of seven accreting T\,Tauri stars.  We aim to improve upon the results obtained in previous studies by combining the high angular and spectral resolution offered by GRAVITY to present spectrally dispersed measurements of the \bg emission region across the sample, capable of identifying broader trends and categories between the different objects. We contextualise our findings by drawing upon the latest results in radiative transfer modelling of magnetospheric accretion in \bg, deriving sets of synthetic observables to which our observations can be compared. \\
Sections~\ref{sec:2} and \ref{sec:3} describe the sample, observations, and data reduction. Sections~\ref{sec:4} to \ref{sec:6} describe the modelling and the results. Comparisons with synthetic data obtained from the models can be found in Section \ref{sec:7}, along with a wider discussion on magnetospheric accretion and the origin of \bg in T~Tauri stars

\section{Scientific sample and observations}\label{sec:2}
\begin{table*}[t]
\tabcolsep=0.09cm
\footnotesize
\caption{Properties of the stars of the T~Tauri sample.}\label{acctab}
\centering
\begin{tabular}{cccccccccc} 
\hline \hline  
 Name   &       Sp.~Type        &$ \text{T}_{\rm eff}$ & $d$  & $M_*$   &       $L_*$   & $R_*$ & $\dot{M}_{acc}$ &       Age & References        \\
        &               & [K] & [pc] &  $ [M_{\odot}] $ &       [$L_{\odot}$] & [$R_{\odot}$]&        [10$^{-8}$\,$M_{\odot}$yr$^{-1}$]       &       [Ma] &  \\ \hline \\

 AS\,353        &       K5      & 4450  & 399.6 $\pm$ 4.2       &       1 $\pm$ 0.2               &       3.9 $\pm$ 0.67 & 3.32 $\pm$ 0.29        &       7.9     - 45 & $\leq$ 1& \citet{Rei2018},\\ & & &  & & & & & & \citet{Prato2003}  \\
 RU\,Lup        &       K7      & 4050  & 157.5 $\pm$ 1.3       &       0.6 $\pm$ 0.2       &       1.64 $\pm$ 0.22& 2.60 $\pm$ 0.17        &       1.8 - 30    & 1-2   & \citet{Herczeg2008},\\ & & &  & & & & & & \citet{Siwak2016} \\
VV\,CrA SW      &       K7      & 4050  &156.6 $\pm$1.2         &       0.6 $\pm$ 0.2       &       1.73 $\pm$ 0.24 & 2.67 $\pm$ 0.19 &     3.33 - 22       &       1-2     & \citet{Sullivan2019}\\ \\
 TW\,Hya        &       K6      & 4200  & 60.14 $\pm$ 0.07&     0.9 $\pm$ 0.1     &       0.41 $\pm$ 0.05& 1.21 $\pm$ 0.07        &       0.13 - 0.23     &       4-12    & \citet{Donati2011}, \\ & & &  & & & & & & \citet{Venuti2019}\\
 S\,CrA N       &       K5-K6   & 4300          & 160.5 $\pm$ 2.2       &       0.9 $\pm$ 0.2       &       3.82 $\pm$ 0.54 & 3.52 $\pm$ 0.25       &       7.8 - 50    &       $\leq$ 1        & \citet{Sullivan2019}\\ & & &  & & & & & & \citet{Gahm2018},\\
 DG\,Tau        &       K6      & 4200  & 125.2 $\pm$ 2.3       &       0.8 $\pm$ 0.2       &       1.58 $\pm$ 0.37 & 2.37 $\pm$ 0.28       &       4.6 - 74    &       1-2     & \citet{White2001},\\ & & &  & & & & & & \citet{White2004} \\
 DoAr\,44       &       K2-K3   & 4840  & 146.3 $\pm$ 0.6       &       1.5 $\pm$ 0.2       &       1.61 $\pm$ 0.21& 1.80 $\pm$  0.12       &       0.63 - 0.9 & 3-6     & \citet{Manara2014PhDTh}, \\ & & &  & & & & & & \citet{Espaillat2010} \\
\hline 
\end{tabular}    
\tablefoot{References are referring to the lower and upper limit $\dot{M}_{acc}$ literature values. We note that $R_*$ was computed from effective temperature and stellar luminosity. Other properties were adopted from \citet{Perraut2021}. }
\end{table*}

\subsection{Sample}
We have studied a total of seven T~Tauri sources, which in the context of this work will be referred to as the GRAVITY T~Tauri sample. The targets in the sample consist of a number of mostly strong accretors with mass-accretion rates of the order of $10^{-8} M_{\odot} \ yr^{-1}$, masses of less than 1.5 $M_{\odot}$, and stellar luminosities ranging up to 4 $L_{\odot}$. The majority are mid K-type stars with surface temperatures in the span between 4000 K and 4900 K and ages of 1 to 2 million years or younger. Their individual properties are summarised in Table~\ref{acctab}.

\subsection{Observations}

The individual sources of the sample were observed between 2016 and 2021 with VLTI GRAVITY, the latest generation K-band beam combiner based at the ESO facilities of the VLTI at Cerro Paranal, Chile. The analysis of the \bg emission region relies exclusively on observations taken with the four Unit Telescopes (UTs) with their  8.2 m primary mirrors. The Auxiliary Telescopes (ATs), while providing better coverage in the uv plane and potentially higher spatial resolution at the longest baselines, are usually not sensitive enough to detect the relatively faint emission signal of \bg originating from T~Tauri type stars. \\
The standard UT configuration combines the four telescopes to a total of six baselines, spanning up to 130 m in length on the ground and providing a spatial resolution of up to $\frac{\lambda}{2B}$, which translates into 1.6 to 1.7 mas when observing in the near infrared between 2 and 2.4 micrometres. With the exception of S CrA N, all of the sources were observed in single field mode, meaning the fringe data were simultaneously recorded on the separate detectors of the instrument's fringe tracker (FT) and science channel (SC).
\\ The FT data were recorded at low spectral resolution (R=22) in six spectral channels over the range of the entire K-band. While the fringe tracker data in its own right can be utilised in the analysis of the K-band continuum, its main purpose with respect to the investigation of the hydrogen emission lines is to allow for longer integration times in the SC than would otherwise be feasible by stabilising the fringes against small perturbations of the atmosphere \citep{Lacour2019}. We made use of the FT capability to take the SC observations in GRAVITY's high resolution mode (R=4000) with a width of $\sim$ 5 $\AA$, or 75 km/s, per channel, enabling us to spectrally resolve the \bg feature and trace the gas kinematics over a range of up to $\pm$ six channels from the centre of the line, depending on the strength of the feature in the individual sources.  \\
The observations of the different targets are subdivided into a set of short term runs of about 5 minutes in length, each of which was saved as a separate file, stretching over a total observation time of around 1 1/2 hours for most of the objects, although the observation of DoAr 44 went on for notably longer with a total time of slightly over 4 hours. Measurements of suitable calibrator stars in a sufficiently close part of the sky were predominantly taken either up to 45 minutes before the beginning or after the end of the telescope run and thus potentially probing the sky under close but not entirely identical atmospheric conditions. For the 2021 epoch of the RU Lup observation, no calibrator close to the target was available for the night, so that an object in a significantly different part of the sky had to be used for data calibration. \\
Weather conditions for the relevant nights were mostly stable with typical seeing values of around 0.5 to 0.7 arcseconds, however, due to the relatively long observation of DoAr 44, we note a larger fluctuation of the seeing throughout the night and deteriorating atmospheric conditions towards the end of the run.  \\
The GRAVITY data on TW Hya, S CrA and DoAr 44 was previously published in \citet{GarciaLopez2020}, \citet{GarciaLopez2017}, and \citet{Bouvier2020a}, respectively.
A log of the observations of the T~Tauri sample is given in Table \ref{obslog}.

\section{Data reduction and spectral calibration}\label{sec:3}

\begin{table*}[!h]
\caption{Log of observations.}\label{obslog}
\centering
\begin{tabular}{cccclcccc}
\hline \hline
 Source      & Date       & Time (UTC)   & Configuration   & Calibrator   &   Airmass &   Seeing [arcsec] &   Files \\
\hline
 DG Tau      & 21.01.2019 & 01:31 - 03:06        & U1-U2-U3-U4     & HD 37491     &      1.59 &              0.65 &      10 \\  
 TW Hya      & 21.01.2019 & 06:59 - 08:24        & U1-U2-U3-U4     & HD 95470     &      1.03 &              0.51 &      13 \\ 
 AS 353      & 21.04.2019 & 08:44 - 10:22        & U1-U2-U3-U4     & HD 183442    &      1.31 &              0.34 &      12 \\ 
 RU Lup 2018 & 27.04.2018 & 03:23 - 04:30         & U1-U2-U3-U4     & HD 142448    &      1.27 &              0.57 &       8 \\ 
 RU Lup 2021 & 30.05.2021 & 01:58 - 03:19       & U1-U2-U3-U4     & HD 99264     &      1.31 &              0.64 &       8 \\ 
 S CrA N     & 20.07.2016 & 05:30 - 06:13       & U1-U2-U3-U4     & HD 188787    &      1.10  &              0.56 &       6 \\ 
 VV CrA      & 20.06.2019 & 02:53 - 04:04        & U1-U2-U3-U4     & HD 162926    &      1.31 &              0.77 &       3 \\ 
 DoAr 44     & 22.06.2019 & 02:04 - 05:46         & U1-U2-U3-U4     & HD 149562,   &      1.04 &              0.68 &      28 \\
 & & & &   HD 147701,& & & \\
  & & & &   HD 147578& & & \\
\hline
\end{tabular}   
\tablefoot{Time denotes the beginning of the first and the last file of the observation.}
\end{table*}

\subsection{Pipeline reduction and post-processing}
The data were reduced and calibrated through the use of the standard GRAVITY pipeline software provided by ESO (Lapeyrere et al. 2014). A full account of the calibrated SC data from the observations is given in Appendix \ref{AppendixIII}. \\
In the context of this work we focus on the analysis of the SC data to extract information about the \bg gas emission region, whereas FT data is only used in a supplemental manner as described further below. For an overview of the treatment of the K-band continuum, based on the same data sets obtained with GRAVITY, we refer to the work of \citet{Perraut2021}. \\
Each of the individual files obtained for all of our targets contains four sets of flux data coming from the four Unit Telescopes, six sets of normalised visibility amplitude data coming from the six UT baselines, six corresponding sets of visibility phase data, and four sets of closure phase data from the four unique closed baseline triangles that can be formed by the UTs. \\
The individual files were merged before further analysis using the post-processing routine of the GRAVITY pipeline, effectively time-averaging our observables with a weighted mean over the 1 1/2 hours of observation. The data were globally shifted to the local standard of rest (LSR) to eliminate observational effects due to the earth's rotation around the sun and the radial velocity of the source. \\
The flux data were treated by first merging the four individual telescope spectra and then normalising the \bg peak to the surrounding continuum via a polynomial fit in the region between 2.1 \mm to 2.2 \mm. The normalised mean spectrum was further fitted with a Gaussian function in a region (2.15 \mm to 2.18 \mm) closer to the \bg peak to smooth out local fluctuations when determining the line-to-continuum ratio on a channel to channel basis. For those objects in the sample which exhibit asymmetric line shapes, a combination of two single Gaussians was used to fit the spectrum. We then consider the function value for a given wavelength as the true line to continuum flux ratio for that wavelength and use this value in our computations concerning the size and photocentre shift as detailed in Appendix \ref{sec:5}. In order to ensure the interpretation of the data remains meaningful, we implemented a selection criterion based on the strength of the flux ratio in each channel relative to the continuum dispersion so that only those channels that lie above a 2$\sigma$ threshold are taken into account for further analysis.   \\
For the visibility amplitudes, which we take as the root of the $V^2$, we used FT data to normalise the science channel continuum around the \bg line to the continuum value measured by the fringe tracker for the corresponding baseline. To this end, we took an average continuum value from the FT channels, discarding the channel at the lowest wavelength as it is affected more strongly by atmospheric conditions and the meteorology laser operating at 1.908 \mm. To determine the visibility value in each of the considered spectral channels, a single peaked Gaussian function was then fitted to the data, similar to the treatment of the flux. The visibility phase data received an analogous treatment after a polynomial function was first fitted to and then subtracted from the data in order to obtain the differential phase between line and continuum channels. \\
We attempted to correct for the telluric effects from atmospheric absorption lines by using the calibrator spectra, but found that many of the calibrators were not suitable for this method due to large photospheric absorption features broad enough to cover the telluric features. An alternative attempt, using an atmospheric model spectrum, to remove the telluric lines proved also to be problematic. Instead, we decided against a full correction and opted to restrain our analysis to a range of channels that falls inside the two telluric features at 2.163446 \mm and 2.168645 \mm. We proceeded to remove the channels affected by those two features from the spectrum before attempting to fit the Gaussian profile as described above. Whilst we recognise that the flux ratios obtained close to the telluric features obtained in this manner are potentially incorrect, we propose that the overall impact of this discrepancy on our results is negligible as low signal strengths in those areas means we do not consider those channels in our analysis for the majority of our targets. \\

\subsection{High-precision spectral calibration}

\begin{figure}
\centering
\begin{minipage}{1\linewidth}
\centering
\includegraphics[width=1\linewidth]{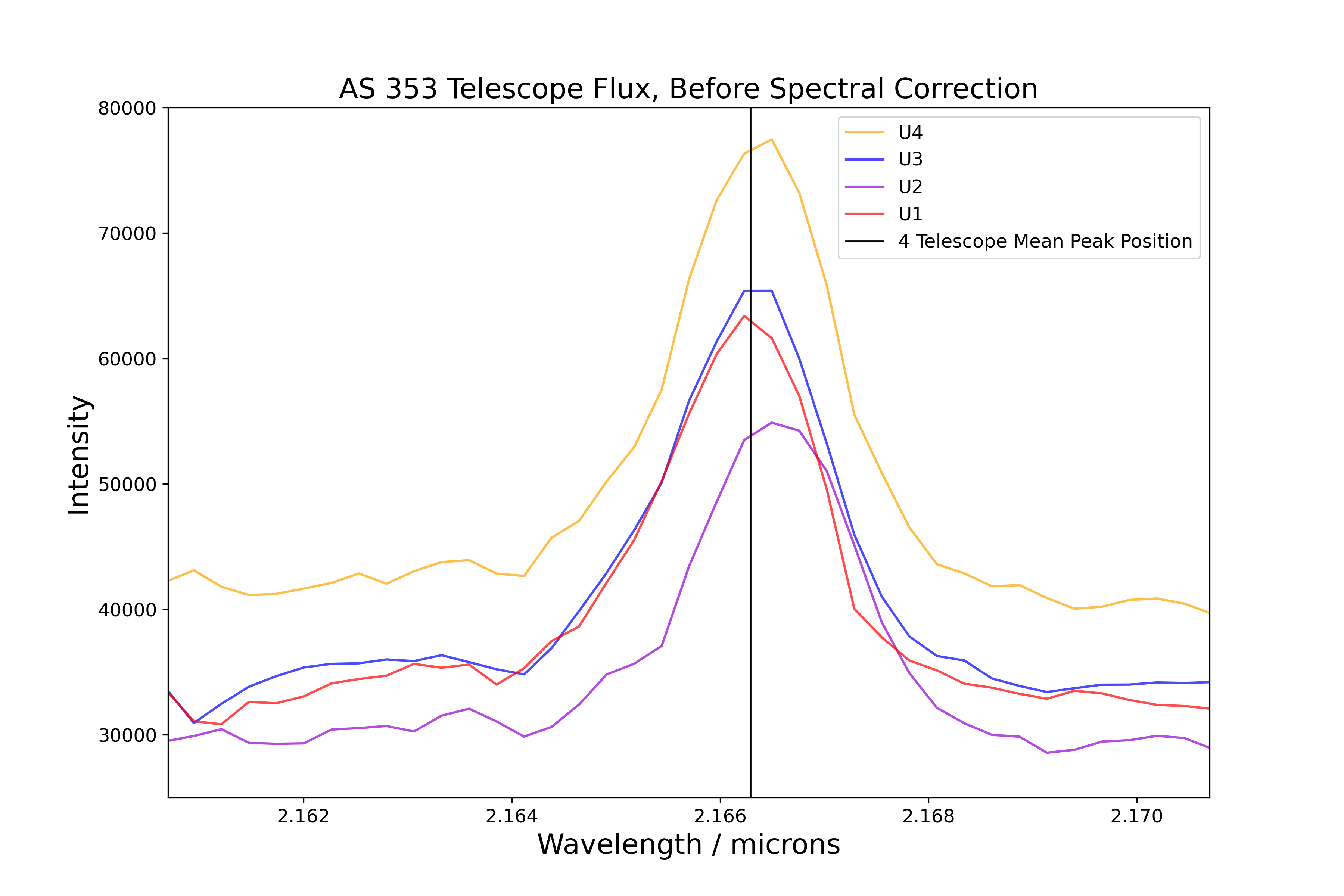}
\end{minipage}

\begin{minipage}{1\linewidth}
\centering
\includegraphics[width=1\linewidth]{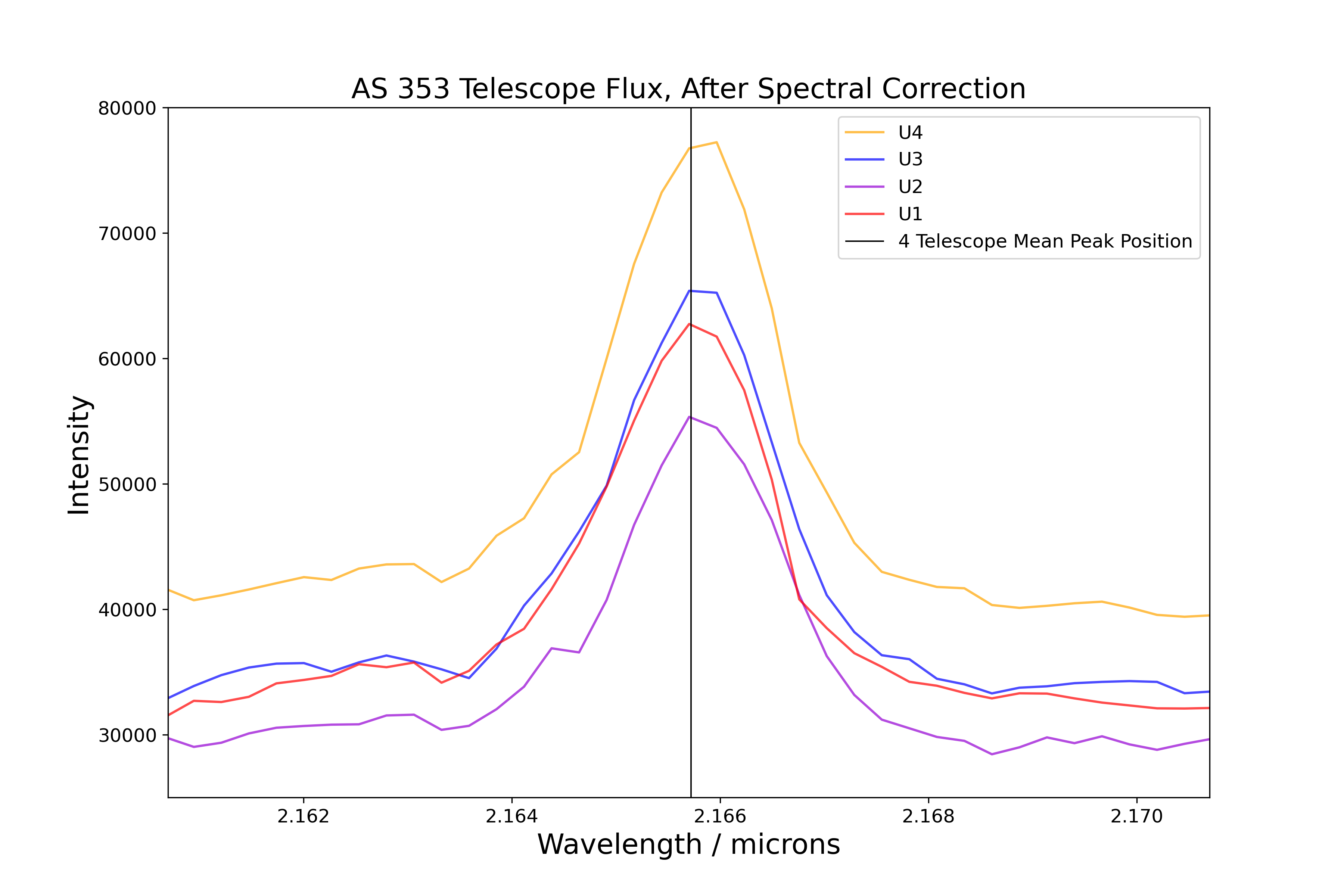}
\end{minipage}

\caption{\bg spectra of the AS 353 observation, taken with the four unit telescopes of the VLTI. \\\textbf{Top:} The spectra before the application of the spectral calibration. The vertical line indicates the mean position of the line peak. Before the correction, the individual spectra for most of our objects were misaligned both with respect to the telluric lines and also between each other. 
\textbf{Bottom:} Same spectra after they were brought into alignment through the spectral calibration. The corrected spectra and recomputed observables were then further shifted to bring them to the local standard of rest (LSR).}
\label{correctioncomp}
\end{figure}

A more detailed analysis of the flux recorded by the 4 UTs has shown a systematic spectral shift of the spectra with respect to known telluric absorption lines in the K-band. This shift, which is on average of the order of 5 \AA, is not affecting the individual UT spectra equally, leading to a misalignment between the 4 \bg lines recorded with the UTs. Such a misalignment not only affects the flux values in the merged spectrum, but also the computation of the complex visibilities by the pipeline during the reduction process. This leads to a distortion of the \bg line signature in both the visibility amplitudes and differential phases for the different baselines, depending on the UTs involved. As a practical consequence of this effect we noticed that the spectrally resolved line features in the visibility amplitudes, which are generally expected to be single peaked, were not well aligned with the peak of the mean spectrum and instead were for some baselines shifted to the left or right, thus affecting the results of the channel-by-channel analysis of those observables. The line feature in the differential phases is equally affected, as S-shape signatures can be effectively suppressed and appear single peaked.\\
We adjust for this effect by introducing a correction based on a fit of the 24 individual spectra of each observational run to the known telluric lines close to \bg. If the telluric lines are too weak or the data too noisy to produce a proper fit, a master correction based on an average shift is  applied.
An example of this, showing the individual spectra before and after the correction, is shown in Fig. \ref{correctioncomp}.\\ 
We found the overall effect of the correction to be more pronounced in the photocentre shift profiles derived from the differential phases, whereas the visibility data appeared to be relatively less affected. The general impact is highest on the pure line quantities (see Appendix \ref{sec:5}), which is shown in Fig. \ref{pp1}. While the marginally changed shape of spectra and visibility signals after the correction also affect the pure line observables, the dominant effect here stems from the relative shift between the mean spectrum and the interferometric observables. When computing $\mathrm{\VL}$ from Eq. \ref{eq:vline}, a different combination of $\mathrm{\FLC}$ and $\mathrm{\VT}$ values in each channel is used after the correction, as now the visibility peaks are better aligned with the mean spectral peak. The difference is immediately obvious when comparing the pure line visibilities in Fig. \ref{pp1} between top and bottom. Before the correction, the peak of the visibility was shifted to the right with respect to the four-telescope mean spectral peak, resulting in comparatively large values for $\mathrm{\VT}$ at much weaker flux ratios in the blue wing. Consequently, the pure line visibilities are increasing as we move to higher wavelengths. After the correction, the decreasing values for $\mathrm{\FLC}$ in the blue wing are paired with decreasing values in in the visibilities, leading to a peak in pure line visibilities that is roughly aligned with the peak in total visibilities. \\
The telluric shift has been detected in all of the data sets except for the 2021 epoch of RU Lup. The effect is currently understood to be connected to the old grism of the science spectrometer and was consequently resolved with the grism upgrade in October 2019, meaning all GRAVITY data taken after this month should no longer be affected.

\begin{figure}
\centering
\begin{minipage}{1\linewidth}
\centering
\includegraphics[width=\linewidth]{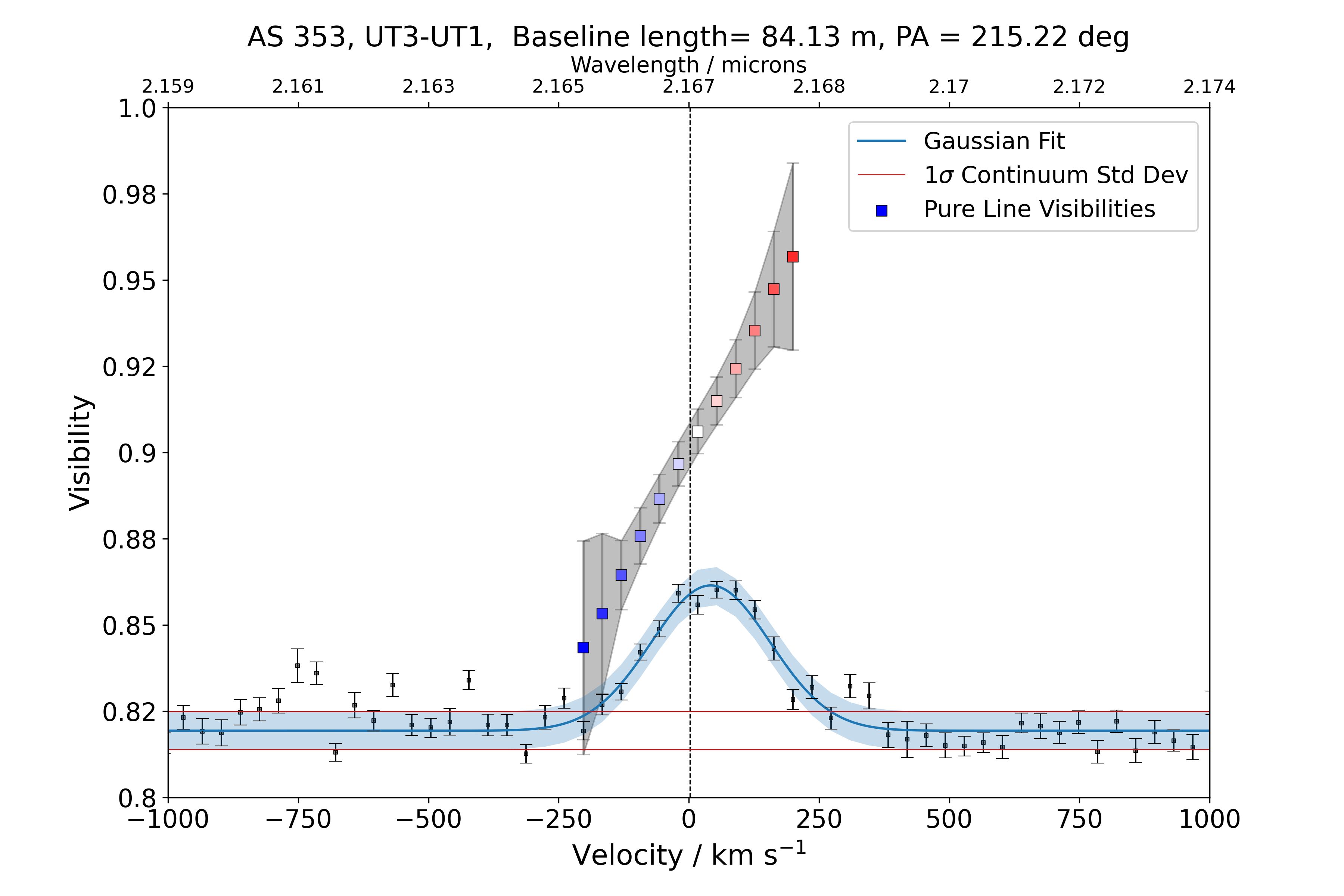}
\end{minipage}

\begin{minipage}{1\linewidth}
\centering
\includegraphics[width=\linewidth]{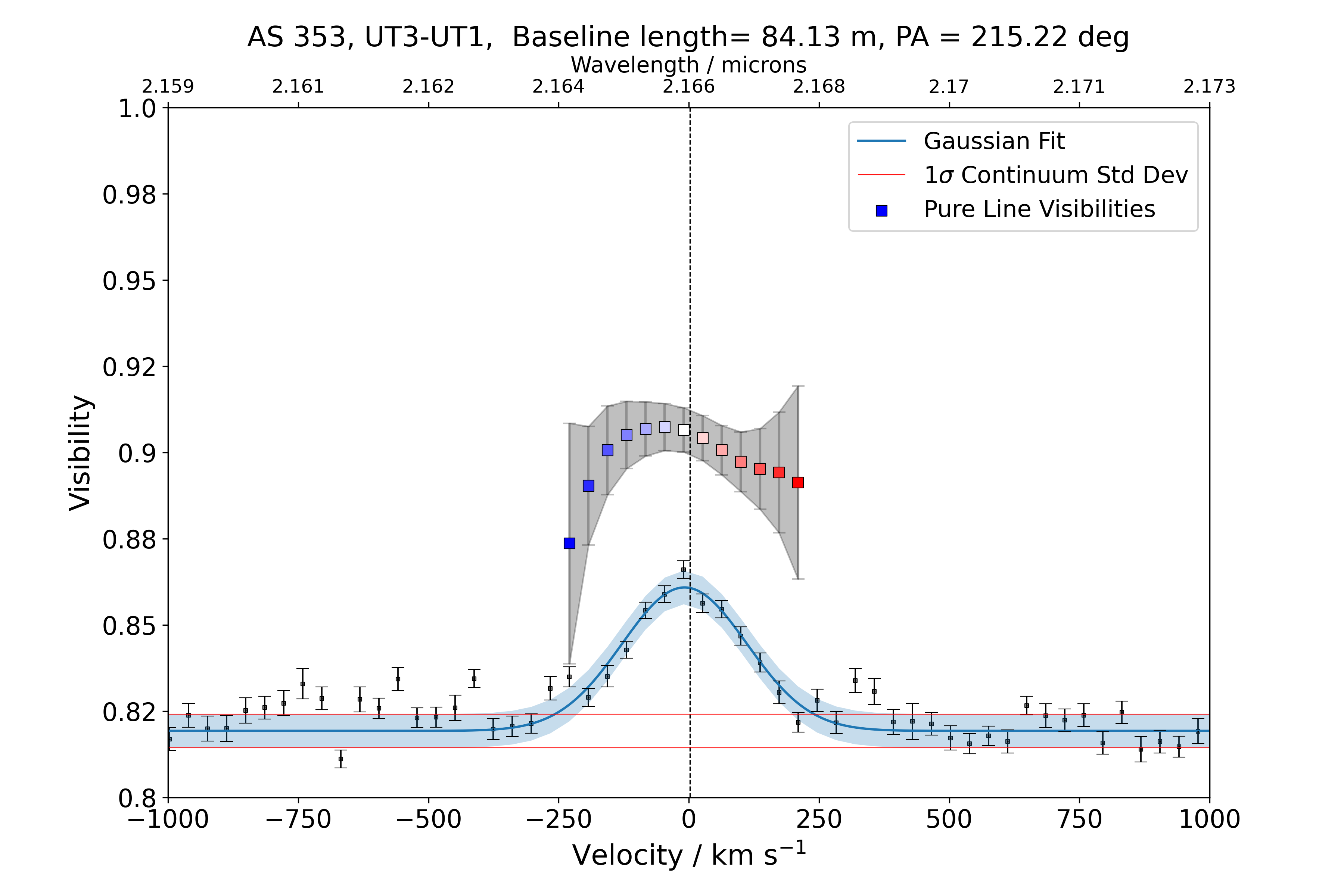}
\end{minipage}

\caption{Total and pure line visibilities for an exemplary baseline from the AS 353 observation. \\ \textbf{Top:} Visibilities before the spectral correction was applied. Blue and grey shaded regions indicate the uncertainties on the total and pure line visibilities, respectively. \textbf{Bottom:} Same baseline after the correction. Especially with respect to the computation of the pure line visibilities, the change appears significant in the red wing of the signal. }
\label{pp1}
\end{figure}
\section{Data and methodology overview}\label{sec:4}

Through our GRAVITY observation, we obtained K-band spectra between 2 and 2.4 microns, showing clear \bg features at around 2.16 microns with peak line-to-continuum flux ratios spanning a range of values from 1.91 in the extreme case of AS 353 down to only 1.19 for the faintest line we observed for DoAr 44. The uncertainties on the flux ratios, derived from the instrumental error bars given by the pipeline, are of the order of 0.01 for the majority of the T~Tauri sample, although in the case of VV CrA we note a severe increase to around 0.03, even though the dispersion of the continuum flux values remains comparable to the dispersion found for the other sources.\\
The continuum visibility data in the vicinity of the \bg feature for the different objects in the sample ranges from visibilities of about 0.7 to amplitudes close to 1, indicating that the total line emission region and the surrounding continuum were partially resolved over the entire K-band at baseline lengths between 40 m and 130 m for all of the targets. The \bg visibilities signals are single peaked and in most cases well aligned with the position of the feature in the corresponding spectrum, see, for instance, Fig. \ref{fig:RULup18Data} in Appendix \ref{AppendixIII}. \\ 
There are clear distinctions in terms of visibility signal strength at the \bg line between the different sources. Particularly clear and strong features were observed in both epochs of RU Lup, where the difference between \bg peak and surrounding continuum can be as high as 0.12 (Fig. \ref{fig:RULup21Data}), compared to the relatively faint signals detected at the position of \bg for DoAr 44 with a line to continuum difference in visibility of only about 0.01 at even the longest baselines. \\
Noise levels can equally vary greatly, with low continuum dispersion in wavelength regions adjacent to the \bg line again for RU Lup, where we measure a continuum standard deviation of 0.0023, compared to the very noisy data obtained for VV CrA (Fig. \ref{fig:VVCrAData}) where the continuum standard deviation can go up to 0.035.\\
We found non-zero differential phases at the position of \bg at several baselines for each target in the sample, with typical absolute phase values of around 0.5$^{\circ}$ to 1$^{\circ}$, although the largest measured differential phase goes up to 6.3$^{\circ}$. S-shaped features, as expected for systems in rotation, can be observed for some sources at some baselines (e.g. RU Lup 2021, UT4-UT1, see Fig. \ref{fig:RULup21Data}), but are rather exceptional and notably absent from many other targets. The features appear single peaked in many cases, but the peak is not aligned with the line centre, suggesting that in such cases a potential S-shape might be masked by one of the two constituent peaks being somehow attenuated. \\
The differential phase data is significantly more noisy when compared to the visibilities. Standard deviations in the continuum adjacent to \bg, even for sources with clear differential phase \bg peaks, are on average the order of 0.4$^{\circ}$. For the very noisy data of VV CrA the continuum dispersion even goes up to 2$^{\circ}$ in standard deviation. \\
The closure phases at the position of \bg are consistent with the closure phases of the nearby continuum throughout the sample with no clear \bg signal visible for any of the sources. The continuum closure phases  for most of our objects are close to 0$^{\circ}$, although small offsets of around 1$^{\circ}$ can be detected at certain baseline triangles for AS 353, RU Lup, and DoAr 44 (Fig. \ref{fig:AS353Data}, \ref{fig:RULup21Data}, and \ref{fig:DoAr44Data}, respectively). For VV CrA (Fig. \ref{fig:VVCrAData}) we found again a comparatively large dispersion of closure phase values, with standard deviations across the different triangles reaching up to 4$^{\circ}$. \\
The data were treated by using the total line to continuum flux ratio and continuum quantities to remove the influence of the continuum on the interferometric observables. This way a 'pure line' \bg visibility and differential phase was extracted from the total observables in each of the chosen spectral channels and then further used to determine the size and photocentre of the \bg emission region. Based on a simple geometric Gaussian disk model \citep{Berger2007}, we extracted the half-flux radius (or half width half maximum, HWHM) in each channel, as well as the relative offset of the photocentre with respect to the position of the continuum photocentre.  This treatment was applied to a number of spectral channels based on the previously explained flux selection criterion to analyse the change of those derived properties across the different velocity components of the \bg line, see, for example, Fig. \ref{fig:AS353}. A full explanation of the process is given in Appendix \ref{sec:5}.

\section{Results}\label{sec:6}
\subsection{General results}

\begin{figure}[!h]

\begin{minipage}{\linewidth}
\includegraphics[width=\linewidth]{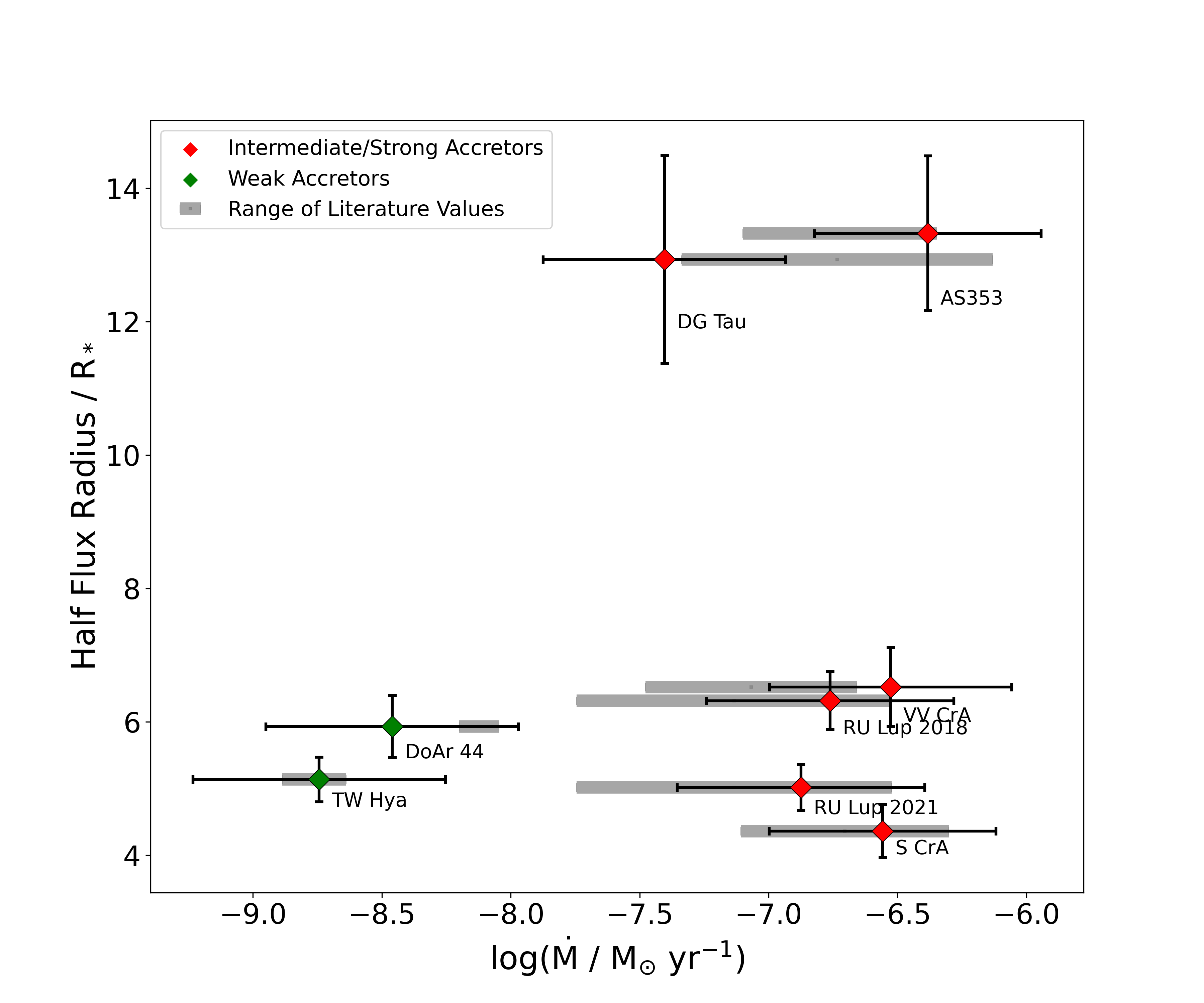}
\end{minipage}

\begin{minipage}{\linewidth}
\includegraphics[width=\linewidth]{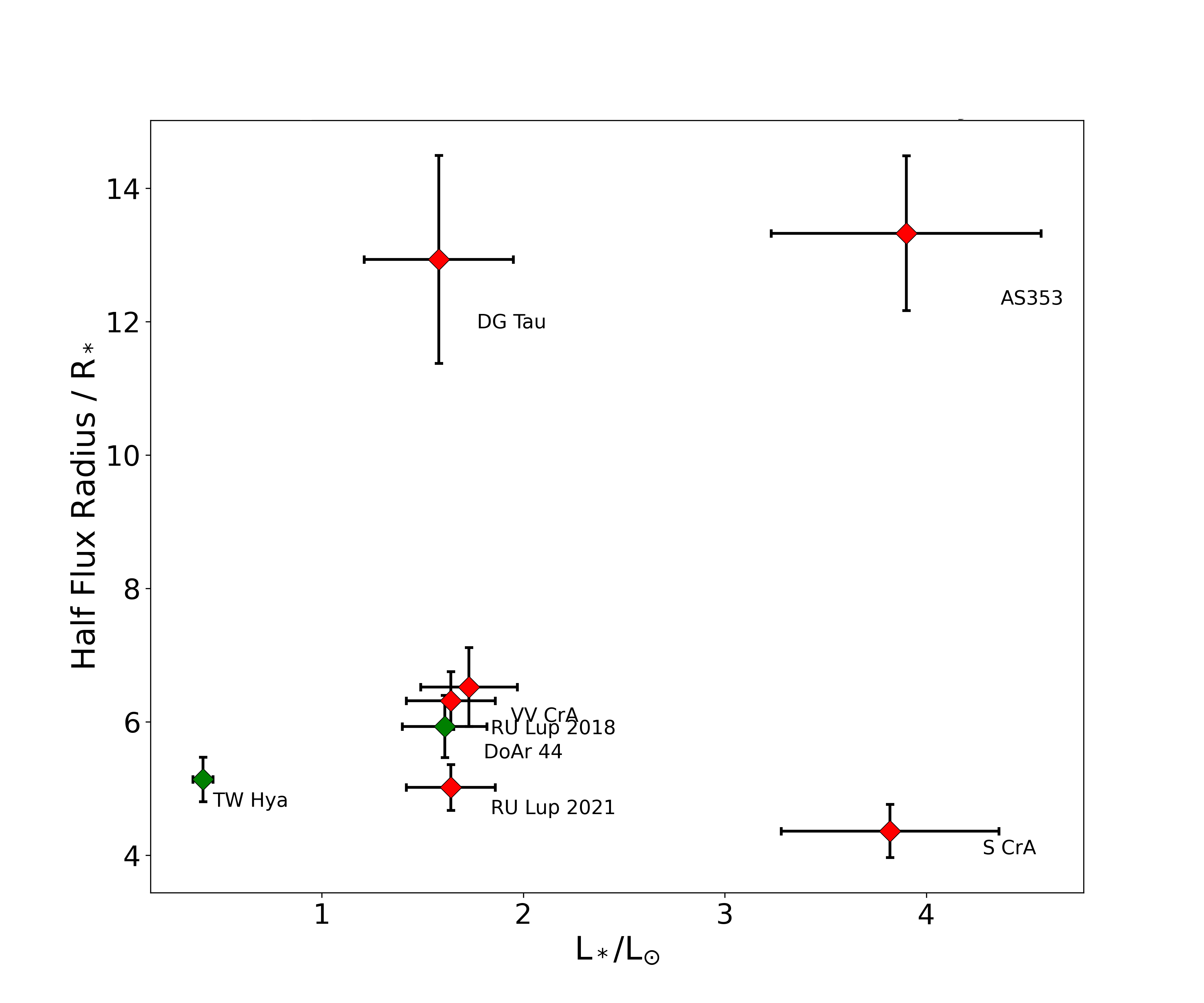}
\end{minipage}
\caption{Relationships between \bg emission region sizes and selected stellar properties. \\ \textbf{Top:} Fitted HWHM of the \bg pure line emission region of the central channel versus the mass-accretion rate. The latter is depicted in two forms: Once as a grey bar which denotes the range of $\dot{M}$ values found in the literature on the respective source, then as the instantaneous mass-accretion rate, which was derived from the equivalent width of the GRAVITY \bg spectra. The colour coding represents the classification of the sample objects as either weak (green) or intermediate to strong accretors (red). Most of the T~Tauri stars cluster in a 4-7 R$_*$ range, which is of the order of typical magnetospheric sizes. \protect\\
\textbf{Bottom:} \bg HWHM versus accretion luminosities. The colour coding corresponds to the upper plot.}
\label{sizelum1}
\end{figure}

\begin{table*}[t!]
\tabcolsep=0.07cm

\footnotesize
\caption{Results for equivalent widths, mass-accretion rates, and characteristic sizes of the \bg emission region.}\label{restab}
\centering
\begin{tabular}{ccccccccc} 
\\
                &       AS353                   &       RU Lup 2018                     &       RU Lup 2021                        &       TW Hya                  &       DG Tau                     &       DoAr 44                 &       S CrA                   &       VV CrA                     \\ \hline \hline \\
                                                                                                                                                                                                                                                                                
$\mathrm{F_{L/C}}$              &       1.91                    &       1.86                    &       1.83                    &       1.61                    &       1.29                    &       1.19                    &       1.53                    &       1.65                    \\[0.1cm]
$\mathrm{FWHM\ [km \  s^{-1}]}$         &       233                     &       183                     &       187                     &       121                     &       224                     &       137                     &       175                     &       150                     \\[0.1cm]
$\mathrm{W10\%  \ [km \ s^{-1}]}$               &       594                     &       607                     &       420                     &       241                     &       445                     &       274                     &       524                     &       416                     \\[0.1cm]
$\mathrm{\mu_{HV} \ [km \ s^{-1}]}$             &       -97                     &       -55                     &       -50                     &       -183                    &       -                       &       -                       &       -273                    &       -75.49                  \\[0.1cm]
$\mathrm{\mu_{LV} \ [km \  s^{-1}]}$            &       3                       &       2                       &       6                       &       1                       &       -                       &       -                       &       3                       &       0.84                    \\[0.1cm]
$\mathrm{FWHM_{HV}\ [km \  s^{-1}]}$            &       520                     &       586                     &       295                     &       482                     &       -                       &       -                       &       549                     &       598.54                  \\[0.1cm]
$\mathrm{FWHM_{LV} \ [km \ s^{-1}]}$            &       204                     &       154                     &       154                     &       116                     &       -                       &       -                       &       170                     &       141.32                  \\[0.1cm]
EW [$\AA$]              &       -19.62  $\pm$   0.91    &       -16.82  $\pm$   1.76    &       -13.47  $\pm$   0.43    &       -7.11   $\pm$   0.76    &       -5.80   $\pm$   0.35    &       -2.65   $\pm$   0.38    &       -8.71   $\pm$   0.66    &       -10.09  $\pm$   0.99    \\[0.1cm]
$\mathrm{L_{acc} \ [L_{\odot}]}$                &       3.07    $\pm$   0.17    &       0.98    $\pm$   0.12    &       0.76    $\pm$   0.03    &       0.03    $\pm$   0.00    &       0.33    $\pm$   0.02    &       0.07    $\pm$   0.01    &       1.74    $\pm$   0.15    &       1.64    $\pm$   0.19    \\[0.1cm]
$\mathrm{log\dot{M}_{acc,inst} \ [M_{\odot} yr^{-1}]}$          &       -6.38   $\pm$   0.44    &-6.76  $\pm$   0.48    &       -6.87   $\pm$   0.48    &-8.74  $\pm$   0.49    &       -7.40   $\pm$   0.47    &       -8.46   $\pm$   0.49    &       -6.56   $\pm$   0.44    &       -6.52   $\pm$   0.47    \\[0.1cm]
$\mathrm{ \dot{M}_{acc,inst} \ [10^{-8} \ M_{\odot} yr^{-1}]}$  &       41.48$_{-20.16}^{+41.35}$       &       17.31$_{-8.54}^{+20.78}$        &        13.38$_{-6.66}^{+15.79}$        &       0.18$_{-0.1}^{+0.32}$   &       3.95$_{-2}^{+4.44}$         &       0.35$_{-0.19}^{+0.38}$  &       27.72$^{+28.96}_{-13.57}$         &       29.77$^{+34.39}_{-14.59}$        \\ \\ \hline \\
$\mathrm{PA \ [^\circ]}$                &       173     $\pm$   3       &       99      $\pm$   31      &       101     $\pm$   31      &       130     $\pm$   32      &       143     $\pm$   12      &       137     $\pm$   4       &       1       $\pm$   6       &       91      $\pm$   6       \\[0.1cm]
$\mathrm{i \ [^\circ]}$         &       41      $\pm$   2       &       16$^{+6}_{-8}$                  &       20$^{+6}_{-8}$                  &       14$^{+6}_{-14}$                 &       49      $\pm$   4       &       32      $\pm$   4       &       27      $\pm$   3       &       32      $\pm$   3       \\[0.1cm]
$\mathrm{Continuum \ HWHM \ [au]}$              &       0.28    $\pm$   0.05    &       0.21    $\pm$   0.06    &       0.21    $\pm$   0.06    &       0.042   $\pm$   0.003   &       0.13    $\pm$   0.01    &       0.16    $\pm$   0.02    &       0.17    $\pm$   0.02    &       0.16    $\pm$   0.01    \\[0.1cm]
$\mathrm{\bg \ HWHM \ [au]}$            &       0.206   $\pm$   0.003   &       0.076   $\pm$   0.001   &       0.061   $\pm$   0.001   &       0.029   $\pm$   0.001   &       0.143   $\pm$   0.004   &       0.050   $\pm$   0.002   &       0.071   $\pm$   0.011   &       0.081   $\pm$   0.005   \\[0.1cm]
$\mathrm{R_{tr} \ [au]}$                &       0.04    -       0.08    &       0.04    -       0.07    &       0.04    -       0.08    &       0.05    -       0.07    &       0.05    -       0.09    &       0.09    -       0.11    &       0.05    -       0.10    &       0.04    -       0.07    \\[0.1cm]
$\mathrm{R_{co} \ [au]}$                &       0.10                    &       0.04                    &       0.04                    &       0.05                    &       0.06                    &       0.05                    &       0.04                    &       0.10                     \\ \\ \hline \\
$\mathrm{Continuum \ HWHM \ [R_*]}$             &       18.13   $\pm$   3.59    &       17.36   $\pm$   5.10    &       17.36   $\pm$   5.10    &       7.47    $\pm$   0.70    &       11.32   $\pm$   1.61    &       19.07   $\pm$   2.69    &       10.38   $\pm$   1.42    &       12.64   $\pm$   1.04    \\[0.1cm]
$\mathrm{\bg \ HWHM \ [R_*]}$           &       13.32   $\pm$   1.16    &       6.32    $\pm$   0.43    &       5.01    $\pm$   0.35    &       5.13    $\pm$   0.33    &       12.93   $\pm$   1.56    &       5.93    $\pm$   0.47    &       4.36    $\pm$   0.75    &       6.56    $\pm$   0.59    \\[0.1cm]
$\mathrm{R_{tr} \ [R_*]}$               &       2.83    -       5.30    &       3.28    -       6.14    &       3.53    -       6.61    &       8.31    -       12.35   &       4.50    -       8.42    &       10.62   -       12.68   &       3.36    -       6.29    &       2.86    -       5.36    \\[0.1cm]
$\mathrm{R_{co} \ [R_*]}$               &       6.47                    &       3.31                    &       3.31                    &       8.89                    &       5.44                    &       5.96                    &       2.44                    &       7.25                    \\\\
\hline 
\end{tabular}   
\label{table:results}
\tablefoot{F$_{L/C}$ is the total line to continuum flux ratio at the centre of the line. FWHM and W10$\%$ are the line widths at 50\% and 10\% of the peak flux, respectively. $\mu$ is the position of the high (HV) or low velocity (LV) component from the double Gaussian fit to the spectrum.  \\
The \bg half-flux radius (HWHM) is given for the central channel. R$_{co}$, continuum HWHM, PA, and i were adopted from \citet{Perraut2021}. Accretion luminosities were estimated from the line luminosities based on the empirical relationship in Eq. \ref{eq:alcala}. The instantaneous mass-accretion rates were then computed as in Eq \ref{eq:macc}. R$_{tr}$ is given based on Eq. \ref{eq:hartmann16} for magnetic field strengths between 1 kG to 3 kG and using the mass-accretion rates derived from the equivalent widths (EW). The asymmetric error bars on the absolute values of $\mathrm{\dot{M}_{acc,inst}}$ represent the 25th and 75th percentiles of the distribution. Other error bars denote 1$\mathrm{\sigma}$ uncertainties.}
\end{table*}

\begin{figure*}[!h]

\begin{minipage}{\linewidth}
\includegraphics[width=\linewidth]{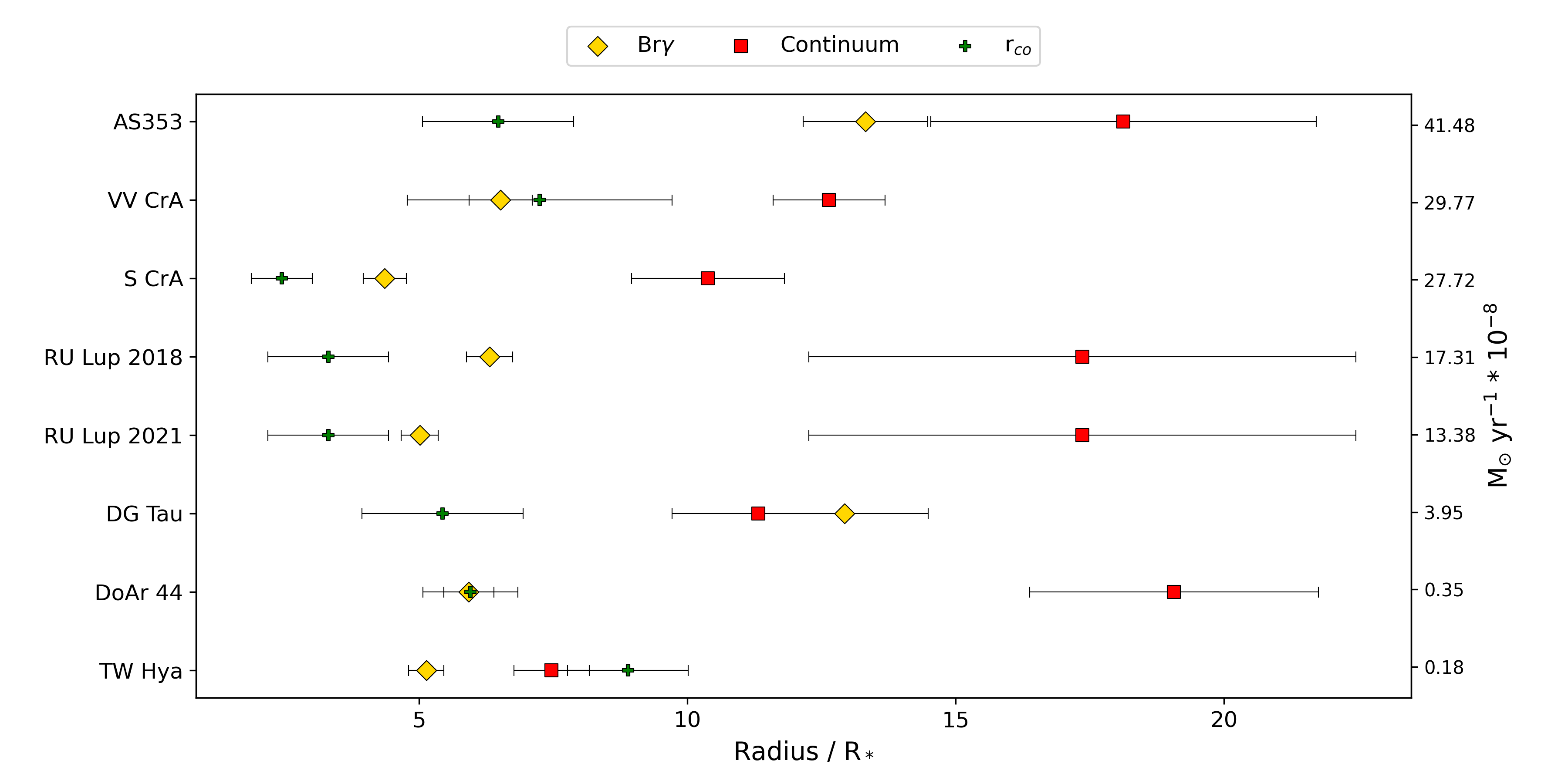}
\end{minipage}

\begin{minipage}{\linewidth}
\includegraphics[width=\linewidth]{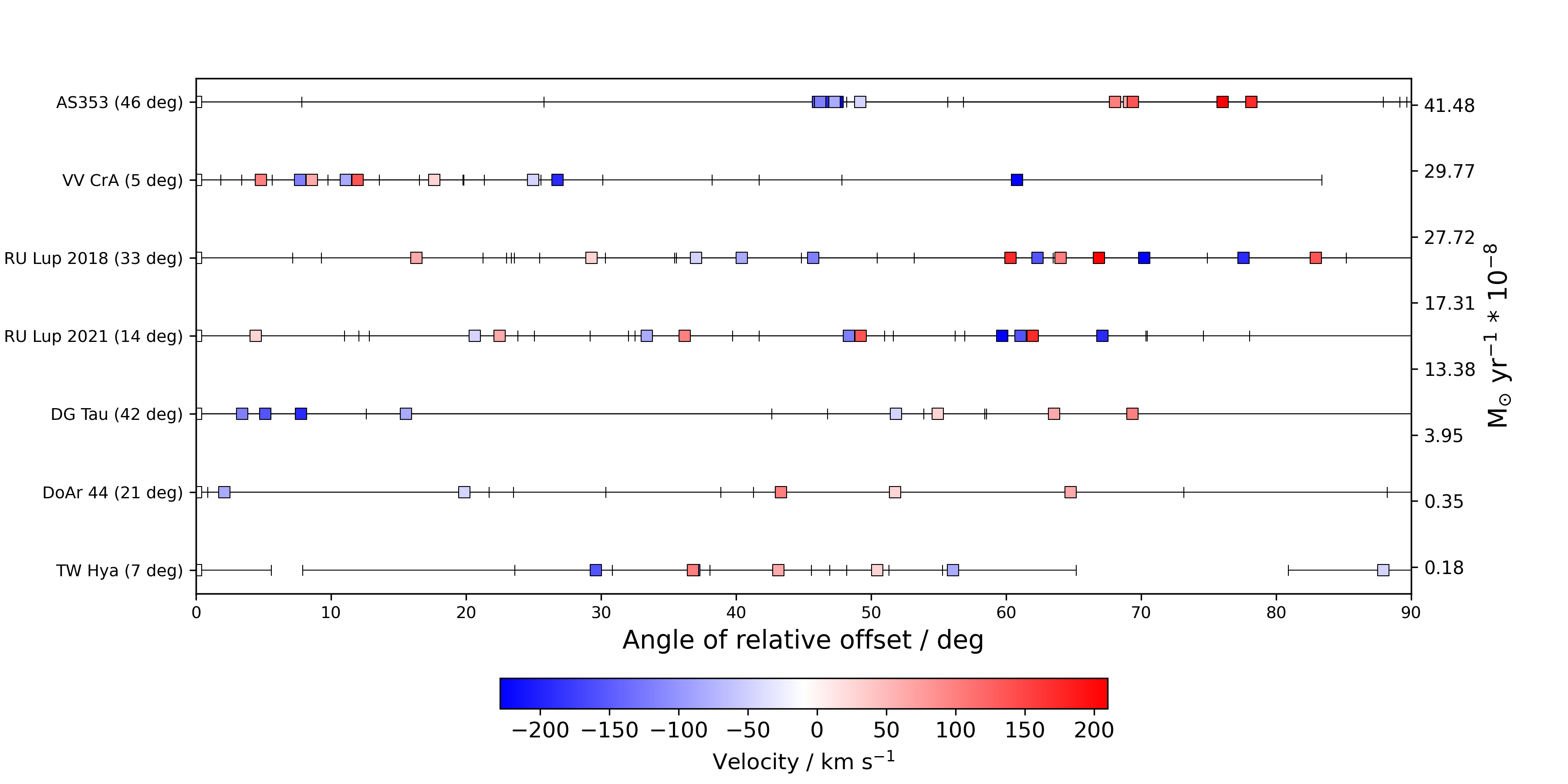}
\end{minipage}

\caption{ Sample overview for emission region sizes and photocenter shifts. \\
\textbf{Top:} \bg half-flux radii (yellow), compared to K-band NIR continuum half-flux radii (red) and co-rotation radii (green) across the sample. We note that \bg region and continuum HWHMs are based on a Gaussian disk and a Gaussian ring model, respectively, which should approximately reflect the differences in the morphology of both regions. The objects are given in order of descending mass-accretion rate.
\protect\\
\textbf{Bottom:} The distribution of photocentre shift angles across the line for the entire sample. The x-axis shows the minimum difference angle between the relative shift vector and the position angle of the continuum disk. The white channel serves as the point of reference for the relative shift vector. 
An angle of 0$^{\circ}$ means the photocentre shift vector is aligned with the disk axis, as would for example be expected for a disk in rotation. A difference angle of 90$^{\circ}$ indicates that the shift vector is perpendicular to the disk axis. A clustering of points close to a certain difference angle indicates an alignment of the photocentre shifts along a preferential axis. VV CrA shows a monoaxial distribution of photocentres, while for other targets more complex profiles with multiple alignments are indicated. The number stated in brackets after the object name gives the average error on the angle for the five central channels. }
\label{sampleplot}
\end{figure*}

In this section we first present a breakdown of the general results obtained from the GRAVITY T~Tauri data, introduce some additional concepts relevant to the interpretation, and give an overview over the global trends we were able to identify across the entire sample. In the following sections, we then proceed to go over each source individually in detail and contextualise our findings by drawing upon relevant references, if and where available. Table \ref{table:results} summarises key results concerning the characteristics of the \bg emission region and accretion properties, respectively. Figures \ref{sizelum1} and \ref{sampleplot} serve as a visual overview of the sample results with regards to the most important criteria we use to interpret our results. \\
The sample consists of mostly strong \bg emitters, with normalised peak line to continuum flux ratios between 1.91 and 1.53 for six out of our eight data sets and ranging down to 1.19 for the weakest emitter DoAr 44. Many of the objects exhibit line shapes that are asymmetric, featuring an excess of blueshifted emission. We approximate this asymmetry by fitting a superposition of two Gaussian functions to the data, basically dividing the feature into a narrow centred low velocity component and a broader offset high velocity component. Whether these Gaussian components can necessarily be directly attributed to specific physical phenomena, such as additional outflows on top of an accreting magnetosphere, is not certain. Inverse P~Cygni features, as expected to be seen when observing magnetospheric accretion columns at low inclination, are generally absent from the line profiles of the sample, even for those objects observed at a close to pole-on configuration, such as TW Hya. 
The width and offset of the broad component, and thus of the total emission line, can vary significantly throughout the sample, even between two epochs of the same target (RU Lup), although the overall strength of the asymmetry correlates with the peak flux ratio and is also reflected in the equivalent widths.
We computed the line luminosities for each object from the equivalent widths and K-band magnitudes from 2MASS \citep{2mass}. We then used the empirical relationship from \citet{Alcala2014},
\begin{align}
\label{eq:alcala}
log\left( \frac{L_{acc}}{L_{\odot}} \right)&= a \cdot log \left( \frac{L_{Line}}{L_{\odot}}\right) +b
\end{align}
with a=1.16 $\pm$ 0.07  and b=3.6 $\pm$ 0.38, to estimate the accretion luminosities and subsequently determined the corresponding mass-accretion rates via \citep{Hartmann1998}
\begin{align}
\label{eq:macc}
\dot{M}_{acc} & = \left( 1 - \frac{R_{*}}{R_{in}}\right)^{-1}  L_{acc} \frac{R_*}{G M_*}\approx 1.25 L_{acc} \frac{R_*}{G M_*},
\end{align}
which describes the relationship between the gravitational energy released and radiated away by the mass falling in from the truncation radius. For the right hand side approximation we follow the common convention of using R$_{in}$ $\approx$ 5 R$_*$ as a typical size for truncation radii. \\
These results are summarised in Table \ref{table:results}. They are also shown in Fig. \ref{sizelum1}, where we present both the instantaneous mass-accretion rates as well as a range based on the lowest and highest previous estimates found in the literature. Based on this comparison, we found that our mass-accretion rates tend to fall into the range established by past observations, even if mostly towards the higher end for the stronger accretors. The uncertainties on the coefficients a,b of Eq. \ref{eq:alcala} translate into relatively large and asymmetric error bars on the absolute value of $\mathrm{L_{acc}}$ and consequently also $\mathrm{\dot{M}_{acc}}$. 
We used a Monte-Carlo approach to determine the final uncertainties on the accretion rate, rather than Gaussian error propagation not applicable here.
\\
We fitted the visibility data  with a 2D Gaussian in order to find the half-flux radii (HWHM) of the pure \bg emission region, using the inclination and position angles of the K-band continuum emission region presented in \citet{Perraut2021} to constrain those two parameters in our fit.
The HWHMs we obtained through this process are, across the entire sample, of the order of less than 0.21 au at their greatest extent, ranging down to regions as compact as 0.03 au. In terms of stellar radii, the central channel sizes vary from up to 13 R$_*$ to less than 5 $R_*$, although the majority of objects show clustering (see Figures \ref{sampleplot}, \ref{sizelum1}) in a range from 4 to 7 R$_*$, which is fairly typical for magnetospheric radii \citep{Bouvier2007}. This clustering can be observed across the entire range of mass-accretion rates of our small sample, so that a correlation between those sizes and accretion rates cannot be established here. Equally, within our sample we do not detect a clear correlation of those \bg sizes with the luminosity of the objects, as opposed to the NIR continuum sizes, which were shown to correlate as per R $\propto$ L$^{1/2}$ \citep{Perraut2021}. However, it is apparent from Fig. \ref{sampleplot} that the two weakest accretors in the sample (TW Hya and DoAr 44) also show most clearly an emission region smaller than the co-rotation radius, whereas most of the other objects show \bg emission coming from beyond the co-rotation radius. \\
Changes in HWHM in the spectral channel across the \bg line can be significant, with largest to smallest size-in-channel ratios being around 1.2 to 1.5 for many of the objects, although in extreme cases this can go up to almost 2, whilst others show a relatively flat profile relative to their error bars. Under a magnetospheric accretion scenario, a decrease of size towards the edges of the line (i.e. higher velocities) would be naively expected, as the gas at the highest velocities should be in free fall close to the star. In our sample, only one source (TW Hya) clearly shows this decrease, whereas most of the other objects show either a straight increase with a minimum at the centre channel or even a mixed profile with a decreasing size at low velocities and then an increase as we move to higher velocities. These mixed profiles could be indicative of a multi-component \bg emission region, although in many cases the extent of these more complex signatures is within the uncertainties.\\
An important point of reference in the search for the origin of \bg emission is the co-rotation radius, since we know that the inner disk must be truncated inside the R$_{co}$ so that stable magnetospheric accretion columns can form. If the dominant driver of \bg emission is magnetospheric accretion, we expect that the \bg emission region would be of the order of the co-rotation radius or even slightly more compact. The co-rotation radius is defined as the radius at which the angular velocity of the rotating disk matches the angular velocity of the star $\Omega_*$ \citep{Bouvier2013}:
\begin{align}
R_{co} &= \left(\frac{G M_*}{\Omega^2_*} \right)^{1/3}.
\end{align}
Stellar angular velocities are determined from measurements of the rotational period of the star or the v sin(i). We refer to \citet{Perraut2021} for the relevant values for our sample objects. The co-rotation radii are given in Table \ref{table:results}.
Of course, the best way to determine whether the size of the \bg emission region corresponds to the size of the magnetosphere in any object would be to compare the region size against the magnetospheric truncation radius directly. The truncation radius can be computed from stellar parameters, if the strength of the magnetic dipole field is known:
\begin{align}
\label{eq:hartmann16}
R_{tr} &= 12.6 \frac{B_3^{4/7}R_2^{12/7}}{M_{0.5}^{1/7}\dot{M}_{-8}^{2/7}}. 
\end{align}
In this notation we follow the convention used in \citet{Hartmann2016}, for example. We note that $M_{0.5}$ is the stellar mass in units of 0.5 M$_{\odot}$, R$_2$ the stellar radius in units of 2 R$_{\odot}$, B$_3$ the surface field strength of the dipolar magnetic field at the stellar equator in kG, and $\dot{M}_{-8}$ the mass-accretion rate in units of $\mathrm{10^{-8} M_{\odot} yr^{-1}}$. The resulting truncation radius is given in units of R$_{\odot}$. The derivation is conceptually discussed in \citet{Bessolaz2008}. \\ 
Measurements of magnetic field strengths in T~Tauri stars are scarce, so that we cannot rely on the truncation radius as a primary criterion to identify strong magnetospheric accretion cases. Even so, we can still use general estimates of R$_{tr}$ as complementary information. Building upon the approach laid out in \citet{Perraut2021}, we set a range of 1 - 3 kG for B and compute the truncation radii for this range as presented in Table \ref{table:results} from the relevant stellar properties.\\
An overview of the emission region sizes in the central line channels, compared to the respective co-rotation radii and the NIR K-band continuum sizes taken from \citet{Perraut2021}, can be found in Fig. \ref{sampleplot}. The sample can generally be divided into targets with emission regions of the order of the co-rotation radius (DoAr 44, TW Hya), those with more extended \bg emission than the co-rotation radius, but still close to typical magnetospheric radii around 5 R$_*$ (RU Lup, S CrA N, and VV CrA SW), and those with highly spatially extended \bg regions outside the typical range of magnetospheric sizes (AS 353, DG Tau). This separation partially corresponds to the distinction between weak accretors ( $\dot{M}_{acc}$ < 1$\cdot$ 10$^{-8} M_{\odot} yr^{-1}$, again DoAr 44 and TW Hya) and the intermediate to strong accretors, which we indicated in Fig.\ref{sizelum1} via the different colours. While the weak accretors do appear to conform to the co-rotation criterion with compact emission regions, the situation is more complex for objects with higher accretion rates, as there is no clear correlation between the relative region sizes and the mass-accretion rates for those targets.  \\
Magnitudes for the observed photocentre displacements are of the order of 0.07 au or below. For multiple targets the entire photocentre profile across the line is sufficiently compact to be within or close to 1 R$_*$, although such a compact concentration of photocentres does not necessarily correlate with the size of the emission region.  
The photocentre profiles are complex and not easily classified, although a broad distinction can be made between targets that show quasi-rotational profiles, featuring distinct blue and red arms aligned along a common axis, targets with compact profiles that lack any clear structures, and complex profiles which again appear potentially as a superposition of multiple components. 
In Fig. \ref{sampleplot}  we give an overview over the possible alignment of the photocentre shifts across the line with different axes. To account for the offset of the profiles from the plot centre, we compute the difference vectors between each point and the centre channel (shown as a white square) and then compute the minimum angular difference between the angle of the difference vector and the NIR continuum disk axis position angle. The resulting value is a measure of how well aligned the photocentre offsets are relative to the disk axis. The plot can be interpreted in a meaningful way by considering the degree of clustering, and at which angles the clustering occurs. A concentration of points in a narrow angular range (AS 353) indicates that the photocentre shift is aligned along a single axis. If the clustering range is close to zero, the photocentres are aligned with the disk semi-major axis, which is something that would be expected from a disk in Keplerian rotation. If the points cluster at 90$^{\circ}$, the photocentres might be aligned with some form of polar outflow (red wing in DG Tau). An even spread of points indicates an absence of recognisable structures and a random distribution of photocentres (TW Hya). The figure clearly shows the previously introduced distinction between simple alignments and complex patterns, where we see multiple clusters (DG Tau and RU Lup).
\\It is important to remember that the differential phases track the relative offset of the \bg emission region from the overall photocentre of the continuum. In Fig. \ref{fig:AS353}, for example, the centre of the plot is the combined photocentre of the star and other continuum emitting components such as the dusty disk. If the dusty disk is perfectly centrosymmetric, its own photocentre  coincides with that of the star and the overall continuum photocentre matches the position of the star. If the continuum photocentre, however, is not centrosymmetric, either offset from the star or with an asymmetric brightness distribution, then the zero position in those plots can significantly deviate from the stellar position. We therefore make use of the closure phases to extract information about the continuum asymmetry. For the majority of our sample objects the closure phases themselves are  close to zero and thus suggest a centrosymmetric continuum. We then instead use the uncertainty given by the phase dispersion  to derive an uncertainty on the stellar position. To do this we follow \citet{Perraut2021} by fitting an azimuthally modulated ring model to the continuum closure phase data. The results are depicted as red circles on the photocentre profile plots, showing that indeed in the majority of the cases the centre channel photocentre could conceivably be centred on the stellar position. \\

\subsection{Individual results}
\subsubsection{AS 353}
\begin{figure}[!h]
    \centering
\begin{minipage}{\linewidth}
\centering
\includegraphics[width=\linewidth]{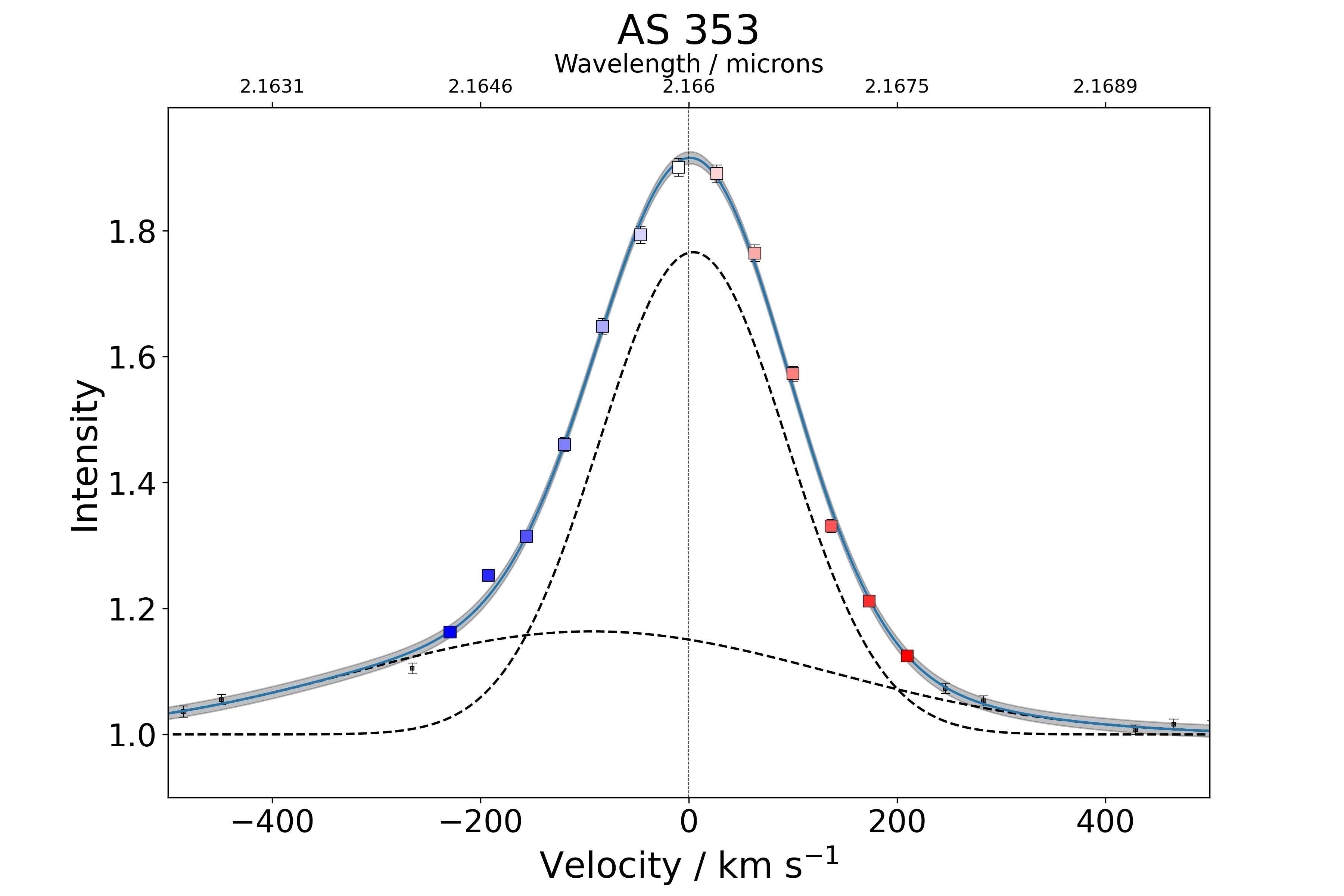}
\end{minipage}
\begin{minipage}{\linewidth}
\centering
\includegraphics[width=\linewidth]{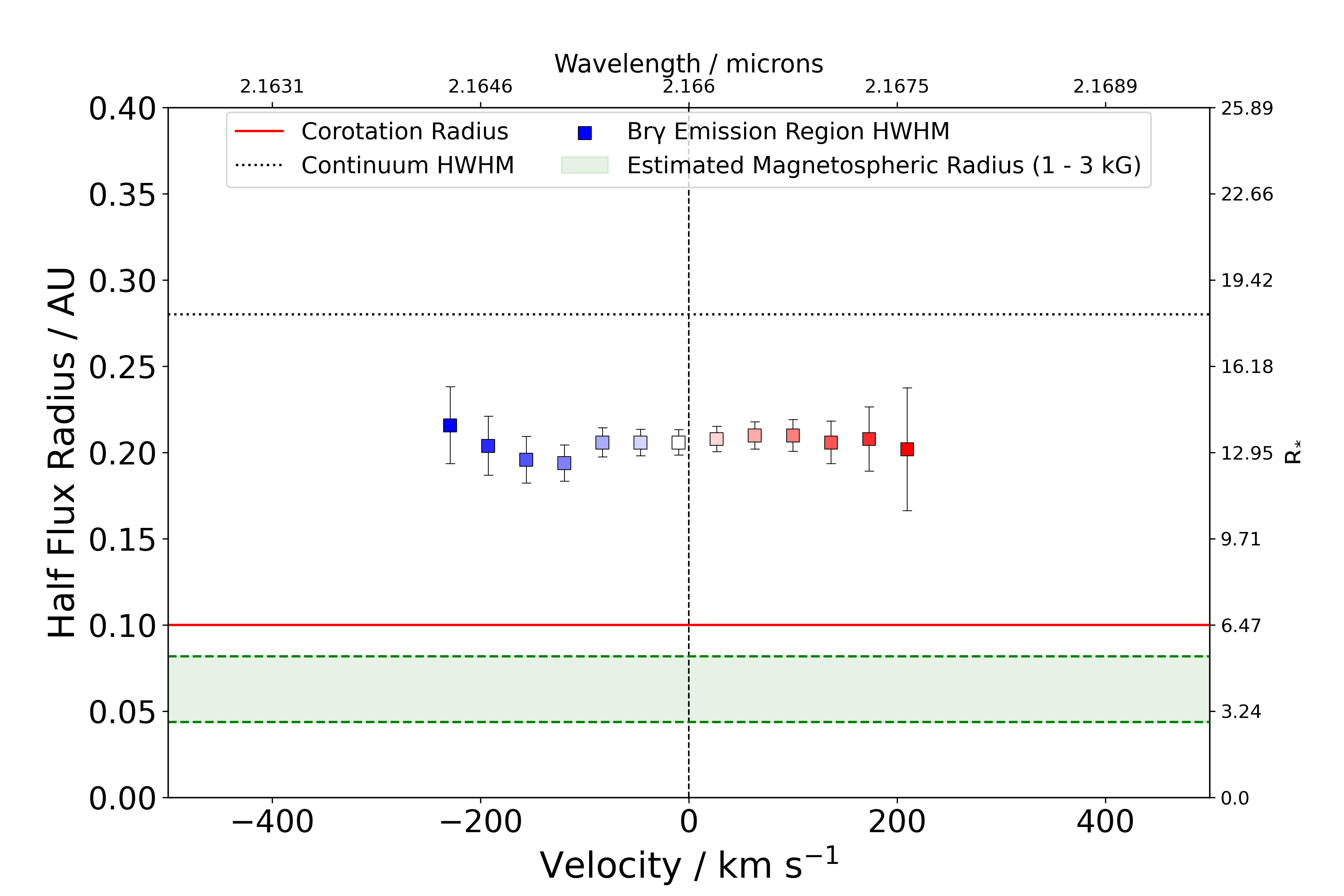}
\end{minipage}

\begin{minipage}{1\linewidth}
\centering
\includegraphics[width=1\linewidth]{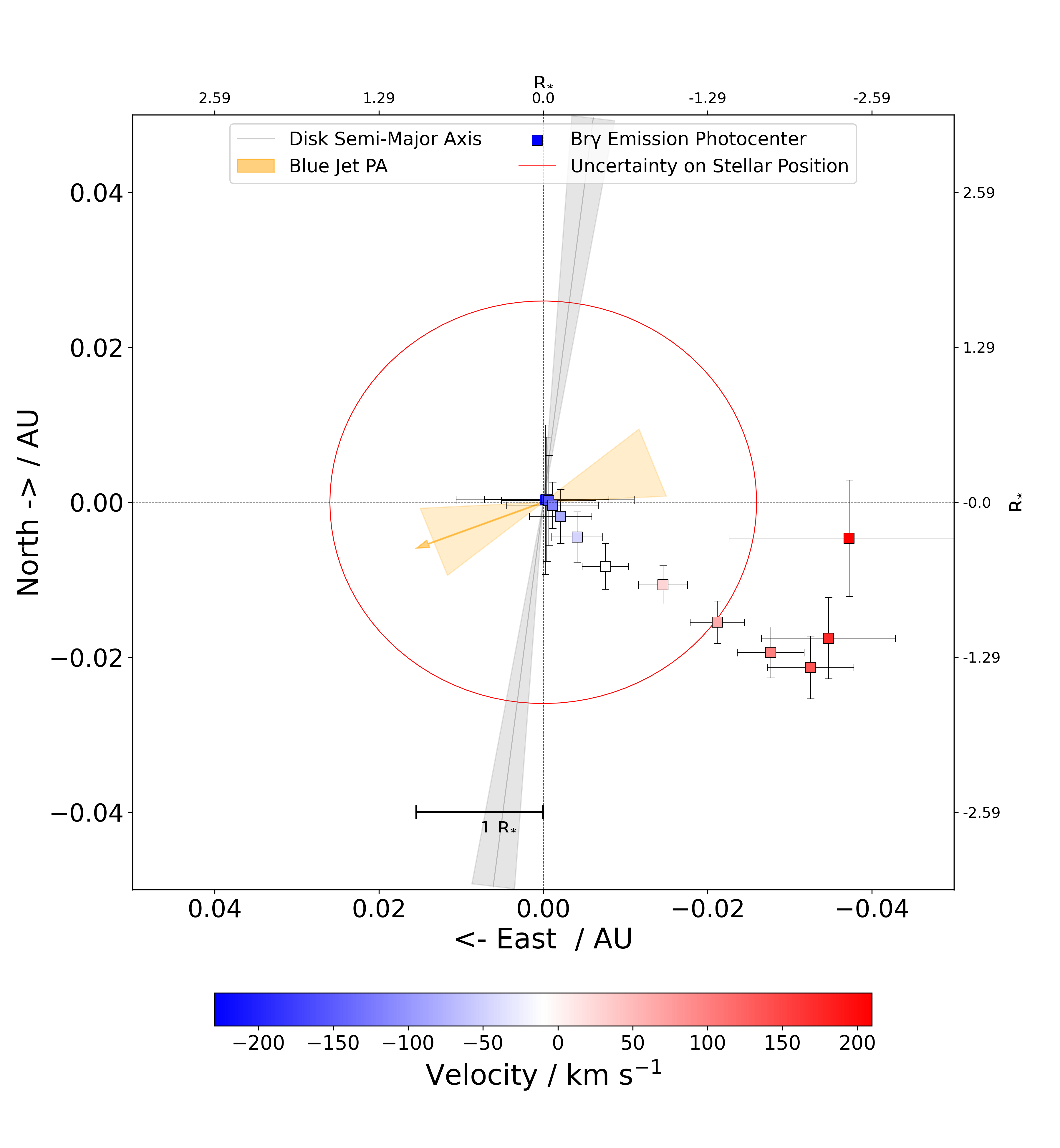}
\end{minipage}
    \caption{Spectrum (top), size (middle), and photocentre shift (bottom) profiles for AS 353. Shaded regions indicate uncertainties.}
    \label{fig:AS353}
\end{figure}

With a peak line to continuum flux ratio of 1.91 and an equivalent width of -19.6 \AA, AS 353 (Fig. \ref{fig:AS353}) is the strongest \bg emitter in the GRAVITY T~Tauri sample. These measurements are consistent with the results of a spectroscopic survey of young binaries done in 1996 by \citet{Prato2003}, who found an equivalent width of -21.1 \AA \ for the \bg emission line of this object. We do not find a significant change in size across the \bg line for the emission region, with all channels appearing consistent within their respective uncertainties with a half flux radius of 0.206 au at the centre of the line. This puts the emission region at about twice the size of the co-rotation radius at 0.1 au, but still well within the NIR continuum HWHM of 0.28 au. However, when comparing these quantities, it is important to point out a certain amount of confusion surrounding the distance estimate to the AS 353 system. Since the system is thought to be connected with the Aquila star forming region \citep{Rice2006} and thus to the complex of dark interstellar clouds of the Aquila rift, accurate measurements of the distance have historically been difficult to obtain. Previous works such as \citet{Edwards1982} refer to a distance of 150 pc, which was also adopted by \citet{Prato2003} with a estimated error of 50 pc based on the idea that AS 353 is in front of the Aquila rift at 200 $\pm$ 100 pc \citep{Dame1985}. More recent high accuracy parallax measurements of AS 353 taken by GAIA and released as part of DR2 and DR3 put the binary at a comparatively remote 400 pc \citep{GaiaDR3}, which is the value we adopt in this paper. This is supported by other recent studies covering parts of the Aquila rift, such as the work of \citet{Ortiz2017}, which presents astrometric measurements that put the WS40/Serpens cloud complex at 436 $\pm$ 9.2 pc. Still, even relatively recent publications like the interferometric study of \citet{Eisner2014} use the 150 pc estimate to convert their angular scales into physical sizes. For the \bg emission region size \citet{Eisner2014} report an angular ring diameter of 1.06 mas and a subsequent ring radius of 0.08 au, which, if converted to the more recent distance measurements, would translate to a radius of 0.21 au. This, within a 1$\sigma$ error bar, agrees with our findings, despite the fact that their model does not take into account the relatively high inclination of 41$^{\circ}$ of the inner disk of AS 353. Conversely, if our results were to be converted into a physical size based on a 150 pc distance, we would obtain a half flux radius of  0.077 au, meaning something more compact than the co-rotation radius. However, based on the fact that the most recent distance measurements for AS 353 independently came to similar results, we consider the 400 pc estimate to be robust. As such, the very large extent of the emission region indicates that the \bg emission is likely produced through some form of disk wind outflow launched from the inner gaseous disk. \\
This could be supported by the spatial profile of the \bg line photocentres, which shows similarity to a rotational profile, albeit not aligned with the disk and severely truncated in the blue wing. It is possible that such an asymmetric structure, with a maximum photocentre displacement of 0.039 au, or 2.5 $R_*$, in the red wing, is caused by the interplay of disk and jet. The AS 353 system is connected to the Herbig-Haro object HH32 and drives a large-scale outflow at a very high inclination of 70$^{\circ}$ \citep{Curiel1997} with a fainter blue component and multiple more prominent redshifted knots. At a smaller scale, H$\alpha$ emission originating from a blueshifted jet was reported at a position angle of (111 $\pm$ 18)$^{\circ}$ by \citep{Takami2003}. This is, within the respective uncertainties, close to perpendicular to the NIR continuum disk axis of (173 $\pm$ 3)$^{\circ}$ \citep{Perraut2021}.  \\ 
Assuming that the weaker emission in the blue knot observed in HH32 translates to a similar asymmetry for the small-scale jet, a superposition of two distinct photocentre shift profiles, one aligned with the jet axis and the other coming from the rotating base of a disk wind and thus aligned with the disk semi-major axis, could lead to the observed distribution of photocentres. This is also supported by the closure phase data, which indicates that the position of the star is consistent with the centroid of the central velocity channel within an uncertainty of 0.026 au.

\subsubsection{RU Lup}

\begin{figure*}[h!]

\begin{minipage}{0.49\linewidth}
\centering
\includegraphics[width=\linewidth]{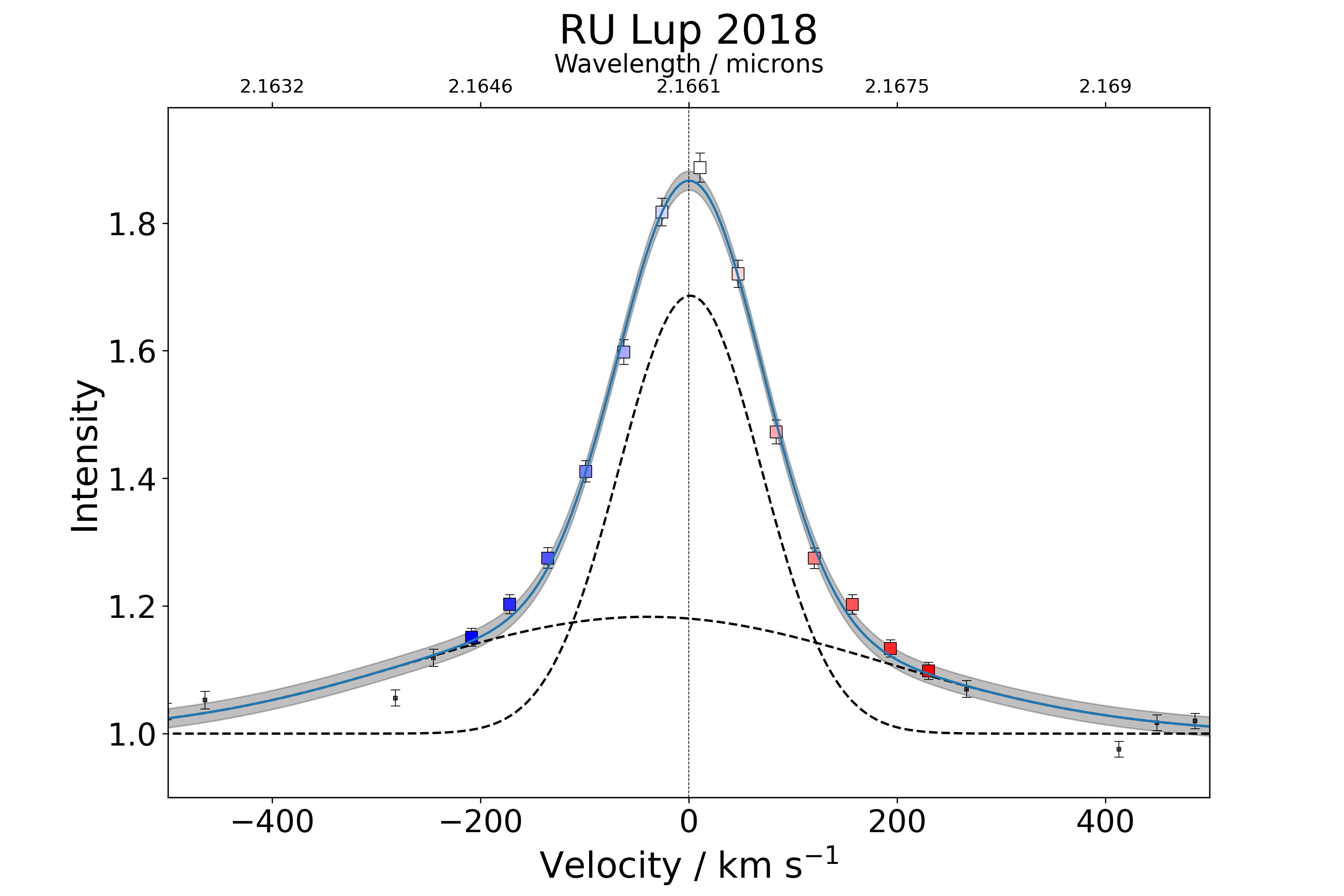}
\end{minipage}
\hfill
\begin{minipage}{0.49\linewidth}
\centering
\includegraphics[width=\linewidth]{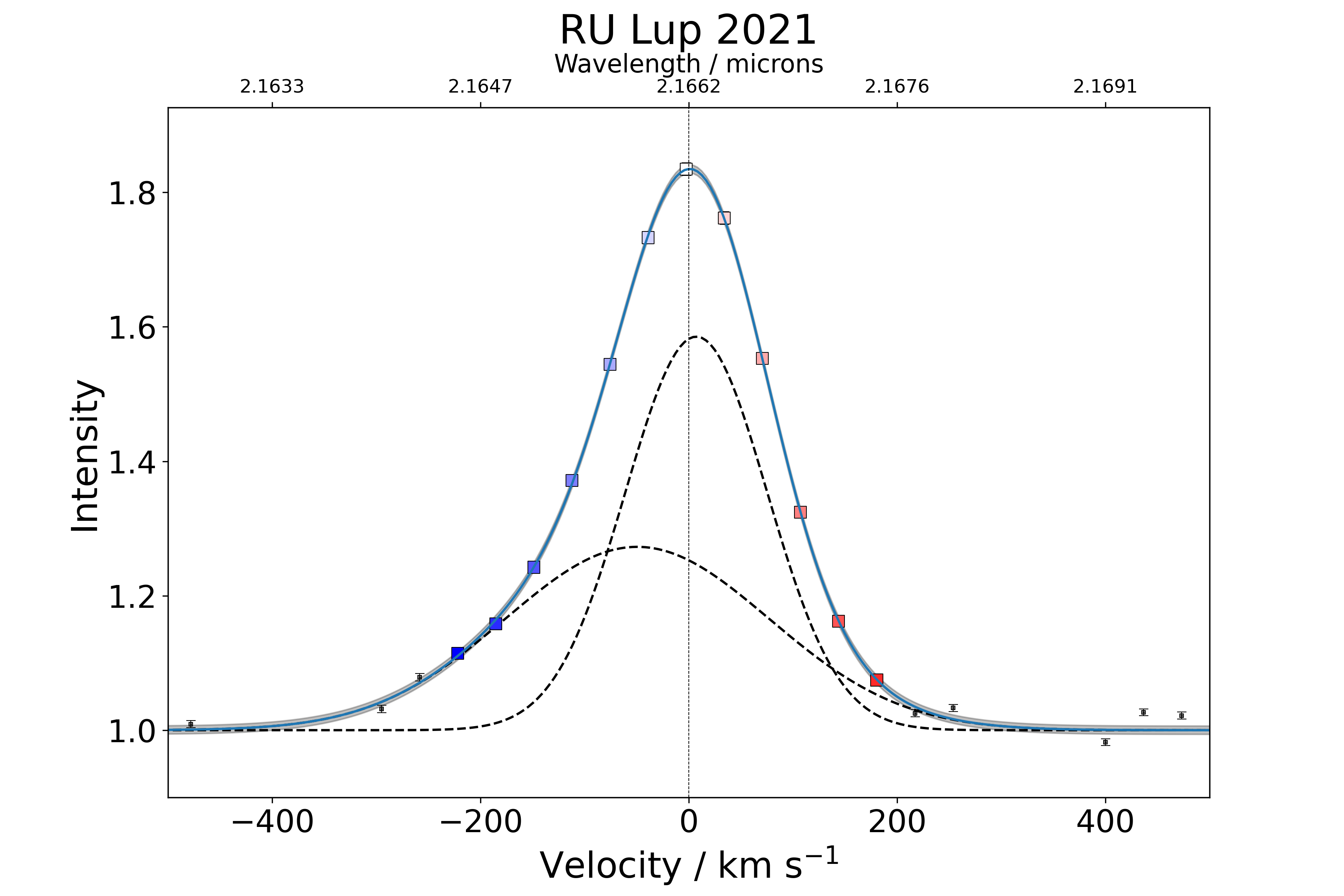}
\end{minipage}

\begin{minipage}{0.49\linewidth}
\centering
\includegraphics[width=\linewidth]{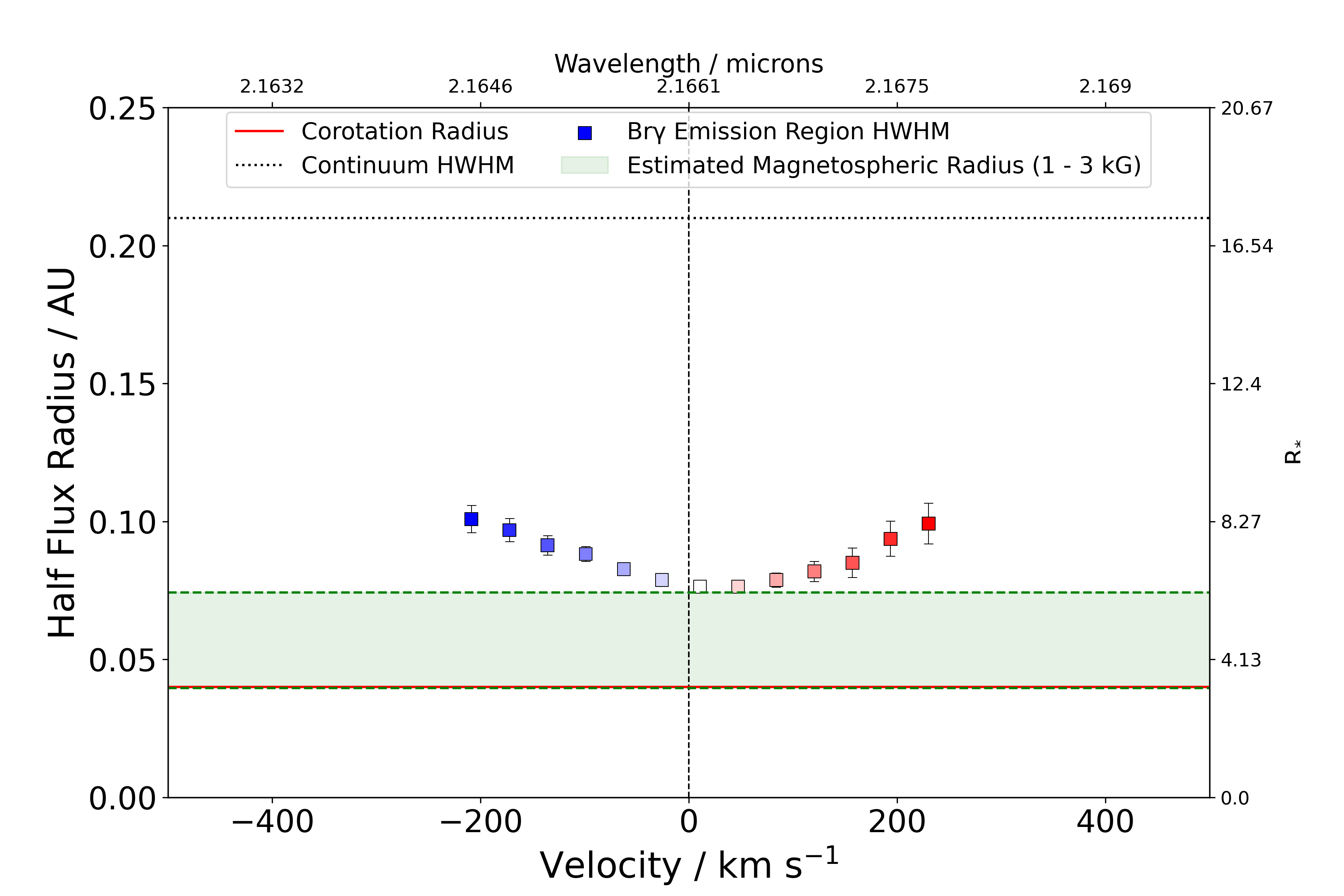}
\end{minipage}
\hfill
\begin{minipage}{0.49\linewidth}
\centering
\includegraphics[width=\linewidth]{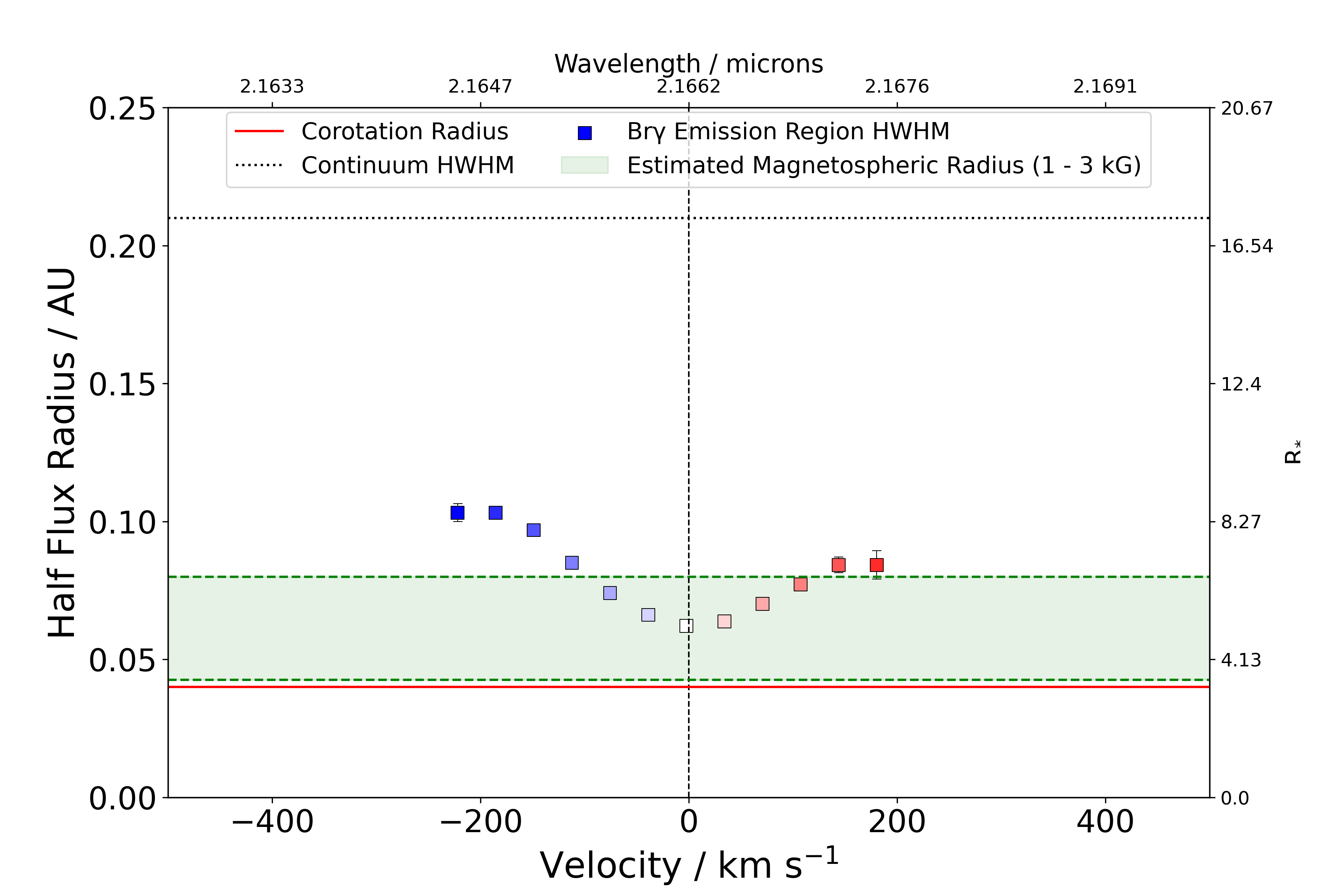}
\end{minipage}

\begin{minipage}{0.49\linewidth}
\centering
\includegraphics[width=\linewidth]{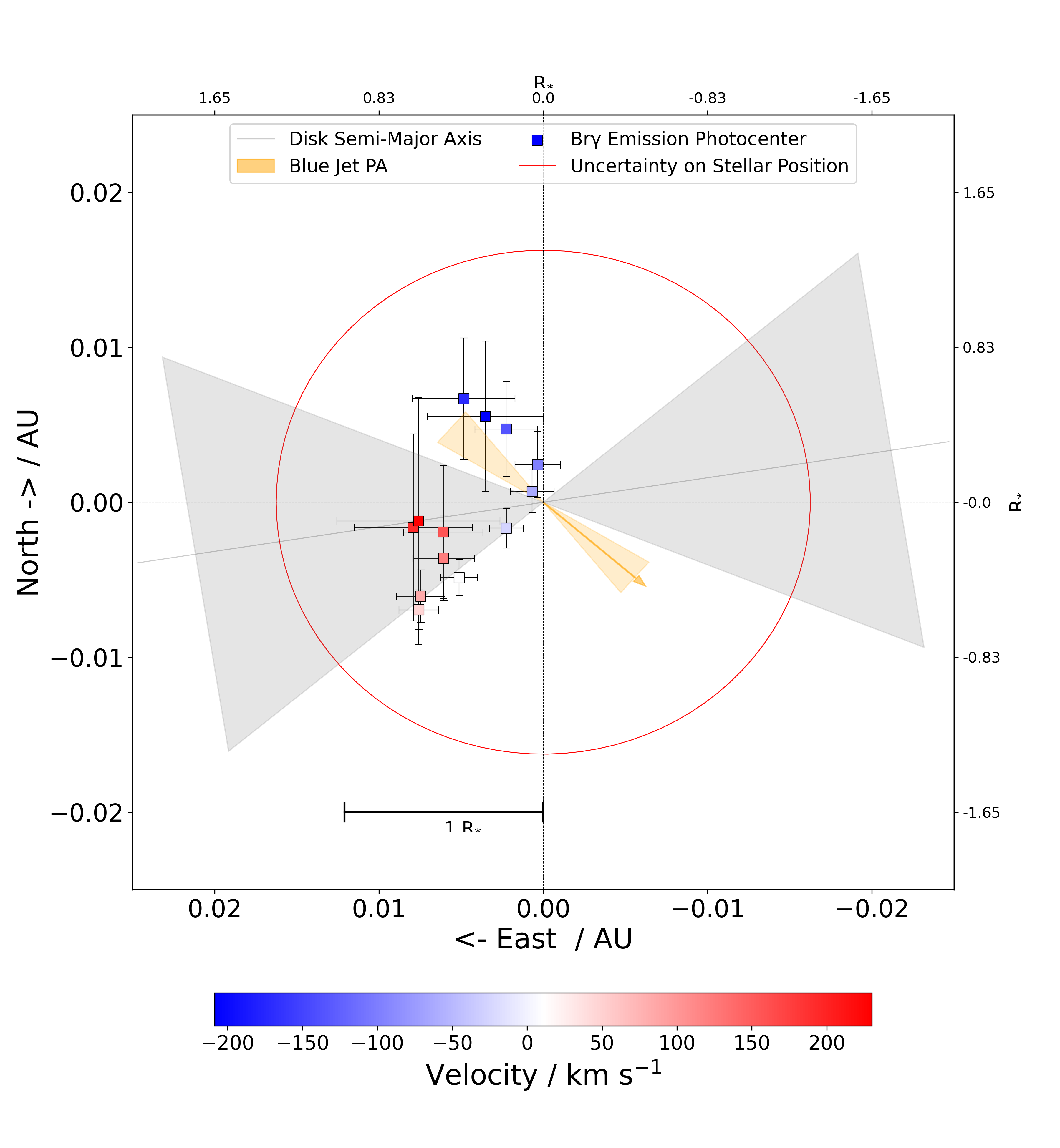}
\end{minipage}
\hfill
\begin{minipage}{0.49\linewidth}
\centering
\includegraphics[width=\linewidth]{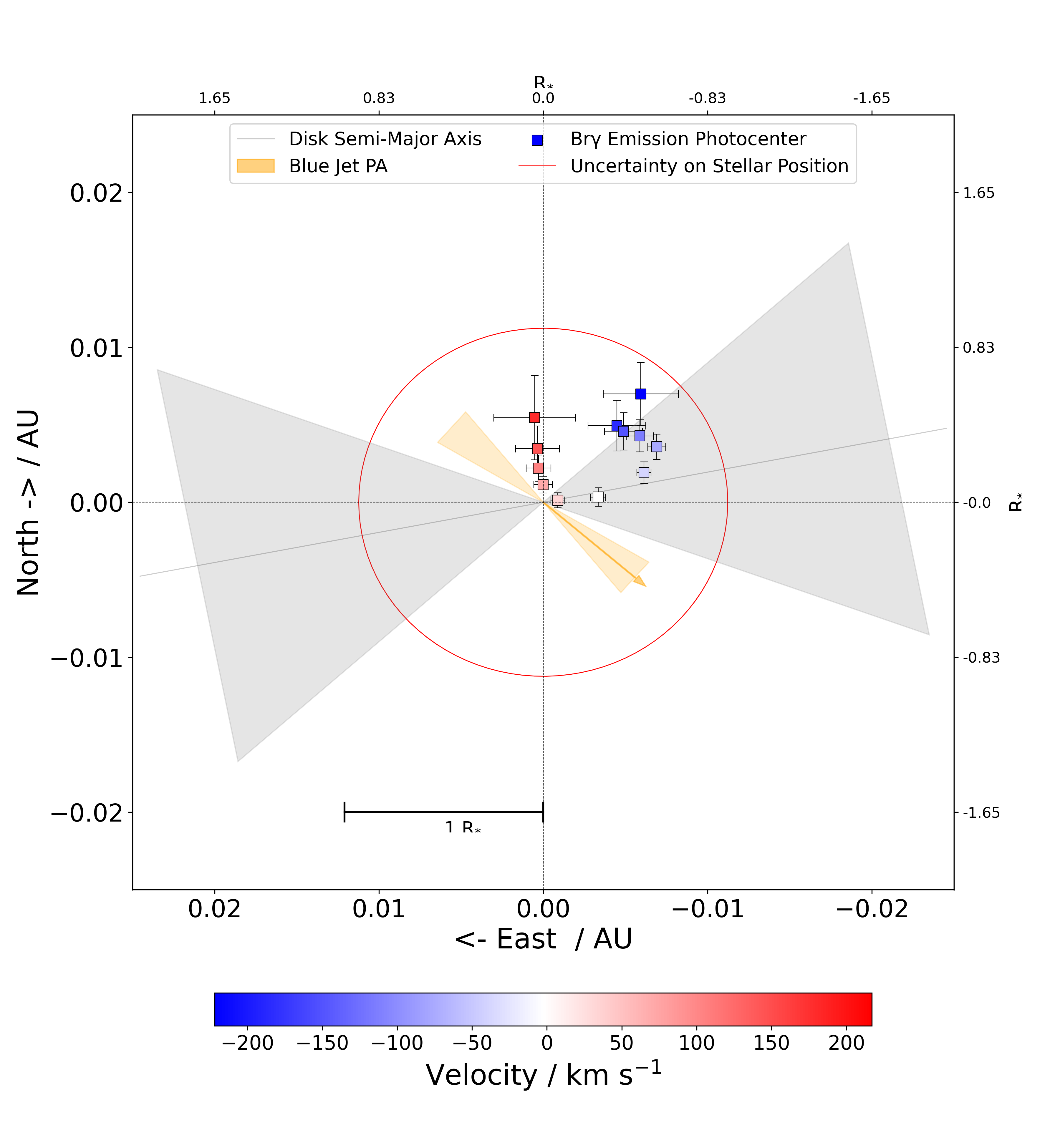}
\end{minipage}
    \caption{Spectrum (top), size (middle), and photocentre shift (bottom) profiles for the two epochs of RU Lup in 2019 and 2021. Shaded regions indicate uncertainties.}
    \label{fig:RULup2021}
\end{figure*}

RU Lup was observed in two distinct epochs in 2018 and 2021 (Fig. \ref{fig:RULup2021}). It appears as the second strongest \bg emitter of our sample of T~Tauri YSOs, exhibiting a peak flux line to continuum ratio of 1.86  and 1.83, respectively. 
From the \bg line we compute an equivalent width of 16.8 $\AA$ in 2018 and 13.5 $\AA$ in 2021, which reflects the narrower shape of the feature in the more recent epoch. Using the empirical relationship presented before, we use the \bg EW to estimate the accretion luminosity to be (0.98 $\pm$ 0.12) L$_{\odot}$ and (0.76 $\pm$ 0.03) L$_{\odot}$, respectively. By comparison, the X-Shooter measurements from 2012 presented in \citet{Alcala2014} put the equivalent width at 9.6 $\AA$ and the accretion luminosity of the RU Lup system at 0.5 L$_{\odot}$. Using the accretion luminosity, stellar radius and stellar mass of RU Lup, we compute the instantaneous mass-accretion rate to be 17.31$^{+20.78}_{-8.54}$ and 13.38$^{+15.79}_{-6.66}$ $\cdot \ 10^{-8} \ \mathrm{M_{\odot}yr^{-1}}$, while they arrive at a more moderate accretion rate of 4.3 $\cdot \ 10^{-8} \ \mathrm{M_{\odot}yr^{-1}}$ largely due to their significantly higher estimate of the stellar mass. Similar accretion rates can be found in \citet{Manara2014PhDTh}, although higher rates similar to the ones we derived were also presented in the past \citep{Siwak2016}.\\
The \bg emission region was marginally resolved across all channels at even the shortest baselines, giving us an extremely clear size profile across different velocity components of the line. For 2018 we detect a minimum half flux radius of 0.076 au (6.3 R$_*$) at the line centre, which increases to 0.1 au  (8.3 R$_*$) at the edges of our selected channel range. A similar size profile presents itself in 2021, where the central minimum is at 0.061 au (5 R$_*$), while in the extreme blue channel the half flux radius increases again to 0.1 au, but only 0.088 au (7.3 R$_*$) au in the red wing. Such changes of the order of 30\% to 50\% are particularly noteworthy when taking into account the high signal to noise in both visibilities and flux data, and the resulting very small uncertainties we obtain for the half flux radii. While the region in both epochs is significantly more compact than the NIR continuum at 0.21 au (17.4 R$_*$), \bg emission does appear to extend beyond the co-rotation radius of 0.04 au (3.3 R$_*$) at both epochs. Both the absolute size, the relative size variation and the pronounced blueshifted excess emission are not characteristic for a scenario where \bg emission is exclusively, or even dominantly, driven by magnetospheric accretion itself, but rather imply the existence of a more extended outflow component, possibly in the form of a disk or stellar wind. \\
The \bg signals in the differential phases are relatively weak and translate into a distribution of different velocity photocentres that is quite complex and not easily interpreted. While the maximum shift is of the order of less than 1 R$_*$, there seem to be substructures in the profile that align themselves along 2-3 different axes. One for the low velocity channels which are, within a 1$\sigma$ uncertainty, broadly aligned with the semi-major axis of the inner NIR continuum disk at (99 $\pm$ 31)$^\circ$ and also the large-scale outer disk as measured with ALMA at 121$^\circ$ \citep{Huang2018}. Then there are two parallel axes which run almost perpendicular to the low velocity axis, although between the two epochs, the red and blue arms are not equally well aligned. The overall profile gives the appearance of a 'crescent' type shape which seems to rotate by about 45$^\circ$ towards the north from 2018 to 2021. We find that in 2018 the blue arm of the profile is almost perfectly aligned with the average PA of the blueshifted jet at (229 $\pm$ 10)$^\circ$ \citep{Whelan2021}, while in 2021 this is true for the red arm. It is noticeable that the arm closest to the jet axis in each case is relatively straight while the other arm appears somewhat distorted in shape, although the effect is within the uncertainty on the photocentre displacements. While it is tempting to conclude that some form of polar outflow is responsible for this behaviour,  a stellar wind would not explain why the blue arm runs anti-parallel to the jet direction in 2018. The complexity of this case would require a better understanding of what a superposition of different photocentre distributions caused by different \bg origin mechanisms would look like in order to comment on this behaviour in a more meaningful way.

\subsubsection{VV CrA SW}

\begin{figure}[h!]
    \centering
\begin{minipage}{\linewidth}
\centering
\includegraphics[width=\linewidth]{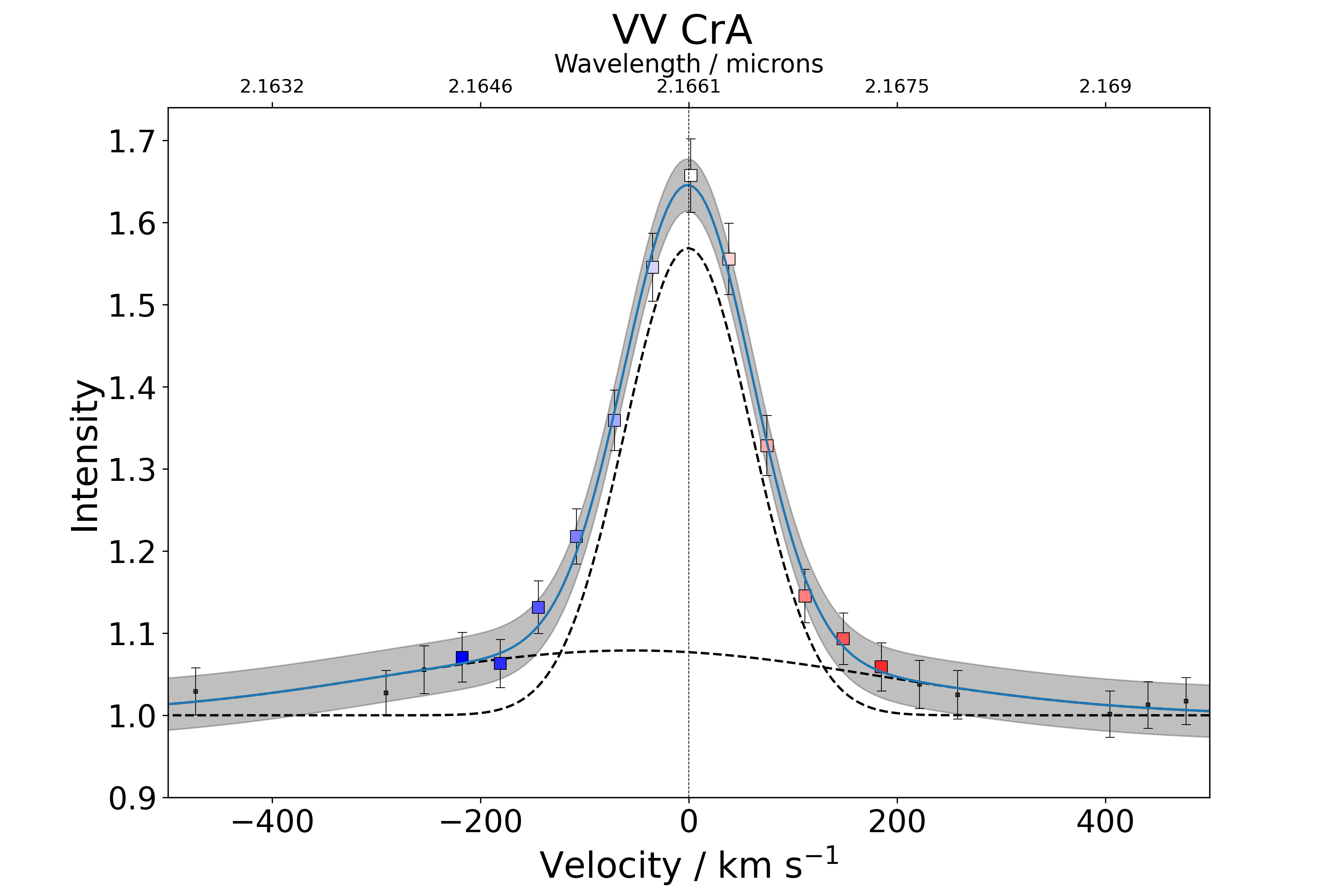}
\end{minipage}
\begin{minipage}{\linewidth}
\centering
\includegraphics[width=\linewidth]{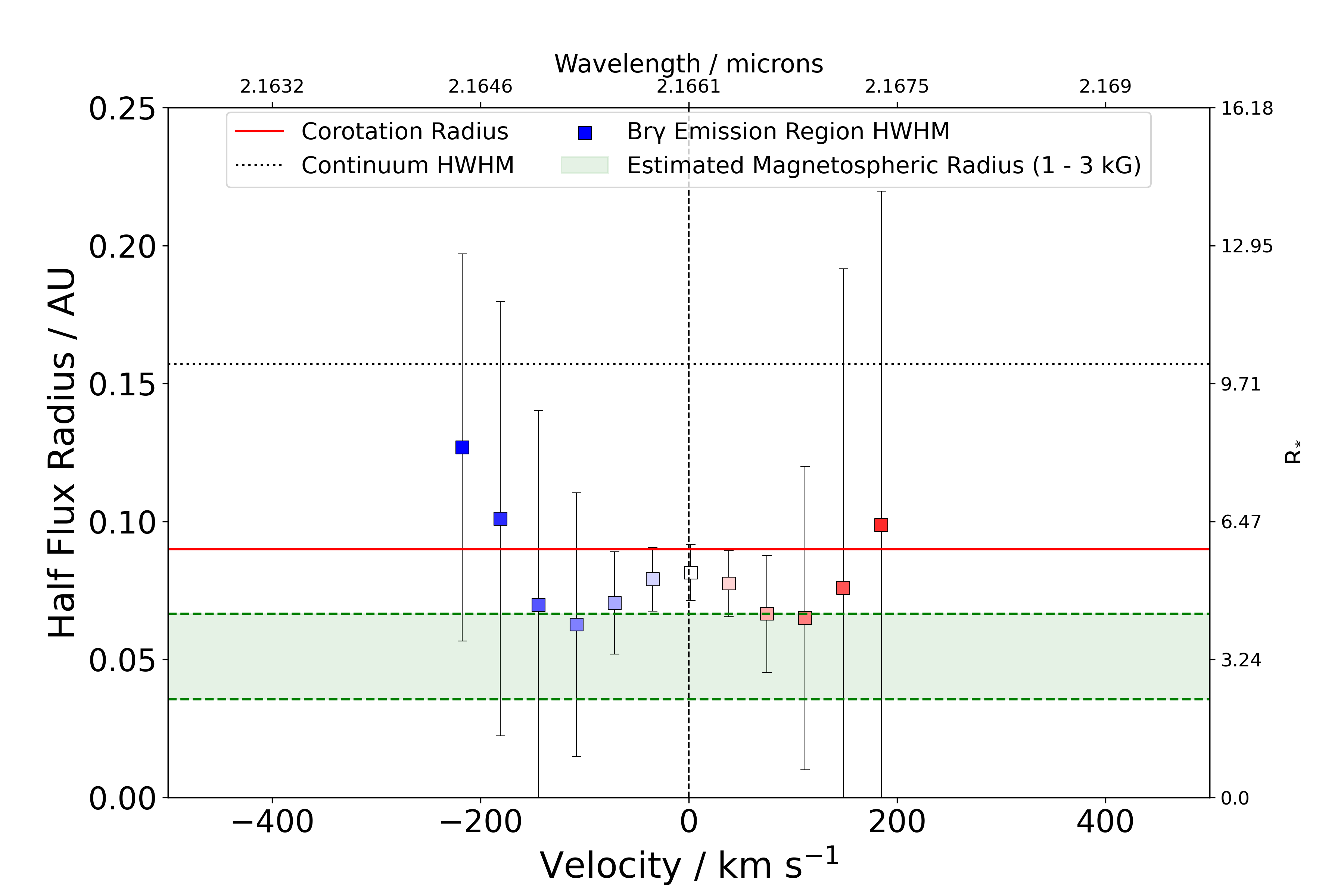}
\end{minipage}

\begin{minipage}{1\linewidth}
\centering
\includegraphics[width=\linewidth]{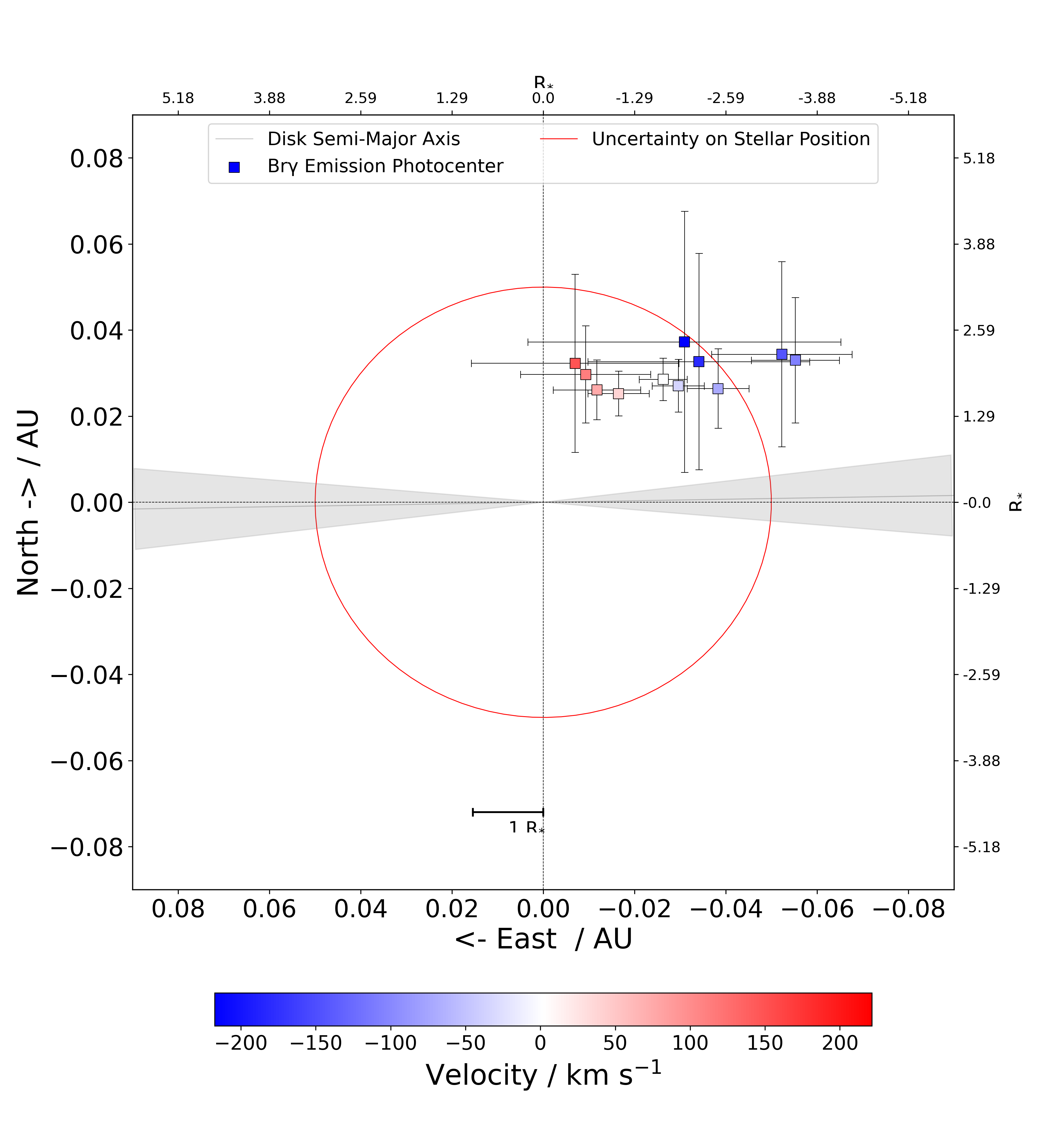}
\end{minipage}
    \caption{Spectrum (top), size (middle), and photocentre shift (bottom) profiles for VV CrA SW. Shaded regions indicate uncertainties.}
    \label{fig:VVCrA}
\end{figure}
The GRAVITY K-band spectrum for the VV CrA (Fig.\ref{fig:VVCrA}) primary shows a \bg emission feature that lacks the pronounced blueshifted excess detected in many of the other objects in the sample. The line shape appears more Lorentzian, featuring broader wings which contain more of the flux than in a Gaussian profile. Flux ratios vary from 1.65 at the centre of the line to 1.05 at the edges of our selected channel range. We computed a \bg equivalent width of 10.1 $\AA$, which appears significantly weaker than than the 30 $\AA$ reported by \citet{Sullivan2019} based on archival K-band NIRSPEC data from 2008. Even by visual comparison, the GRAVITY and NIRSPEC spectra around the \bg line appear quite distinct, as the NIRSPEC data shows both a peak flux ratio of about 2 and a higher excess of blueshifted emission, leading to an asymmetric shape similar to those observed for other objects of the GRAVITY sample. By contrast, \citet{Scicluna2016} present \bg spectra taken with CRIRES in 2013 which are visually closer to ours, also displaying quasi-Lorentzian line shape that is slightly skewed towards the blue wing. Their equivalent width of 9 $\AA$ is also consistent with our findings, although due to their very different stellar parameters, their estimate of the mass-accretion rate is significantly lower (4 $\cdot \ 10^{-8} \ M_{\odot}$yr$^{-1}$) than the 29.77$^{+34.39}_{-14.59}$ $\cdot \ 10^{-8} \ \mathrm{M_{\odot}yr^{-1}}$) we derived from our data. \\
At the centre channel we obtain a size of 0.081 au (6.5 R$_*$) for the emission region, while the change across the line appears to be mostly within the significant uncertainties, which are amplified by the low flux and high visibilities in the wings of the feature. Even so, there is seemingly an increase in size towards the line edges and especially the over blueshifted channels, where the half flux radius goes up to 0.127 au (10.2 R$_*$). In the channels close to the line centre, \bg emission originates from within the co-rotation radius at 0.09 au (7.3 R$_*$). If we were to interpret the two Gaussian components we fitted to the spectrum physically, this could signify the presence of an additional emission source at scales beyond the co-rotation radius, which is responsible for the broad component, while the narrow component, which is dominant in the centre, traces the emission coming from the magnetosphere. This is further supported by the fact that within the range of the narrow component, we first see a decrease in size as would be expected for the case of magnetospheric accretion. A combination of a disk wind launched from the inner disk and a system of accretion columns would be a suitable candidate to explain the radius profile, although the large uncertainties do not allow us to conclude this definitively. \\
A contributing wind could also explain the distribution of photocentres across the line, as the separation between red and blue arm is similar to a rotational profile which is aligned with the position angle of the NIR continuum disk at (91 $\pm$ 6)$^{\circ}$. Such an alignment is typical for a disk in rotation, and a wind launched from such a rotating disk might be expected to display a similar displacement of photocentres, see for example \citet{Goto2012}. Again, the low signal to noise ratio in our differential phase data results in very large error bars, making any kind of commentary ultimately speculative, but it is notable that the maximum shift (3.8 R$_*$)  we detect for a photocentre is significantly larger both in absolute terms and as a fraction of the size of the \bg emission region than what we find for any of the other targets in the sample. This implies that there is some influence which distorts the symmetry of the \bg brightness distribution at larger scales than what we usually see.  \\

\subsubsection{DG Tau}

\begin{figure}[!h]
    \centering
\begin{minipage}{\linewidth}
\centering
\includegraphics[width=\linewidth]{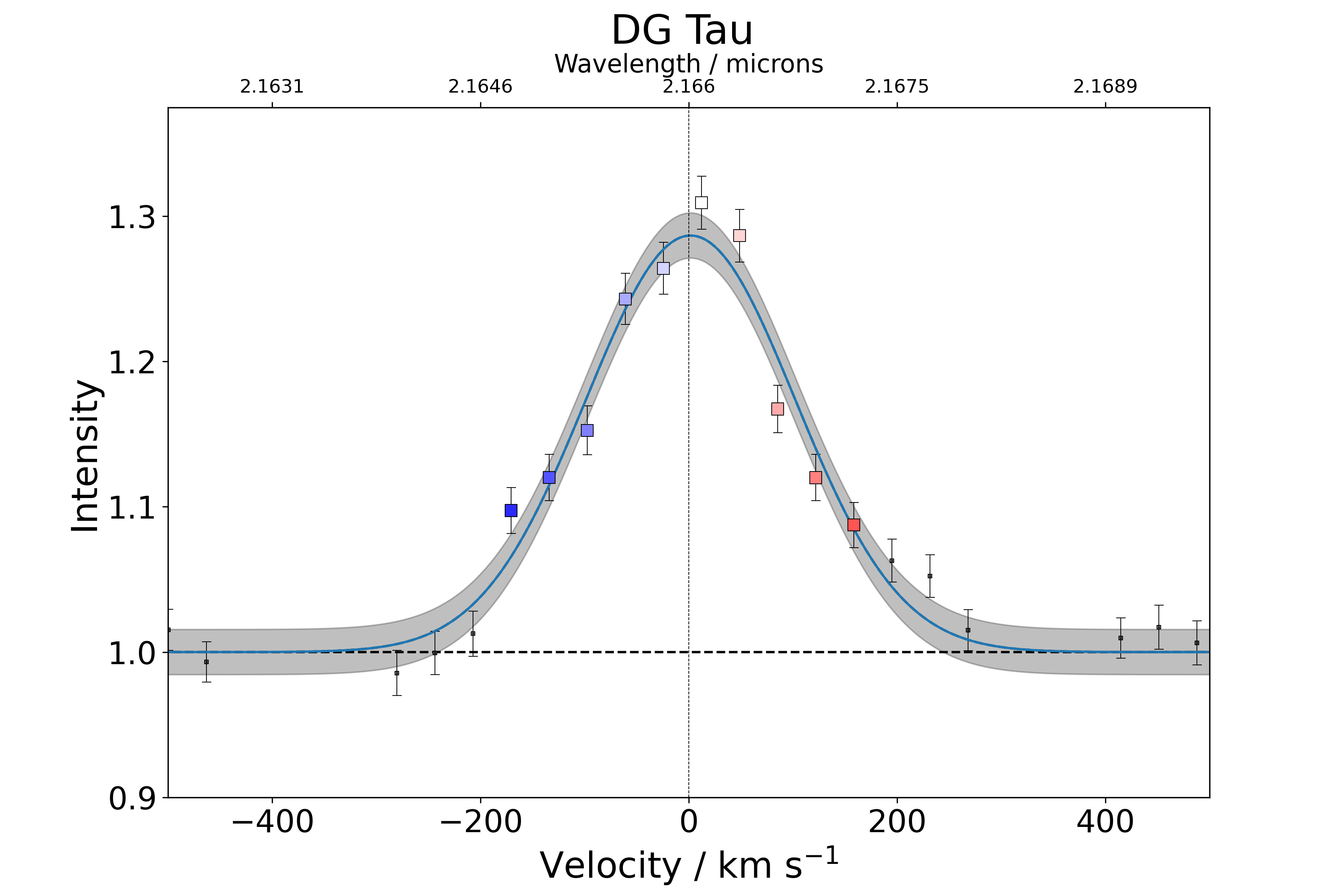}
\end{minipage}
\begin{minipage}{\linewidth}
\centering
\includegraphics[width=\linewidth]{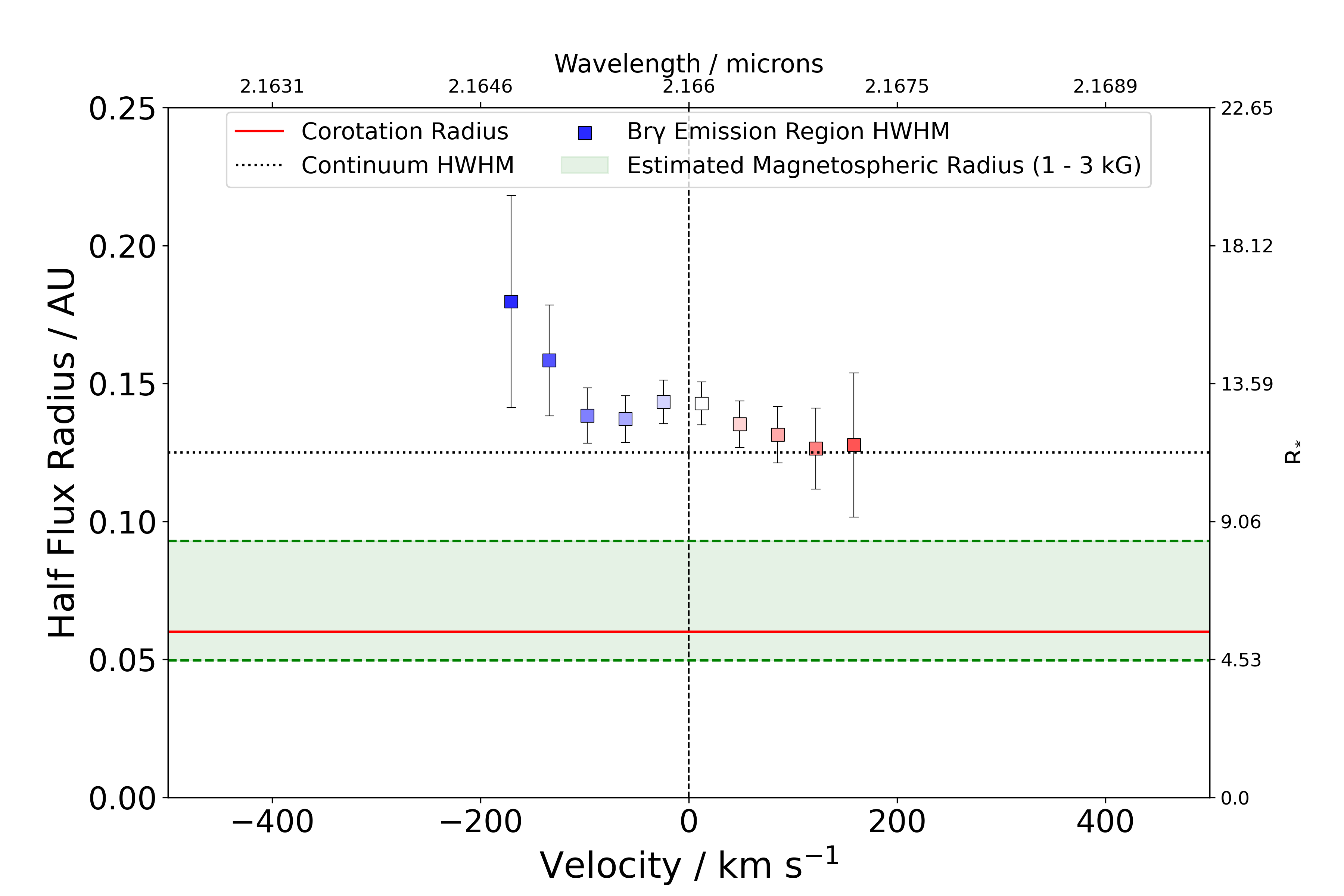}
\end{minipage}

\begin{minipage}{1\linewidth}
\centering
\includegraphics[width=\linewidth]{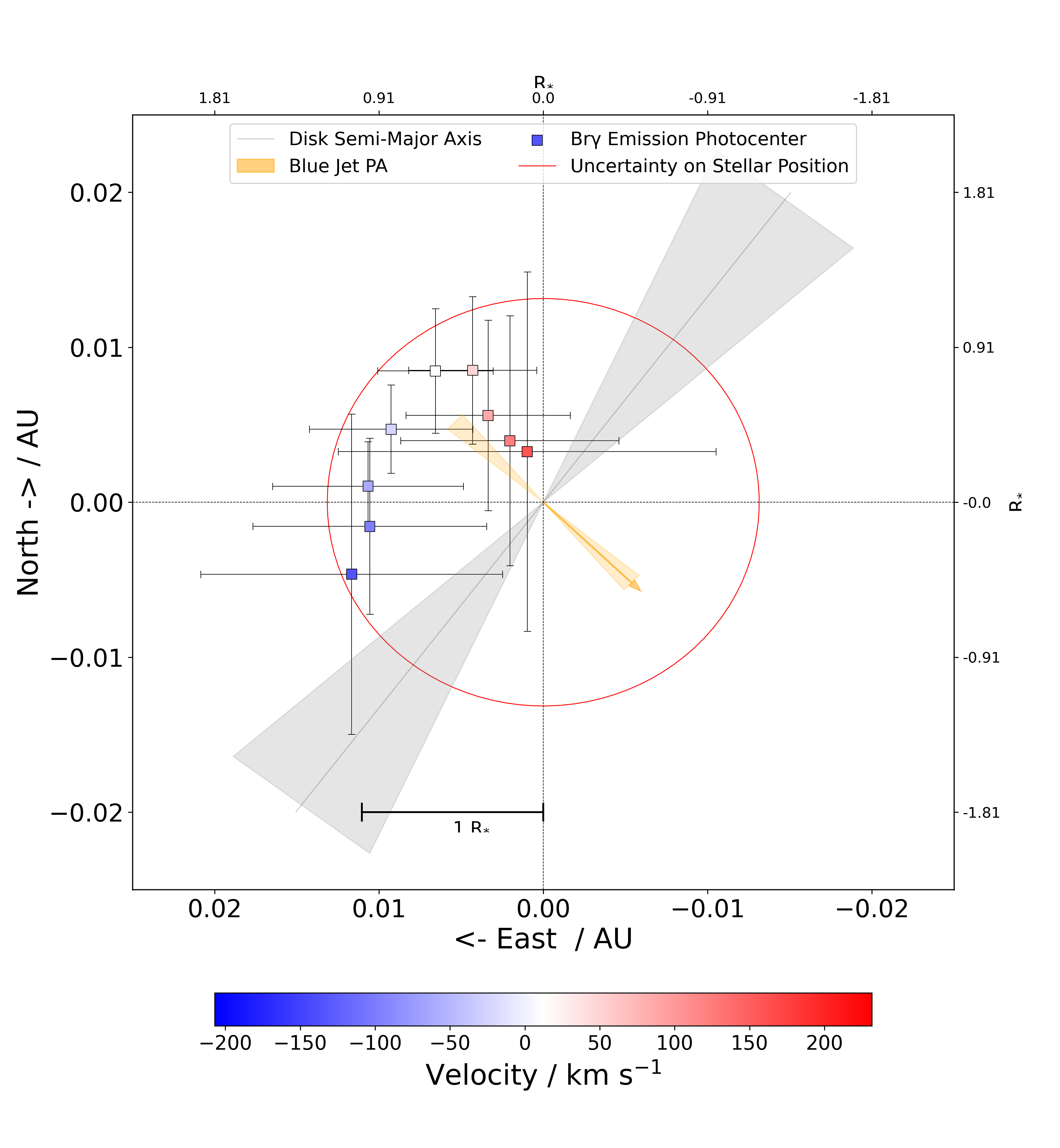}
\end{minipage}
    \caption{Spectrum (top), size (middle), and photocentre shift (bottom) profiles for DG Tau. Shaded regions indicate uncertainties.}
    \label{fig:DGTau}
\end{figure}
GRAVITY observations of the DG Tau (Fig. \ref{fig:DGTau}) \bg flux show a weak spectral line, with a peak flux ratio of only 1.29, the second lowest in the sample, and a FWHM of about $\pm$ 224 km/s. 
The original assumption of DG Tau as one of the strongest accretors in the sample appears at odds with the small equivalent width of only 5.79 $\AA$, from which we estimate a mass-accretion rate of just $\mathrm{3.95^{+4.44}_{-2}}$ $\cdot \ 10^{-8} \ \mathrm{M_{\odot}yr^{-1}}$. This puts this observation at the very low end of a range of previously reported accretion rates, some of which can surpass the current result by an order of magnitude (see e.g. \citet{Iguchi2016}. Such a degree of variation among past results indicates either high variability in accretion itself, some form of temporary obscuration of the central star and the \bg emission region, or a combination of similar factors leading to an attenuated \bg feature. \\
Fits of the continuum data with a Gaussian ring model have constrained the inner dust disk PA at (143 $\pm$ 12)$^\circ$ \citep{Perraut2021}, so that we were not able to detect a misalignment between the K-band continuum and the outer disk at 135.4$^\circ$ position angle as measured by ALMA at 870 $\mu m$ \citep{Gudel2018}. The semi-major axis of the inner dusty disk appears almost perpendicular to the reported jet PA \citep{Solf1993} of 226$^\circ$. By contrast, there appears to be a significant divergence in terms of inclination between the outer disk at 37$^\circ$ and the inner disk at (49 $\pm$ 4)$^\circ$. The known jet inclination of 46$^\circ$ \citep{Eisloeffel1998} agrees with the GRAVITY result, which suggests that the inner disk is aligned with the magnetic field. \\
For DG Tau, \bg visibility signals tend to be faint, with multiple baselines showing no recognisable \bg feature at all. The absence of a clear detection for some of the baselines cannot be directly attributed to a lack of spatial resolution as there seems to be overall no clear correlation between strength of signal and baseline length. At a central channel half-flux radius of 0.143 au (12.9 R$_*$), the emission region stretches far beyond the co-rotation radius and even slightly beyond the NIR continuum HWHM of 0.125 au, which makes it a unique case within our sample. \\
Our results appear more extended when compared to those of \citet{Eisner2014}, who examined the \bg feature of DG Tau with the Keck Interferometer (KI) and found a uniform ring radius of 0.12 au in \bg versus 0.17 au in the NIR continuum for an assumed distance of 140 pc. If updated to the more recent distance estimate of 125 pc from GAIA DR3, this would translate into  0.10 au and 0.15 au, respectively. As the \bg line recorded with KI shows an equivalent width of 6.9 $\AA$ and is thus broadly similar in strength and width to the one measured with GRAVITY, this discrepancy is likely the result of differing methods in determining the region size. Due to their reliance on a single baseline, they fit a uniform ring model without taking into account possible inclination effects. While the methodology section of their paper does reference a 'circumstellar emission' which seems similar in approach to the pure line emission used in our analysis, it is not clear whether the reported \bg radius is based on this or the total visibilities. If the latter is the case, this could certainly cause the emission region to appear significantly more compact.\\
Differential phase signals are also relatively faint and translate into photocentre shifts that, even at a maximum magnitude of 0.0125 au, do not extend significantly beyond the stellar radius of 0.0123 au. The photocentre profile shows a significant displacement between the central channel photocentre position and the continuum photocentre at the zero position (i.e. the centre of Fig. \ref{fig:DGTau}). When taking into account the closure phase information, this displacement can be explained in terms of an uncertainty on the relative location of the continuum photocentre with respect to the star. The position of the star, which in Fig. \ref{fig:DGTau} is located somewhere within the red circle determined from the closure phases, is still consistent with the shift measured for some of the photocentres on a 1$\sigma$ level. \\  
The photocentre profile features a blue arm that is extended along the semi-major axis of the continuum disk, whereas the displacements across the red wing appear to be more in line with the axis of the redshifted jet component. However, the scale of the error bars on the individual points in both wings prevent us from definitively associating the profile with either of these alignments. Given the large uncertainties, it is certainly possible to interpret the profile as essentially Keplerian, possibly tracing the \bg emission coming from a wind that was launched from the rotating base of the disk near the inner rim. Not only would such a case be consistent with the large spatial extent of the \bg emission region, but also with the  velocity profiles obtained with ALMA by \citet{Podio2020} and \citet{Gudel2018} from the large-scale CO and other molecular line emission, which also indicate blushifted emission along the south-east axis.

\subsubsection{TW Hya}

\begin{figure}[h!]
    \centering
\begin{minipage}{\linewidth}
\centering
\includegraphics[width=\linewidth]{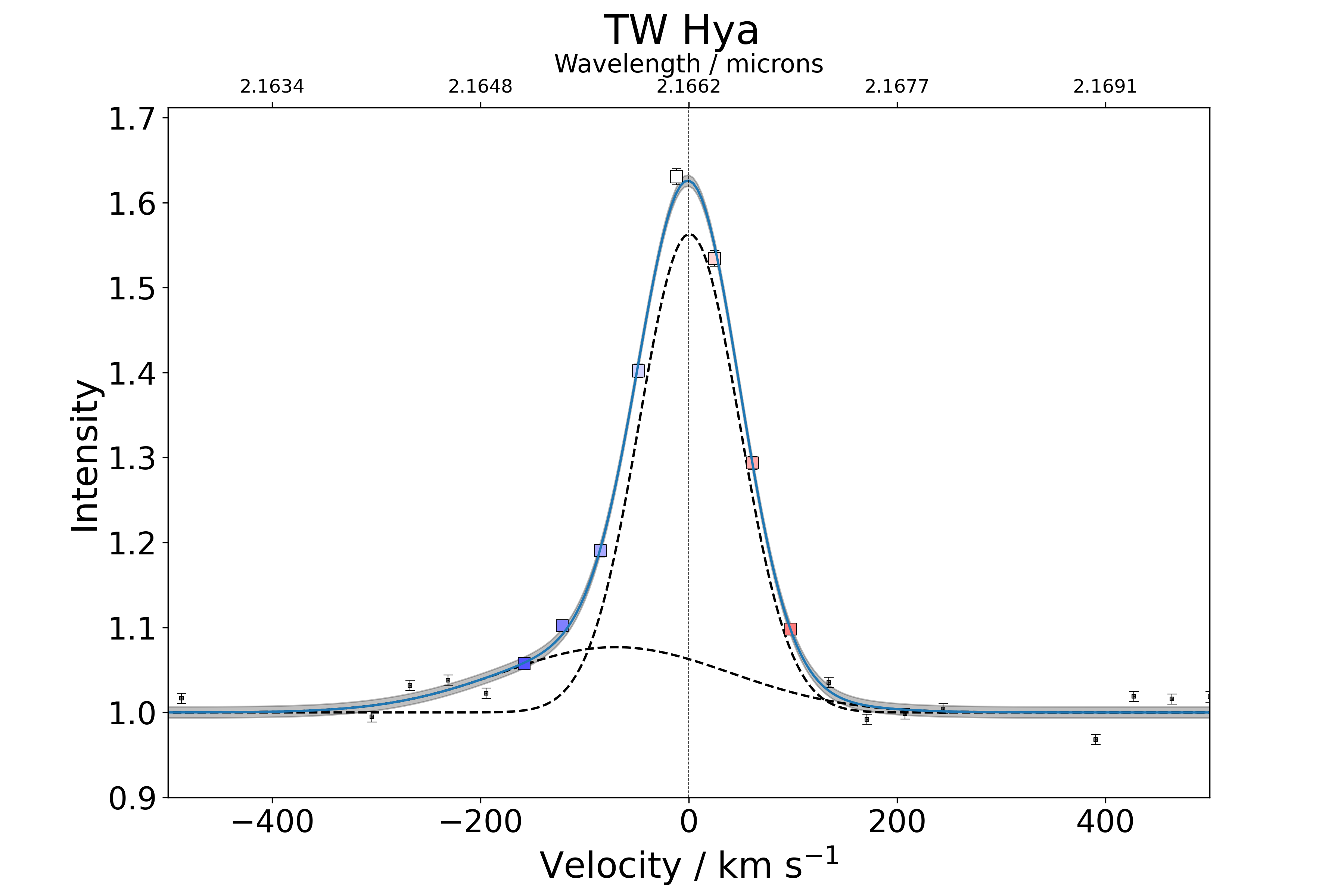}
\end{minipage}
\begin{minipage}{\linewidth}
\centering
\includegraphics[width=\linewidth]{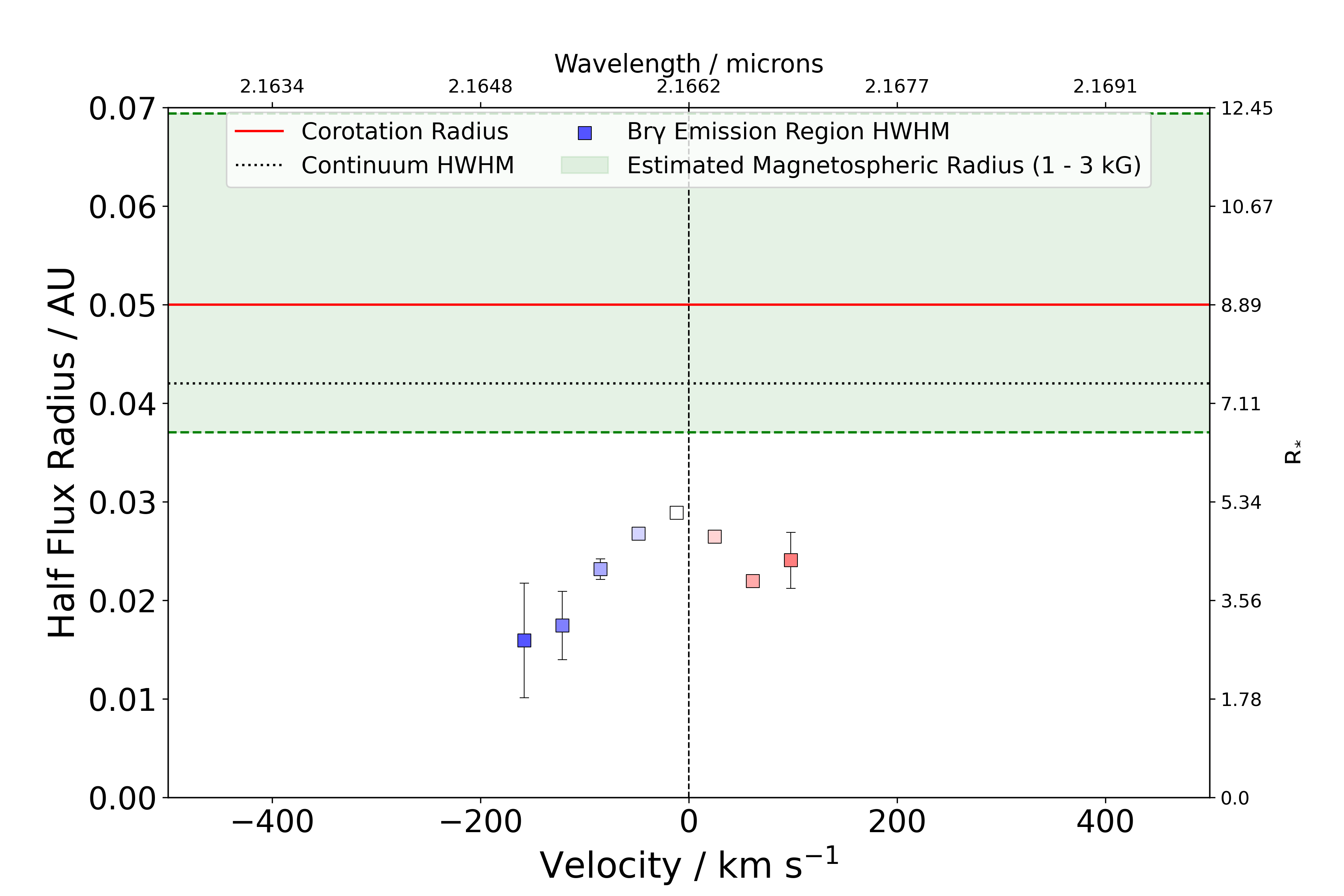}
\end{minipage}

\begin{minipage}{1\linewidth}
\centering
\includegraphics[width=\linewidth]{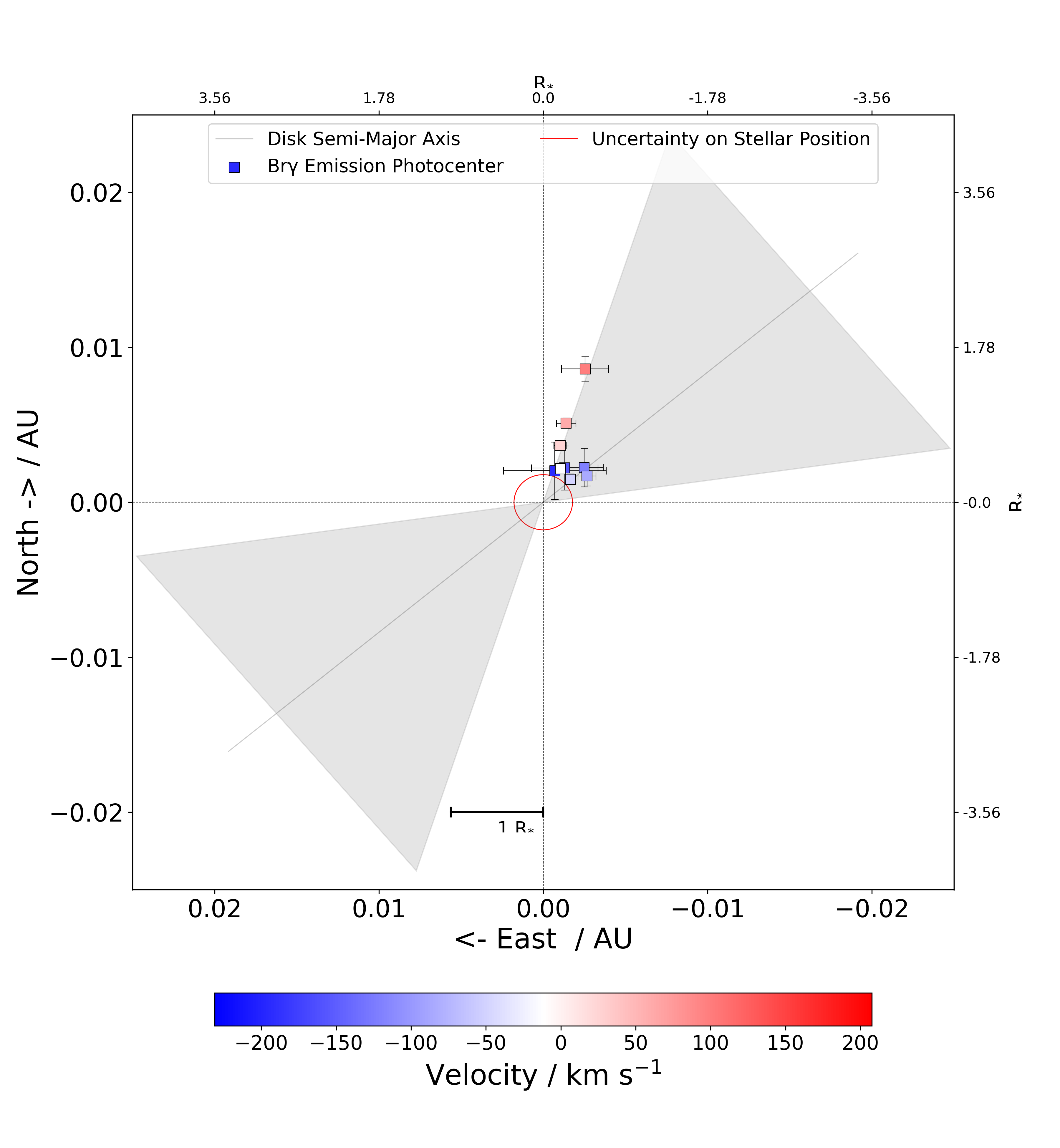}
\end{minipage}
    \caption{Spectrum (top), size (middle), and photocentre shift (bottom) profiles for TW Hya. Shaded regions indicate uncertainties.}
    \label{fig:TWHya}
\end{figure}
The \bg emission line in the GRAVITY spectrum of TW Hya (Fig. \ref{fig:TWHya}) shows an excess of blueshifted emission, which is also seen in a spectrum recorded with SINFONI \citep{Goto2012} with a similar (R=4000) spectral resolution as GRAVITY and further confirmed in high resolution (R=45000) \citep{Sokal2018} observations made with IGRINS between 2015 and 2017. \\ 
At the centre of the feature we find that the \bg emission region originates in area with 0.029 au (5.16 R$_*$)  half flux radius, which falls off to 0.016 au (2.85 R$_*$) in the extreme blue wing and 0.022 au (3.91 R$_*$) at the lowest point in the red wing. As such TW Hya is the only source in the sample that clearly shows a decline in size towards the edges of the line beyond the uncertainty on the individual points. \\ 
We also find that the emission region in all channels is significantly more compact than the co-rotation radius of 0.05 au and slightly smaller than the magnetospheric radius of 0.037 au estimated from stellar parameters and an assumed magnetic field strength of 1 kG.  As most of the observational evidence actually suggests the magnetic field to be of the order of up to 1.5 kG (e.g. \citet{Bouvier2007}, \citet{Donati2011}, \citet{Johnstone2014}), we can adjust the estimate of the truncation radius to 0.046 au and still find it to be less extended, but close to the co-rotation radius, which is expected for magnetospheric accretion. This agreement and the compact nature of the \bg half flux region when compared to the co-rotation radius are making a compelling case for the idea that \bg emission in TW Hya indeed originates from the magnetosphere. This is further supported by the size profile. The decrease in HWHM would be expected under a magnetospheric accretion scenario as line components with the highest velocity are likely tracing gas in free fall towards the stellar surface. \\   
This very important result, presenting the \bg emission region as the first direct measurement of a spatially resolved magnetosphere, was previously reported in \citet{GarciaLopez2020}. Based on the same data, they found a \bg region size of 0.021 au and thus something even more compact, although the deviation ultimately comes down to methodology and data correction. Even though a Gaussian disk was used as a geometric model in both cases, they performed their analysis on an average of the three central channels of the line and used previous ALMA measurements to fix their inclination at the level of the outer disk (7$^{\circ}$, see e.g. \citet{Andrews2016}), while we refer to the GRAVITY measurements of the NIR continuum to perform our fit for a slightly more inclined (14$^{\circ}$) disk. Another key difference is the spectral correction that we applied to align telescope spectra with each other and with the telluric lines. As the interferometric quantities are recomputed based on the corrected spectra, the realignment can have a significant enough impact on the visibilities and differential phases to account for deviations such as this.\\
It is notable that the magnetospheric radius at B=1.5 kG extends beyond the K-band continuum half flux radius, whereas the dust sublimation radius was estimated to fall between 0.02 and 0.04 au \citep{Perraut2021}. This might indicate that dust grains exist up to the truncation radius and that the magnetosphere is playing a part in truncating not just the gaseous disk, but also the dusty disk, which was also suggested by \citet{Eisner2006} based on their Keck Interferometer observations of this target. \\ 
The photocentre profile across the line is highly compact, concentrating in an area within 0.01 au (1.78 R$_*$) around the continuum photocentre. Since the dispersion of closure phases at \bg wavelengths is small, it is reasonable to assert that, given the maximum shift of 0.009 au and an average shift of less than 0.003 for the central channels compared to a stellar radius of 0.006 au, the \bg photocentre across the line coincides with the position of the central star. This suggests that most of the different photocentre velocity components of \bg emission, observed almost pole-on, are well centred on the stellar position. Little can be inferred directly from such a distribution. At very low inclination, any centrosymmetric brightness distribution is likely to lead to a similarly compact profile centred on zero, meaning only in combination with more conclusive evidence, as presented above, does it support the magnetospheric accretion scenario. It is still noticeable that the distribution, and indeed the differential phase signals at the individual baselines, appear to show slightly larger shifts in the red wing of the line. In the case of an inclined magnetosphere with two accretion columns, the opposite could be naively expected in a pole-on configuration, as infalling gas at the far side of the star at the highest blueshifted velocities would be blocked by the star itself, which should lead to a shift away from the centre for the blue components. The absence of a stronger shift in the blue arm could indicate that the configuration of the magnetic field in TW Hya is close to an axisymmetric scenario with a small obliquity angle between dipole and rotational axes. Such a case would be in line with the relatively small difference in inclination between the inner NIR continuum disk and the outer disk and also the lack of blueshifted absorption. However, a previous study by \citet{Johnstone2014} indicates a substantially tilted dipole, with an obliquity between 10$^\circ$  and 40$^\circ$, respectively, and it is unclear whether this still would be consistent with such a scenario.

\subsubsection{S CrA N}

\begin{figure}[h!]
    \centering
\begin{minipage}{\linewidth}
\centering
\includegraphics[width=\linewidth]{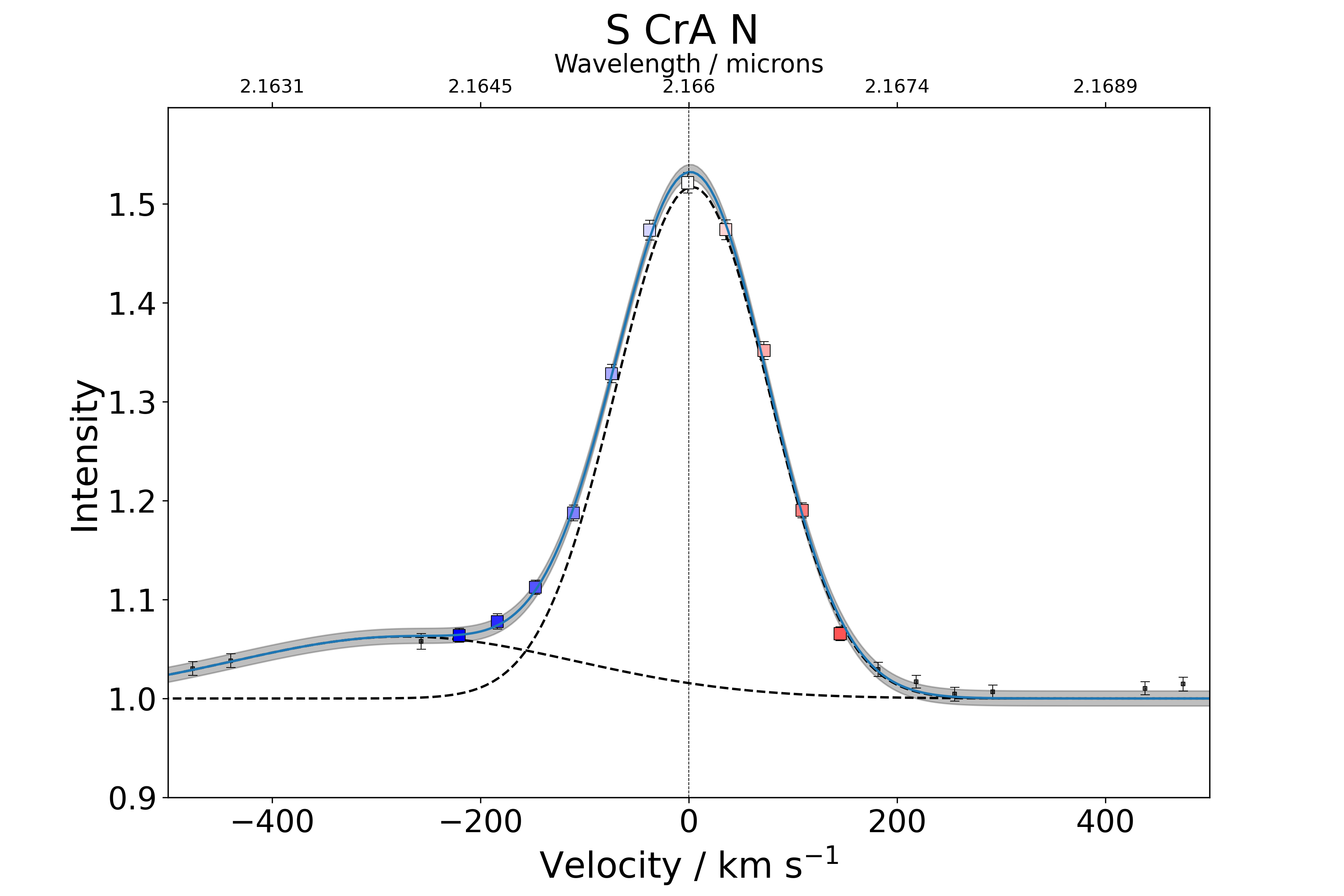}
\end{minipage}
\begin{minipage}{\linewidth}
\centering
\includegraphics[width=\linewidth]{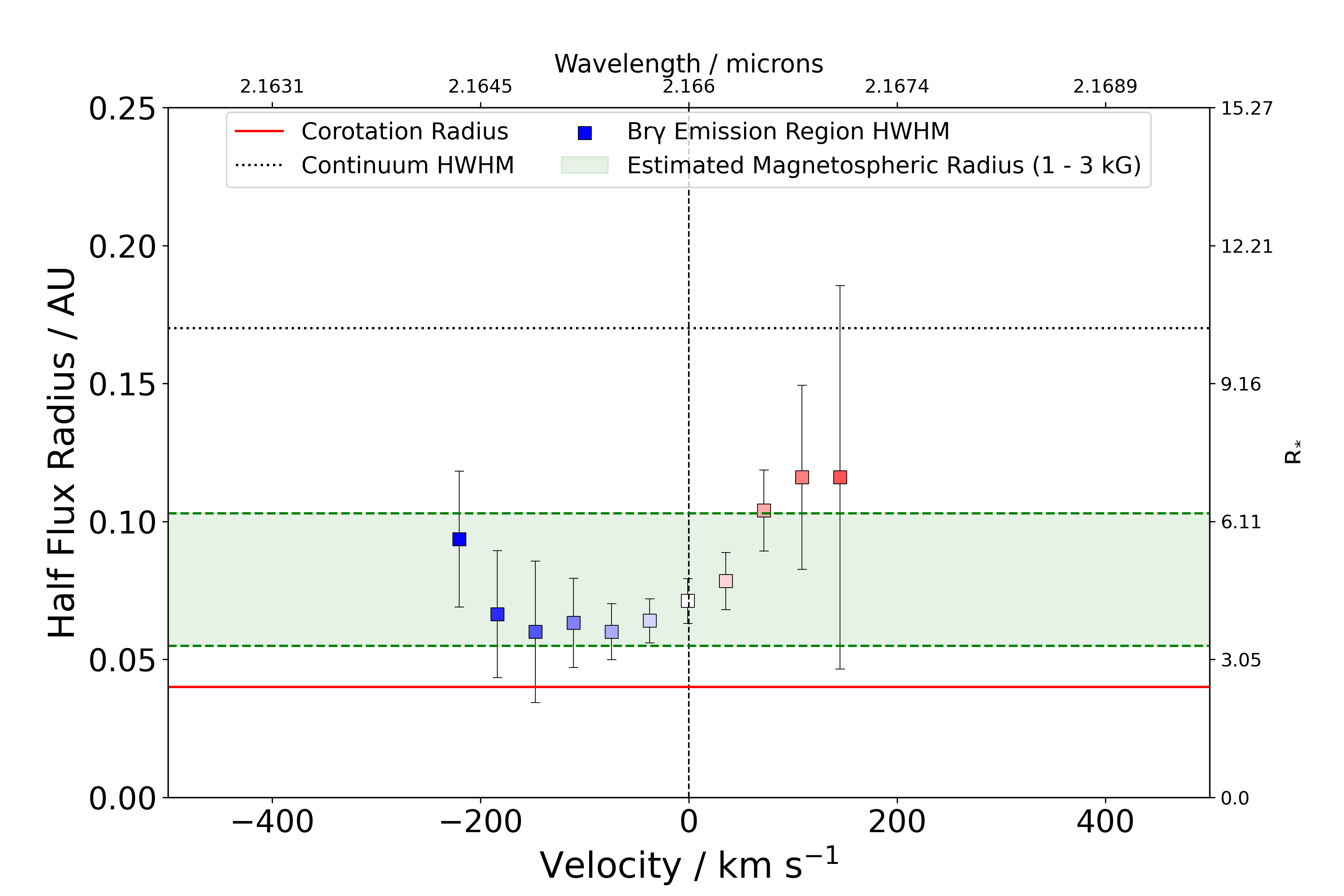}
\end{minipage}

    \caption{Spectrum (top) and size (bottom) profile for S CrA N. Due to the poor quality of the phase data we chose not to include an analysis of the photocentre shift distribution for this object. Shaded regions indicate uncertainties.}
    \label{fig:SCrA}
\end{figure}
We observe a similarly asymmetric spectral line as for most of the previous sources, although for S CrA (Fig. \ref{fig:SCrA}) it appears at a higher blueshift, with a broad high velocity component centred at -273 km/s. \\
The equivalent width of -8.7 $\AA$ is consistent with measurements of the \bg line taken in 2008 with NIRSPEC, as presented in \citet{Sullivan2019}, who reported -7.2 $\AA$. However, they also report a significantly weaker mass-accretion rate of 12.8 $\cdot$ 10$^{-8}$\,$\mathrm{M_{\odot}yr^{-1}}$ when compared to the estimate of 27.72$^{+28.96}_{-13.57}$ $\cdot$ 10$^{-8}$\,$\mathrm{M_{\odot}yr^{-1}}$ we derived from our equivalent width. The NIRSPEC K-band spectrum around the \bg feature is visually consistent in shape with our GRAVITY observations and shows the same asymmetry, arguably even more so than the previously published work \citep{GarciaLopez2017} on the same GRAVITY data, which was at the time not corrected for the misalignment between the four telescopes.  S CrA was also investigated by \citet{Gahm2018} with the UVES spectrograph of the VLT, who fitted a magnetospheric accretion model to their spectroscopic data and found an even higher mass-accretion rate of 50 $\cdot$ 10$^{-8}$\,$\mathrm{M_{\odot}yr^{-1}}$.\\
The spectral correction we applied also affects the shape and position of the \bg signals in the visibilities. In \citet{GarciaLopez2017} their unusual shape and the misalignment between the visibility peaks at different baselines with the position of the line in the spectrum was presented as a possible indicator of the presence of different kinetic signatures in the data. In our spectrally corrected data, we do no longer find this misalignment for most of the baselines to any significant degree. Except for the shortest baseline, the peaks are sufficiently aligned with the spectral line and are well represented by a single peaked Gaussian shape. For the UT3-UT2 baseline with a length of 41 m, we still find a visibility shape that is strongly offset towards the blue wing and decreases much more quickly in the red, imitating the asymmetry in the spectrum. This could be indicative of a large-scale component in the system that is overresolved at the longer baselines, although this in itself would not explain the lack of signal in the red wing at this baseline.  \\
At a number of combinations of channels and baselines, the pure line emission region is completely unresolved, as indicated by the fact that the computation of the pure line visibility yields results that are consistent with 1 within the uncertainties for those wavelengths at those baseline lengths. For the purpose of determining the region size, such points were discarded so that the fits of the ellipsoid are based purely on those baselines and channels for which the region is marginally resolved. While this correction is reflected in the error bars on the region size in each channel, the results in those cases also overestimate the true size and should thus be considered as an upper limit. \\ 
With this in mind it is difficult to comment on the trend of the size across the different velocity channels since for S CrA we find that it is mainly the outer parts of the wing, and the red wing in particular, which are affected by this issue. While the size profile shows an increase from 0.071 au (4.4 R$_*$) at 0 km/s to 0.104 au (6.4 R$_*$)  in the extreme red and 0.094 (5.7 R$_*$)  in the extreme blue channel, the uncertainties and upper limit nature of those results must put some doubt on any possible interpretation. \\ 
Even at the lowest extent in the central channels, the region extends outside the range of the co-rotation radius at 0.04 au (2.4 R$_*$). While the large uncertainties in the line wings  again suggest that it is problematic to overstate this case for the higher velocity ranges, we consider it unlikely that the degree of overestimation is sufficiently large to mask an emission point of origin within the co-rotation radius.  \\
Due to the poor quality of the differential phases, we cannot present a photocentre shift profile as for our other sample objects and have to rely largely on the visibilities for interpretation. As such, even when taking into account the degrading quality of the fits in the edges of the line, neither the absolute size of the emission region nor the size profile across the line appear characteristic for magnetospheric accretion. It is likely that additional \bg emitting components are indeed present in the system, as the peculiar shape of the UT3-UT2 visibility implied. This was in principle also the argument put forward in \citet{GarciaLopez2017}, where they found a size of the semi-major axis of a \bg Gaussian disk of (0.54 $\pm$ 0.07) mas at the centre of the line, which would translate to (0.087 $\pm$ 0.011) au. Given the obvious impact of the spectral correction on the visibilities, this is still very much consistent with the current results, although in the previous work the size of the region declined throughout the line rather than reach a minimum close to the line centre. S CrA is associated with a large-scale jet \citep{Peterson2011} and could also be driving an outflow on smaller scales which contributes to the \bg feature of the system, which might contribute to the blueshifted excess in the spectrum. The -273 km/s at the centre of the broad, blueshifted component fit the  typical velocities found in small-scale jets (see e.g. \citet{Takami2003}). There they also found a positional displacement in H$\alpha$ in S CrA N as part of their spectro-astrometic survey. While they argue that this displacement at a position angle of 153$^{\circ}$ is likely caused by the companion, we do note that baseline UT4-UT3 at a similar PA of 132$^{\circ}$ consistently yields the largest size for the \bg emission region across all of the spectral channels. Given the separation between S CrA N and S CrA S of 1.3 arcseconds, the visibility measurements would not be contaminated by this visual companion.\\
However, differentiating between such stellar outflow or a wind being launched from the rotating inner disk as primary drivers of \bg in S CrA N is challenging given the lack of robust differential phase data to trace the kinematics of the gas.

\subsubsection{DoAr 44}

\begin{figure}[h!]
    \centering
\begin{minipage}{\linewidth}
\centering
\includegraphics[width=\linewidth]{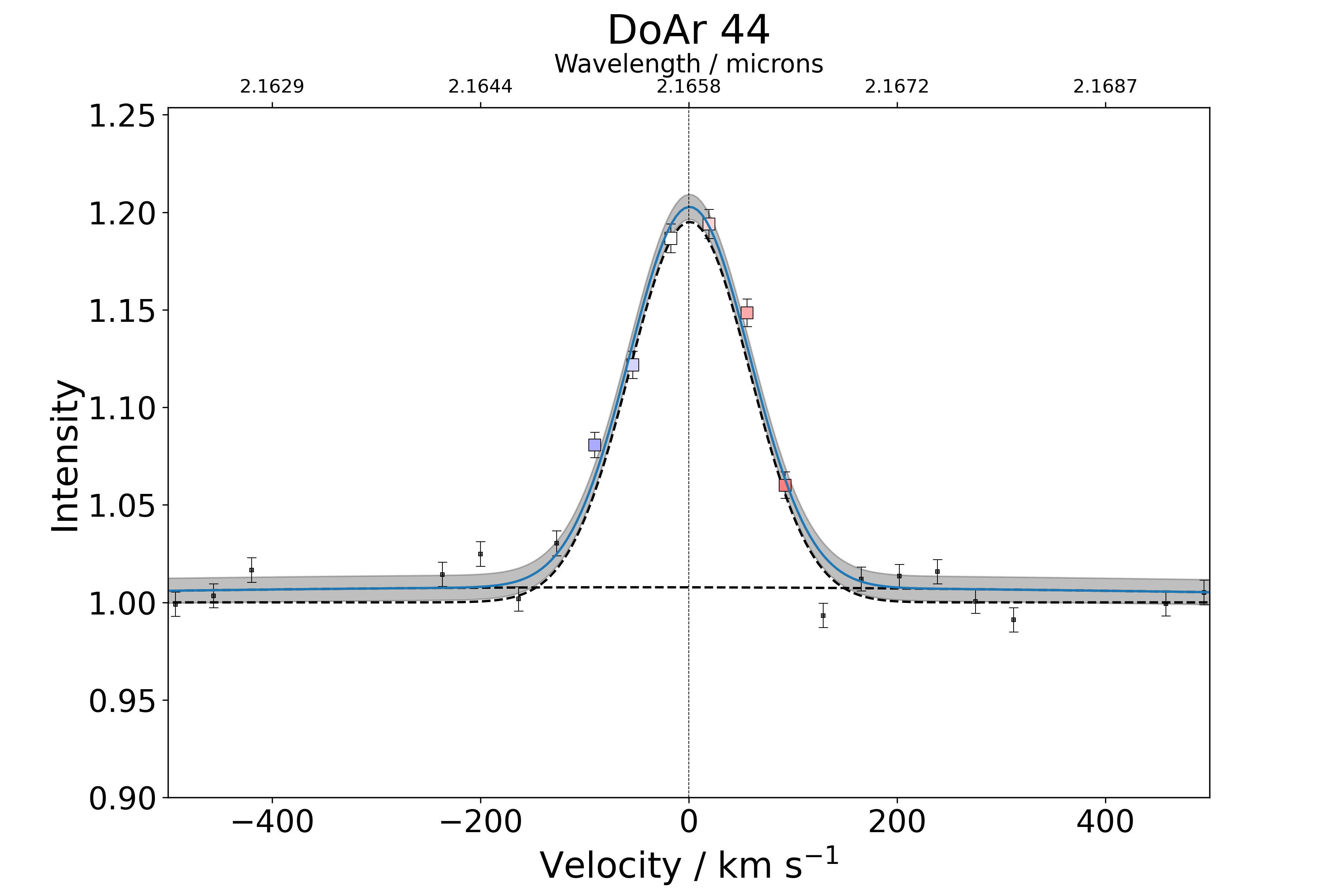}
\end{minipage}
\begin{minipage}{\linewidth}
\centering
\includegraphics[width=\linewidth]{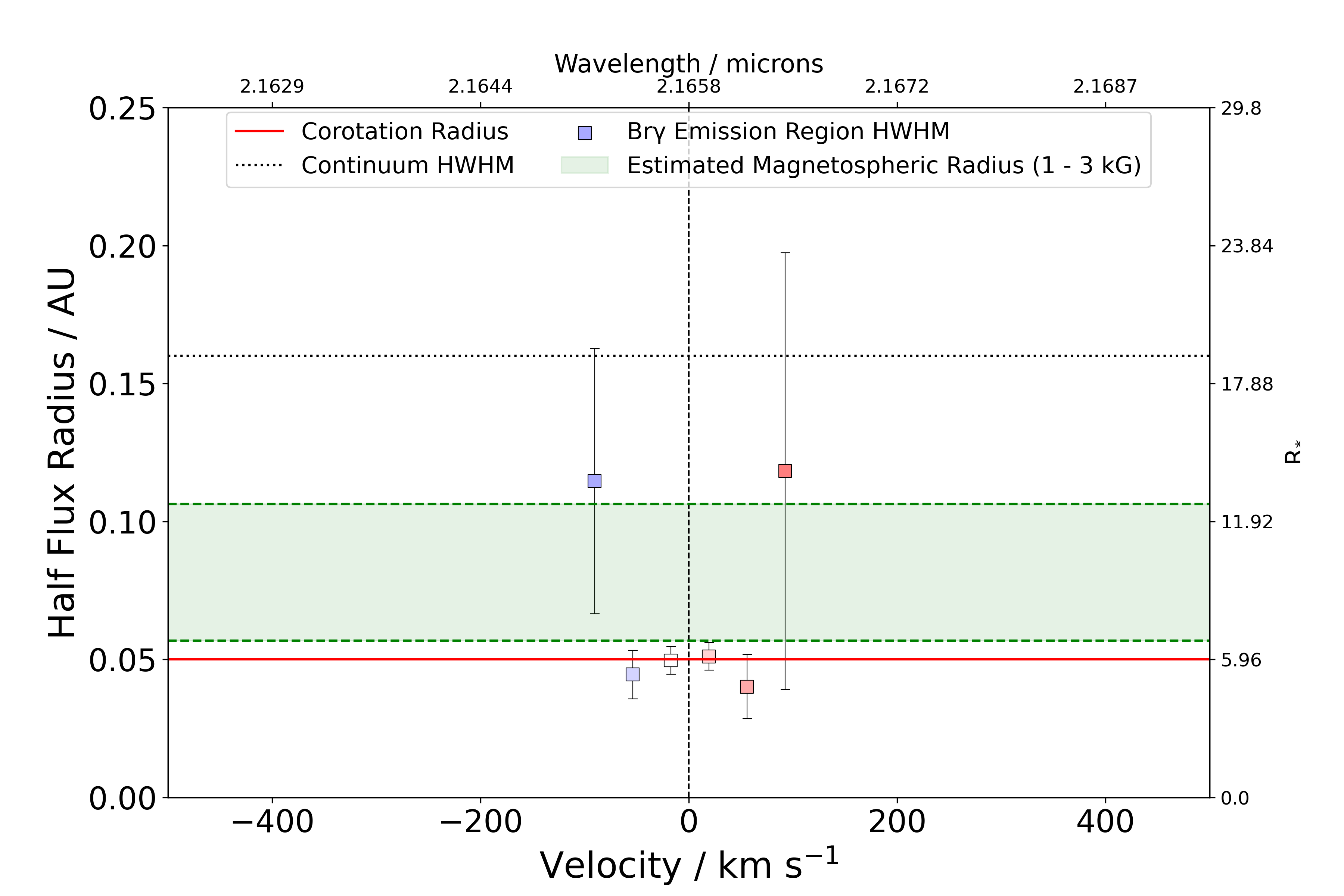}
\end{minipage}

\begin{minipage}{1\linewidth}
\centering
\includegraphics[width=\linewidth]{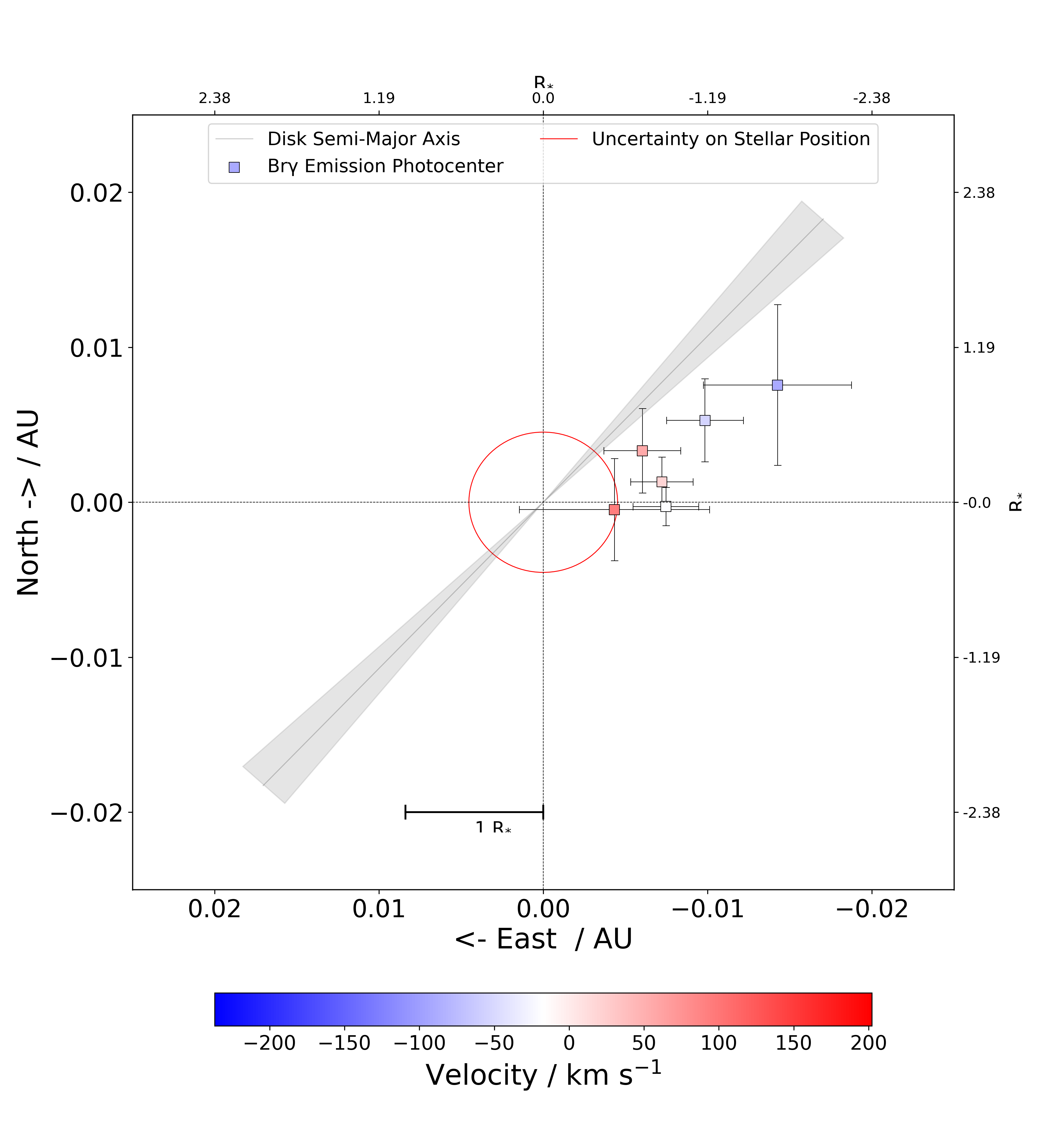}
\end{minipage}
    \caption{Spectrum (top), size (middle), and photocentre shift (bottom) profiles for DoAr 44. Shaded regions indicate uncertainties.}
    \label{fig:DoAr 44}
\end{figure}
DoAr 44 (Fig. \ref{fig:DoAr 44}) exhibits a symmetric line shape with peak line-to-continuum flux ratio of only 1.19,  making it the weakest \bg emitter in the sample. 
Similar to the case of S CrA N, the pure line emission region is completely unresolved at a number of channels for certain baselines, with pure line visibilities reaching values of 1  within the uncertainties. The so affected channel-baseline combinations were disregarded in the ellipse fitting process, as with a geometric Gaussian disk model these can only be associated with FWHMs of 0 arcseconds. The two outermost channels that meet the selection criterion are most affected by the reduced number of data points to fit the ellipse to, leading to disproportionately large error bars compared to the inner four channels. Even then, only for three of those four channels do we resolve the emission region under all six baselines simultaneously, which consequently makes those three channels the most robust estimators of the true region size that should be compared against the co-rotation radius of 0.05 au.\\
 The emission region has a half flux radius of 0.05 au, which remains relatively constant between the centre channels within the uncertainties.  With \bg emission coming from a region similar in extent to the co-rotation radius, it is likely that magnetospheric accretion is a driving mechanism behind a large part of the total line emission and that a significant portion of the emission originates in the accretion columns of the magnetosphere. At about 5.9 R$_*$, this size falls into the range of values typically associated with magnetospheres \citep{Bouvier2007}.\\
Our results are in agreement with those previously reported by \citet{Bouvier2020a}, who published a dedicated paper on a multi-instrument campaign on DoAr 44, including these GRAVITY interferometric observations. There they reported an upper limit to the emission region size of 0.047 au, which, when taking into account the impact of our additional spectral correction, is consistent with what we found.  \\
The differential phase signals are faint, the resulting distribution of \bg component photocentres hints at a separation of blue and red arms similar to a possible rotational profile, but the uncertainties prevent us from showing this conclusively. The maximum shift in the most blueshifted channel is of a magnitude of 0.016 au (1.9 R$_*$), while the red and centre channels cluster around an area at about 0.005 au (0.59 R$_*$) from the continuum photocentre. The closure phases show a weak offset in the continuum of around 0.9$^{\circ}$ which is still consistent with a shifted continuum photocentre of around that magnitude, meaning the observed \bg emission could in truth be centred on the position of the star. This is also in agreement with the findings of \citet{Bouvier2020a}, who reported a continuum offset of 50 $\mu$as, which almost exactly corresponds to the offset of the white channel in Fig. \ref{fig:DoAr 44}. \\
Despite the relatively compact emission region, the photocentre profile appears less concentrated than the similarly compact case of TW Hya, which is likely rooted in the relatively higher inclination (32 $\pm$ 4)$^{\circ}$ of the inner disk when compared to the pole-on case. The profile appears also to be almost perfectly aligned with the semi-major axis of the inner disk (137 $\pm$ 4)$^{\circ}$, which would be typical for a disk in rotation, but is not necessarily associated with a rotating magnetosphere. This could be explained by an additional wind component launched from the disk-magnetosphere interface, but in that case a blue excess in the line shape similar to what was observed in AS 353, for example, would be expected. As it stands, the alignment is either coincidental or the contribution from the wind is too weak to clearly modulate the line shape.

\section{Discussion}\label{sec:7}

Based on the results presented in Section \ref{sec:6}, the objects of our sample can be broken down into three distinct categories: For two of them (TW Hya and  DoAr 44) we found an emission region size that is of the order of the co-rotation radius or smaller, and also either of the order of, or smaller than, the truncation radius we estimated from the equivalent width. \\ 
Two of the other objects (AS 353 and DG Tau) showed emission regions that extend beyond 10 R$_*$, far exceeding both the co-rotation radius and also the range of truncation radii at magnetic field strengths of 1-3 kG. \\
The remaining T~Tauri stars (VV CrA, S CrA N, and both epochs of RU Lup) do not fit clearly into either of those two groups, featuring \bg emission regions with HWHMs either larger than the co-rotation radius, but consistent with the estimated range of truncation radii, or smaller than the co-rotation radius, but larger than even the upper limit of the truncation radius computed for a 3 kG field. \\
In this section, we discuss these results in the context of different \bg origin mechanisms in T~Tauri stars. We begin by first introducing our work on a radiative transfer model of magnetospheric accretion in order to establish a baseline for the expected spectral, size and photocentre profiles for \bg emission originating exclusively from the magnetosphere. We then proceed to consider how deviations from these baseline profiles in the form of additional components of extended \bg emission would be able to account for those results which are not well explained by the magnetospheric accretion scenario.

\subsection{Radiative transfer models of magnetospheric accretion}
\label{sec:6.1}
\begin{figure}[!htpb]
\centering
\includegraphics[width=\linewidth]{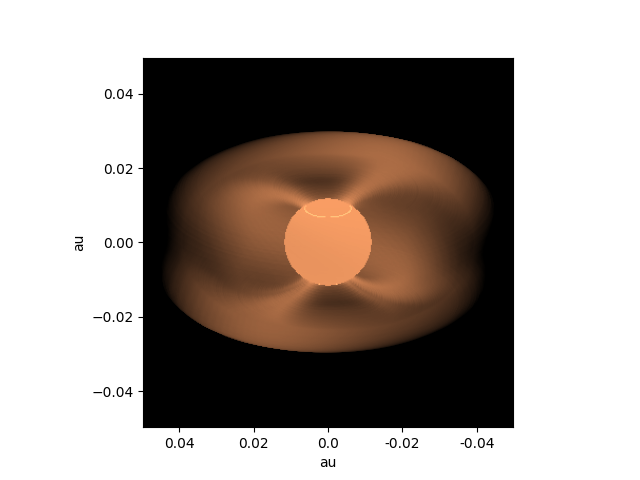}
\caption{\bg emission region image of the magnetospheric accretion region of an RU Lup-like system at the \bg peak wavelength, obtained from radiative transfer modelling with MCFOST. Depicted here are the accretion columns at the line centre, the star and the shock region at high latitude, as well as a contribution coming from a faint magnetospheric continuum. The model is axisymmetric, i.e. the magnetic axis is aligned with the stellar rotational axis, i.e. the vertical axis in the image above.}
\label{fig:modelimage}
   \end{figure}

   \begin{figure}[!h]
\centering
\begin{minipage}{\linewidth}
\centering
\includegraphics[width=\linewidth]{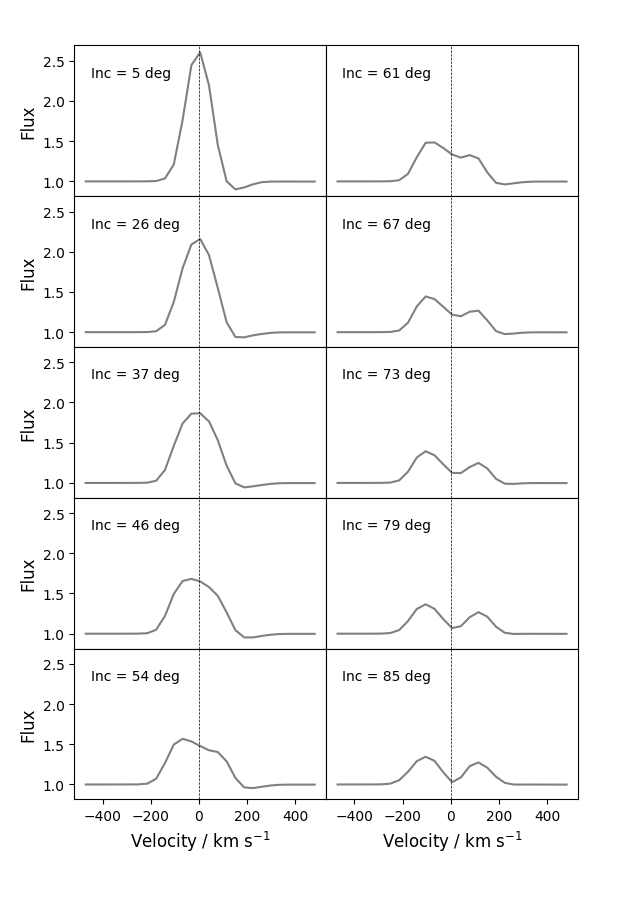}

\end{minipage}
      \caption{\bg spectral line profiles produced by the radiative transfer models at different inclinations of the magnetosphere. As we move to higher inclinations, the peak line-to-continuum flux ratio decreases and the line broadens. The inverse P~Cygni profile with a redshifted absorption feature disappears beyond inclinations of 54$^\circ$, as the accretion columns cover less of the stellar surface and the shock region. The blue- and redshifted velocity components of the rotating magnetosphere are increasingly separated, leading to a double peaked profile at high inclinations. We note that the synthetic data were convolved with a Gaussian and brought to the spectral resolution of GRAVITY.} 
         \label{fig:IncFlux}
        
   \end{figure}

\begin{figure*}[!h]

\begin{minipage}{1\linewidth}
\includegraphics[width=\linewidth]{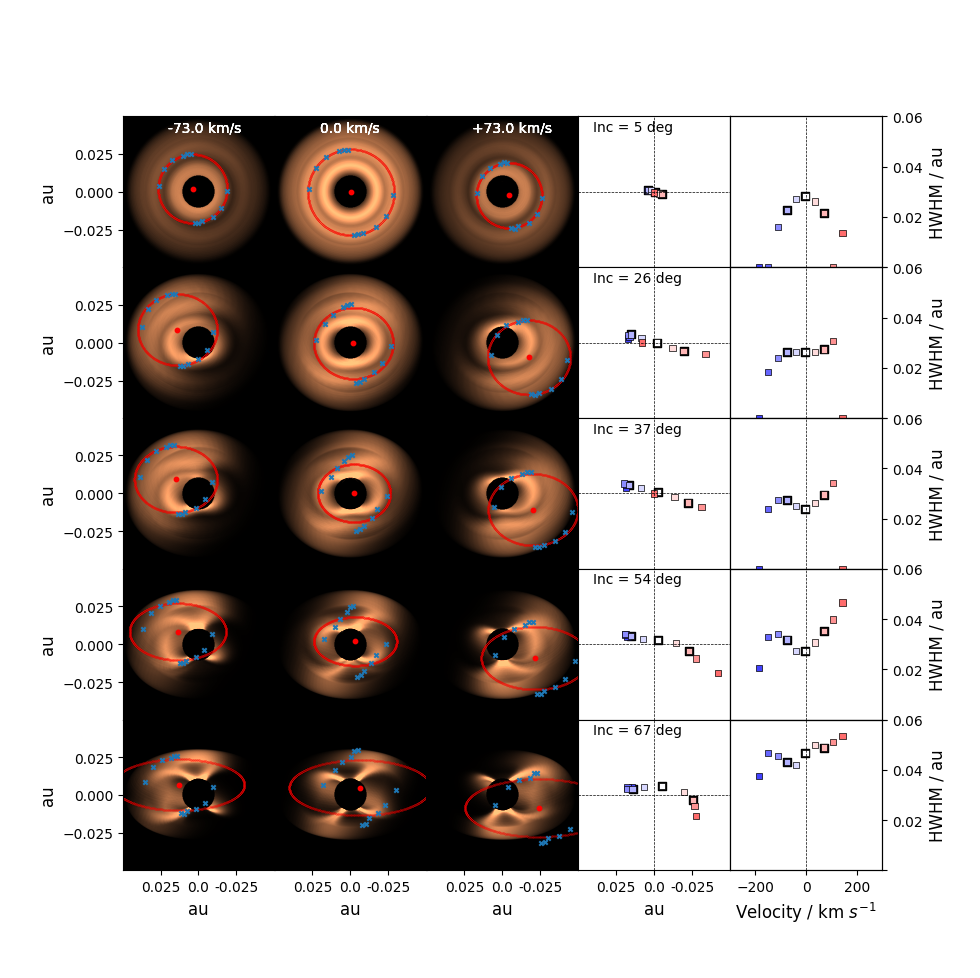}
\end{minipage}
\begin{minipage}{\linewidth}
\centering
\includegraphics[width=0.75\linewidth]{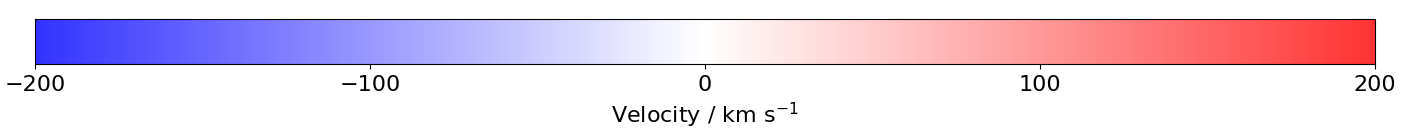}
\end{minipage}

\caption{ Model images and derived quantities of the \bg emission region. \\ The three columns on the left show the continuum-subtracted \bg emission region image at three different velocities, while the two columns on the right show the photocentre shift profile and the size profile across all velocity components of the \bg line, for a range of inclinations between 5$^\circ$ and 67$^\circ$. The model images correspond to the blue, white and red channels marked with a bold black outline, and the photocentres and region sizes in those particular channels are defined by the red dots and the semi-major axes of the red ellipses, respectively. Only the semi-major axis of the ellipse is left as a free parameter of the fit, which emulates the approach taken with the observational data. Fig. \ref{fig:inclination_effects} shows the corresponding profiles if inclination and position angle are also free parameters of the fit. }
\label{fig:inclination_effects_fixed}
\end{figure*}

\begin{figure*}[!h]

\begin{minipage}{1\linewidth}
\includegraphics[width=\linewidth]{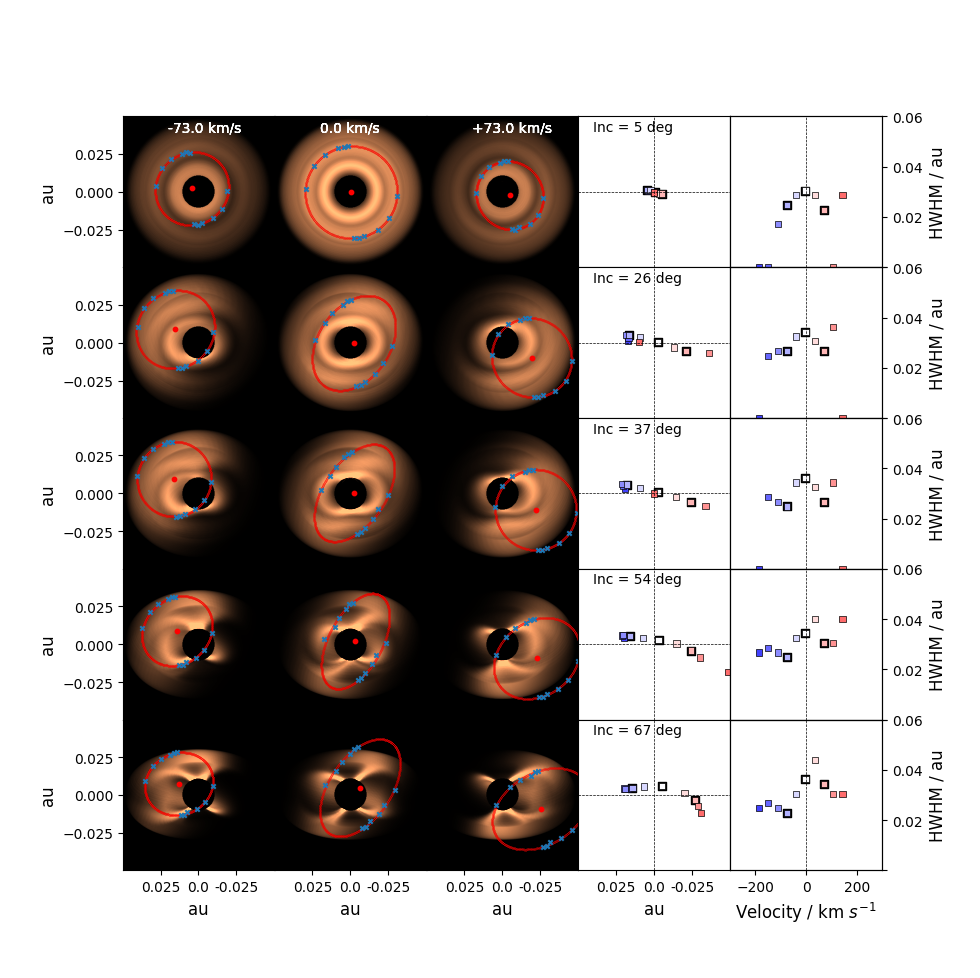}
\end{minipage}
\begin{minipage}{\linewidth}
\centering
\includegraphics[width=0.75\linewidth]{BryinTTauris/colorbar.png}
\end{minipage}

\caption{Model images, photocentre, and size profiles at different inclinations as in Fig. \ref{fig:inclination_effects_fixed}, but in this version the position angle and inclination of the red ellipse are free parameters of the fit. This approach does not match the data treatment and suffers from a lack of constraints along the axis of the uv coverage gap, but finds better fitting solutions at high inclinations. We note that while the size profile is different between the two approaches, the photocentre profile is not affected by this and identical to Fig. \ref{fig:inclination_effects_fixed}.}
\label{fig:inclination_effects}
\end{figure*}

Estimates of the co-rotation and the truncation radii serve as simple criteria to differentiate between objects for which \bg emission originates on spatially compact scales, consistent with the magnetospheric accretion scenario, and those showing an extended emission region, where other mechanisms are likely to contribute to a significant degree. Beyond this, it is challenging to assess the implications of the complete set of interferometric observables without being overly speculative. 
To provide us with a more complete idea of the expected signatures, and to aid us in the interpretation of the line shapes, the size profiles and the photocentre shift profiles, we additionally investigated the predictions of radiative transfer models of the magnetospheric accretion region (Tessore et al., in prep). We explored a number of model images, see Fig. \ref{fig:modelimage}, showing the magnetospheric emission region at different inclinations and different radial velocities across the \bg line. The model features an axisymmetric magnetosphere, meaning the magnetic dipole is aligned with the rotational axis of the star. While it has been shown in the past that a dipole tilt is common in T~Tauris (e.g. \citet{Johnstone2014}), we consider the axisymmetric model still as a first step to establish broad interferometric trends and their dependencies on the model parameters.\\
Each of the image cubes represents the magnetosphere at a specific set of stellar and magnetospheric properties, including the stellar radius, mass-accretion rate, magnetospheric temperature and the temperature of the shock region. We set the parameters to emulate one of our sample targets in order to produce a set of synthetic observables analogous to those obtained from the GRAVITY observations. We subsequently applied the methodology laid out in Section \ref{sec:5} to determine the half-flux radius of a Gaussian disk  and photocentre shift in each channel, so that the synthetic data were treated in the same way as the observational data.  \\
To complement this, we chose the physical model of the magnetosphere to represent an RU Lup-like system at R$_*$=2.5 R$_{\odot}$ and an intermediate accretion rate of $\dot{M}$ = 5 $\cdot$ 10$^{-8}$  $M_{\odot} yr^{-1}$ at a mass of 0.8 $M_{\odot}$  and a rotation period of 3.7 days. The magnetospheric temperature was set to 7500 K, with a shock region heated to 8000 K and a total magnetospheric radius of 4.57 R$_*$ (0.053 au), consisting of a truncation radius of 3.37 R$_*$ (0.039 au) and a magnetospheric width of 1.2 R$_*$ (0.014 au). \\
The model intensity distribution (Fig. \ref{fig:modelimage}) features \bg radiation coming from the emission line itself as well as a continuum contribution consisting of a photospheric component, the shock region and a faint magnetospheric continuum. It does not include an extended continuum component, such as the inner regions of a dusty disk. While this affects the direct comparison between synthetic and observational total observables, the pure line quantities  remain unchanged. As detailed in Appendix \ref{AppendixII}, in depth knowledge about the exact composition of the continuum is not required in order to recover the pure \bg line visibilities and differential phases. \\ 
A model spectrum was computed by integrating the image fluxes at each wavelength.
At low inclinations it shows a single peaked profile with a shallow absorption feature in the red wing, caused by infalling optically thick gas obscuring part of the stellar surface and the shock region. This inverse P~Cygni profile, which is considered to be a characteristic feature of hot hydrogen emission lines produced in accretion columns observed at face-on configurations \citep{Bouvier2007}, is visible in the model up to inclinations of 54$^{\circ}$ at velocities of 150 km/s up to 200 km/s, see Fig. \ref{fig:IncFlux}. 
At the lowest modelled inclination of 5$^\circ$, the emission line extends over a velocity range of about $\pm$ 150 km/s, whereas beyond 54$^\circ$, the profile becomes double peaked and the range expands to $\pm$ 220 km/s. \\
When observed close to pole-on, the image of centrosymmetric accretion flows forms a disk-like structure with an intensity distribution that features rings of variable brightness, decreasing from the stellar surface to the edge of the magnetosphere. At such a low inclination, it is reasonably well approximated in principle by a geometric Gaussian disk model, given that the 'hole' left by the star in the continuum-subtracted images (Fig. \ref{fig:inclination_effects} and \ref{fig:inclination_effects_fixed}, left side panel) is unresolved at typical GRAVITY baselines. As we move to higher inclinations, centrosymmetry is lost and the internal geometry of the intensity distribution becomes complex, although the outline is still well demarcated by an inclined ellipse in the centre channel. In the blue and red wings of the line, \bg emission is concentrated in regions away from the centre, so that the synthetic visibilities and consequently the fitted ellipse are tracing the corresponding substructures within the total magnetosphere.  \\
For an inclination of 5$^{\circ}$, the model system exhibits a decrease in size from the centre channel, with a half-flux radius of 0.032 au (2.75 R$_*$), to higher velocities, where the region can be as compact as 0.016 au (1.38 R$_*$). At higher inclinations, this trend becomes essentially inverted, with the centre channel forming a minimum and an increasing HWHM towards the line edges. In this case illustrated in Fig. \ref{fig:inclination_effects_fixed}, the ellipse position angle and inclination were fixed to be in line with the magnetosphere, following the approach we took with the GRAVITY data when fixing the ellipse fit at the PA and inclination of the NIR continuum disk. We revoked this assumption in Fig. \ref{fig:inclination_effects}. The results indicate that the exact choice of methodology potentially can have a large impact on the recovered trends across the line at the various inclinations. When all the parameters of the ellipse are left free for fitting, the size profile at higher inclinations broadly retains the  original decrease in HWHM seen at 5$^\circ$ inclination, even if some of the individual channels show some fluctuations. \\
The best fit result of the ellipse for the 5$^{\circ}$ inclination suggests that the half-flux radius of the Gaussian disk underestimates the actual spatial extent of the full \bg emission region by a factor of about 2/3 at the centre of the line. There the 0.032 au derived from the geometric model fall short in comparison to a magnetosphere with a defined size of 0.053 au. This discrepancy appears less pronounced at higher inclinations, although the comparison is not necessarily meaningful, as the intensity distribution in those cases is no longer well approximated by a simple geometric model. So while the ellipse fits based on the Gaussian disk do yield size estimates on roughly the appropriate order of magnitude, it is clear that a geometric Gaussian disk or ring model is not well suited to represent the realities of a system of accretion columns at higher inclinations, even for an axisymmetric magnetosphere.
\\
The photocentre in each velocity channel was determined from a fit of the synthetic differential phases to again ensure consistency with the approach taken with the GRAVITY data. As a consistency check, we also computed the photocentre directly by calculating the barycentre of the intensity distribution, and found excellent agreement between the results of the two methods.
The so obtained photocentre shift profiles are highly compact at inclinations of 5$^\circ$ (Fig. \ref{fig:inclination_effects_fixed} and \ref{fig:inclination_effects}, centre panel), as at such a low angle the system of funnel flows is essentially centrosymmetric for the model of an axisymmetric magnetosphere. The maximum magnitude of shift in such a case is close to 0.007 au at 5$^\circ$ , which corresponds to 0.6 R$_*$. As the inclination increases, we see a more extended profile, similar to a disk in Keplerian rotation, which begins to bend inwards at the edges at inclinations above 54$^\circ$. The global maximum photocentre shift is thus found at moderate inclinations and remains relatively stable from 37$^\circ$ to 54$^\circ$ at 0.034 au or 2.9 R$_*$.  At low to intermediate inclinations, the quasi-rotational profile is aligned at an angle of about 60$^{\circ}$ relative to the 'north' (i.e. the vertical) axis. In this the photocentre profile of the rotating magnetosphere is distinct from a rotating disk, as in the latter case the profile would be aligned with the disk semi-major axis. 

\subsection{Comparison of model to observational data}

\begin{figure}[!t]
\centering
\begin{minipage}{\linewidth}
\centering
\includegraphics[width=\linewidth]{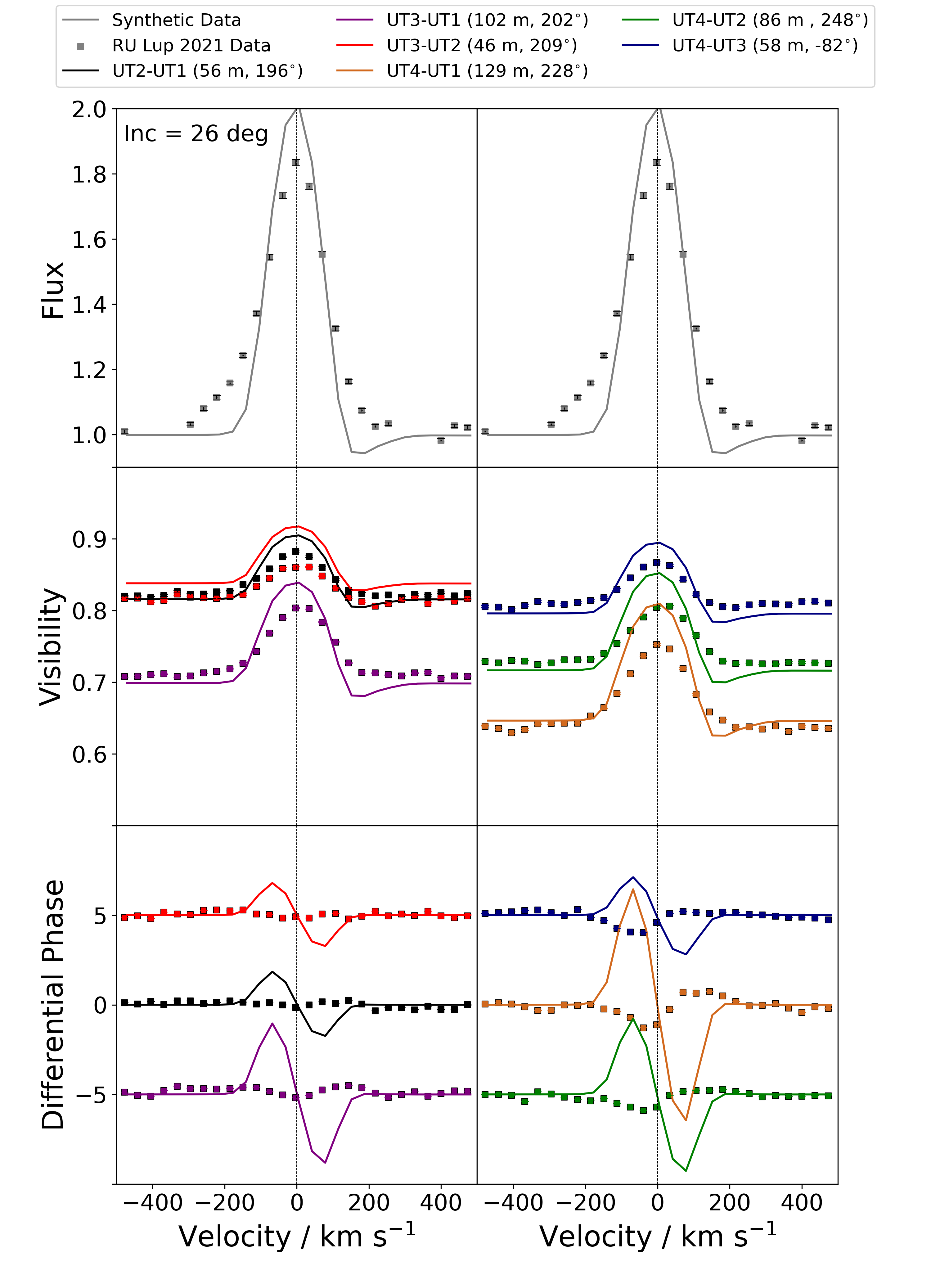}

\end{minipage}
      \caption{Synthetic observables (solid lines) computed from the radiative transfer images of the magnetosphere, compared to the observational data of RU Lup from 2021. A \bg spectrum was computed from integrating the image fluxes at different wavelengths. Synthetic visibility amplitudes and phases were obtained from Fourier transformation of the images. In order to ensure comparability with the total observables recorded by GRAVITY, an additional continuum disk contribution, taking into account parameters like disk size, disk flux and halo flux, was included through a form of post processing. \protect\\
      The figure shows the synthetic data for disk parameters chosen in accordance with the results reported in \citet{Perraut2021}.}
         \label{fig:synth}
        
   \end{figure}

\begin{figure*}[!htbp]
\centering
\begin{minipage}{0.33\linewidth}
\centering
\includegraphics[width=\linewidth]{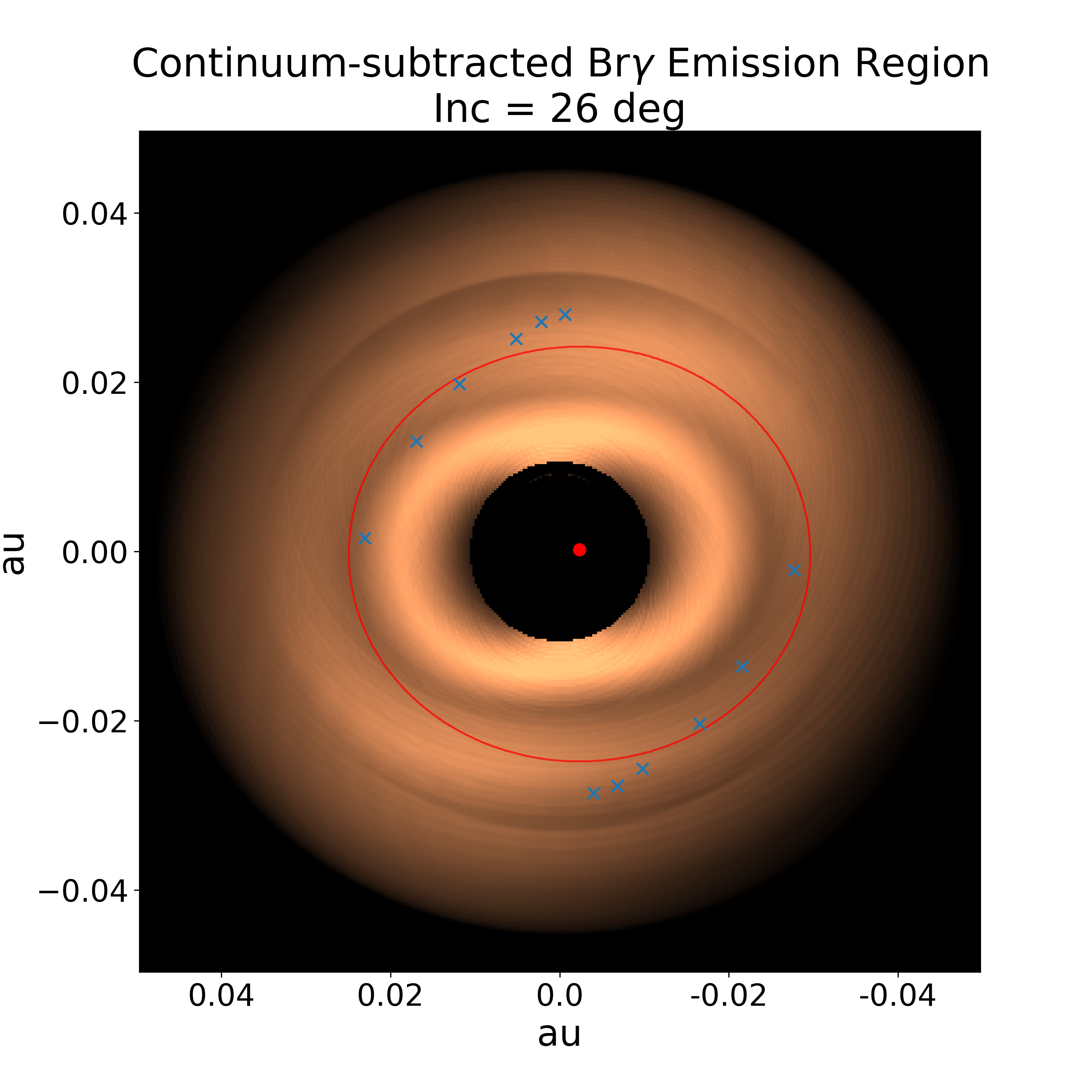}
\end{minipage}
\hfill
\begin{minipage}{0.33\linewidth}
\centering
\includegraphics[width=\linewidth]{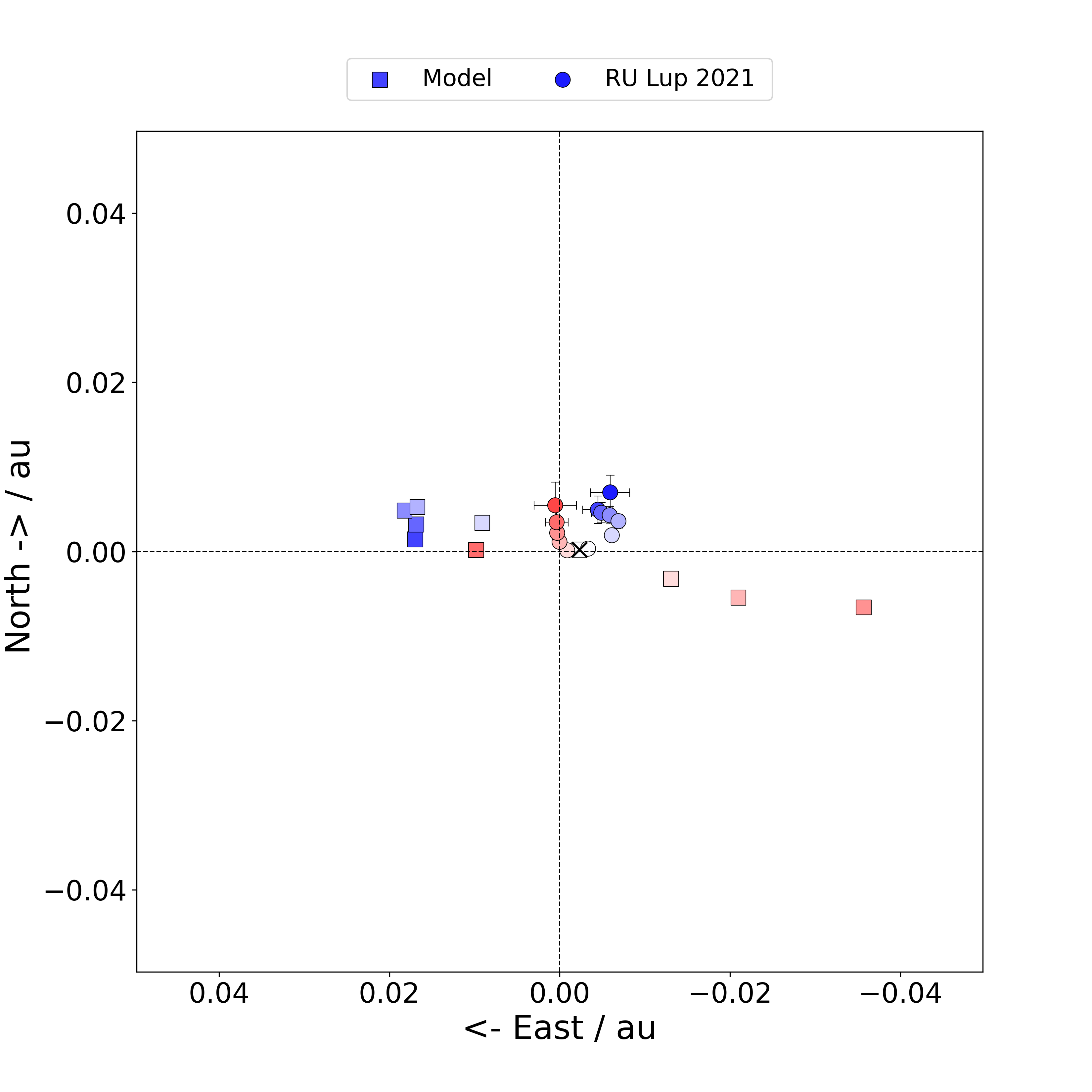}
\end{minipage}
\hfill
\begin{minipage}{0.33\linewidth}
\centering
\includegraphics[width=1\linewidth]{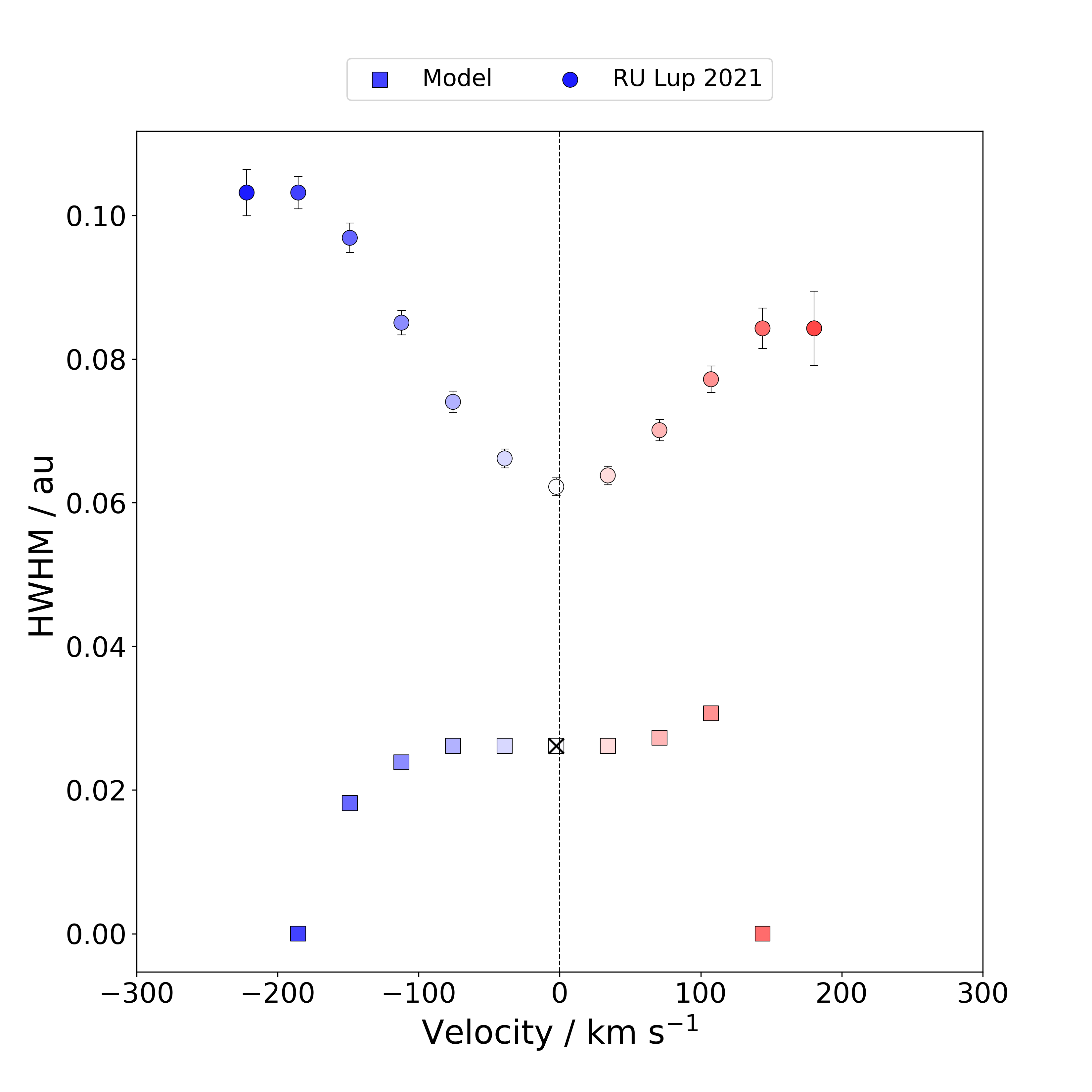}
\end{minipage}

\caption{Model image and derived quantities for an RU Lup-like system. \\
\textbf{Left:} A continuum subtracted model image of the magnetosphere at 26$^\circ$ inclination, representing the pure line emission region at the central channel of the \bg line. The size of the region and its photocentre, depicted here as ellipse and red dot, were both determined from the synthetic observables in a process analogous to the way the GRAVITY data were treated. \protect\\ 
\textbf{Centre:} The model photocentre shift profile across the line, compared to the results for RU Lup 2021. The model profile is largely aligned along a single axis, while the observational profile shows a significantly higher degree of complexity. \protect\\
\textbf{Right:} Variation of emission region size across the line. While the size of the magnetosphere is an input parameter of the model, the change of the half flux radii at different velocities still provides a useful baseline to which the trends in the sample can be compared. The observational profile shows a region more than twice the size of the model HWHM at the central channel, growing larger at the line edges, whereas the model predicts a decline in size in the wings of the line. }
         \label{fig:synthsize1}
        
   \end{figure*}
Whilst the model cube was set up to be comparable to an RU Lup-like system, it is still useful to compare the broader trends against the rest of the sample. The range of inclinations covered by the seven objects starts at the almost face-on case of TW Hya and goes up to (49 $\pm$ 4)$^{\circ}$ for DG Tau, with most of the objects being observed at low inclinations between 20$^{\circ}$ and 30$^{\circ}$. In such cases, the spectral line is single peaked with a slight asymmetry caused by redshifted absorption at low inclinations, and a blueshifted line peak at intermediate inclinations. This behaviour is not consistent with the observed emission lines from the GRAVITY sample, as clearly visible absorption features in the red wing or shifts of the peak emission away from the centre are never detected in the observational data. Conversely, the asymmetric broadening of the base of the emission line, observed in four of the seven targets, is not reproduced by the model. \\
The size profile of the model shows that, at those low to moderate inclinations represented by the majority of sample objects, the HWHM either decreases towards the line edges, or remains relatively flat. The photocentre profile is either very clustered within a radius of 1 R$_*$, if observed face-on, or rotational and well aligned along a specific axis at an angle to the disk PA. 
Both of these trends together are roughly consistent with only the two weakest accretors. For the other objects we measure, within the error bars, either flat or increasing sizes at higher velocities. Only the two epochs of RU Lup exhibit a clear U-shape, such as can be seen in the model. However, in the model this feature only appears at significantly higher inclinations than those determined for the RU Lup system. \\
Most of the objects show photocentre profiles with  higher orders of complexity than a simple alignment along one axis. One notable exception here is AS 353, for which the photocentre profile agrees with the model prediction in that it shows both a single axis rotational profile, but also a misalignment between that photocentre axis and the disk PA. However, for this target the emission region size is highly extended far beyond typical truncation radii, indicating that these photocentre profiles are not unique to the case of a rotating magnetosphere. \\
Overall the model trends are most accurately replicated by the data for TW Hya.  Both the photocentre clustering and the smaller half-flux radii at higher velocities correspond very well to what is depicted in Figures \ref{fig:inclination_effects_fixed} for a 5$^{\circ}$ inclination, further strengthening the argument for a \bg emission region dominated by magnetospheric accretion \citep{GarciaLopez2020}. Still, even for this object, the inverse P~Cygni profile that is predicted by the models is curiously absent from the observation, as is the case across the entire sample. It is possible that in TW Hya the asymmetric line shape is rather an expression of the absorption in the red wing rather than of an excess in the blue wing. The relatively narrow emission line in TW Hya, which matches up well with the range of $\pm$200 km/s predicted by the model, could support this notion. We detect a notably larger range of velocities in emission for the other objects with strong asymmetric features, where the emission region size indicated  spatially extended \bg emission on a scale far beyond the magnetosphere.  \\
DoAr 44 is another candidate for a \bg emission region that is strongly shaped by magnetospheric accretion, although in this case the comparison against the model profiles is not as conclusive as for TW Hya. At the higher inclination of 32$^\circ$ of DoAr 44, the size profile derived from the model is more impacted by the methodology behind the emission region fit, although Fig. \ref{fig:inclination_effects_fixed} shows profiles which are relatively flat for the centre channels at 26$^\circ$ and 37$^\circ$. This would be consistent with what is seen in Fig. \ref{fig:DoAr 44}. The photocentre profile of DoAr 44 can be interpreted as quasi-rotational, although the low number of selected channels with sufficient flux and the relatively large uncertainties on the photocentre positions prevent this from being conclusive.  Whether the GRAVITY spectrum of DoAr 44 shows the expected redshifted absorption is open to interpretation due to the relatively high dispersion among the continuum-level data points. At this level of inclination, the absorption peak would already be significantly less pronounced than in the pole-on case of TW Hya, see Fig. \ref{fig:IncFlux}, so that the discrepancy between observed spectral line and model line shape is not large. The model thus seems to again support the idea that \bg emission in DoAr 44 is indeed largely stemming from magnetospheric accretion processes, as was also proposed by \citep{Bouvier2020a} \\
VV CrA presents itself as something of an edge case here due to the very large uncertainties associated with the HWHMs and photocentre offsets. It appears that at least for the central channels the region is more compact (6.56 R$_*$) than the co-rotation radius (7.25 R$_*$), yet we also find a magnetospheric radius between 2.86 and 5.36 R$_*$ for 1 and 3 kG magnetic fields, respectively. As the emission region extends further than what we would consider a reasonable upper limit on the truncation radius, the \bg emission does not seem to trace the magnetosphere.
For the remaining objects, comparisons between images and observations indicate that \bg emission is not well described by only the magnetospheric accretion model. We see strong asymmetries in the line shapes, with moderate to high amounts of excess blueshifted emission, which are not explained by magnetospheric accretion alone. This is also true for the size profiles across the line, which appears either flat or U-shaped, with a minimum at the centre. Such profiles can be seen in the model when fixing inclination and position angle for the ellipse fit (Fig. \ref{fig:inclination_effects_fixed}), but generally at higher inclinations than those measured for our sample objects. The photocentre profiles equally show a high degree of complexity, with either multiple alignments in different parts of the line or shift magnitudes far beyond the magnetosphere. We also must account for the possibility that some of the discrepancies between models and observations certainly arise from the axisymmetric nature of the model, given that non-aligned cases are common and obliquity angles between stellar dipole and rotational axis are typically on orders of 10$^\circ$ \citep{McGinnis2020}. This would likely affect the photocentre shift profiles the most, whereas the impact on the size profile is not trivial to evaluate intuitively.
\\
To facilitate a more detailed comparison between the model of the RU Lup-like system and the data of the RU Lup observations, we made use of a custom developed tool which allowed us to derive the spectral and interferometric observables at a uv coverage set to correspond to the actual observation of RU Lup in 2021. The tool has the ability to modify the observables in such a way that they reflect the impact of varying the physical and geometric properties of the system, such as the inclination of the magnetosphere. These modifications also include the possible addition of a continuum disk contribution to compensate for the lack of a K-band emitting dusty disk in the original model, see Appendix \ref{AppendixII}. 
The full set of parameters that can be manipulated to produce the spectrum, visibility amplitudes and differential phases are: the common inclination and position angle of a system of continuum disk and magnetosphere, the size of the disk and the flux percentages coming from either the disk or an unresolved halo component. \\
Fig. \ref{fig:synth} shows the synthetic data versus the observational data from 2021. In this case, the continuum disk size, disk flux, halo flux, position angle and inclination were set to the values reported in \citet{Perraut2021}. At those parameters, the peak flux agrees with the observational results, but the line shape is lacking the broad base seen in the RU Lup data. The absorption feature in the red wing of the model spectrum translates to a similar dip in the model visibilities, as the obscuration of the stellar surface  increases the relative weight of flux coming from a more extended area. The synthetic visibilities reproduce the observational continuum visibilities, but overestimate the \bg peak, suggesting that the model is significantly more compact than the emission region that is reflected in the observational data. This is also reflected in the size profile (Fig. \ref{fig:synthsize1}), where we also see that the change across the line is essentially inverted for the model when compared to the observational profile. 
The synthetic differential phases exhibit S-shaped signals at all baselines, with two phase peaks with opposite signs at red- and blueshifted velocities. The amplitude of these peaks ranges from 1$^{\circ}$ to 5$^{\circ}$, which is an order of magnitude larger than the signals we detect in the observational data.  This translates into a far more extended photocentre shift profile than the comparatively compact observed shifts for RU Lup in 2021, and also shows a clear alignment along one axis. The model here fails to reproduce the curved shape of the observational photocentre profile. We also see that the S-shapes produced by the model are essentially inverted in sign when compared to the observational data, indicating that the sense of rotation of the model is also essentially inverted with respect to the observed RU Lup system. \\
We also made an attempt to reproduce the observational data as closely as possible by fitting the model parameters, and found that we were not able to produce synthetic data in agreement with flux, visibilities and differential phases simultaneously. It is possible to reproduce the visibilities quite accurately, but only at a significantly higher inclination than what was found for RU Lup, and in such case the line flux would be severely underestimated. It is possible to find a much better agreement between synthetic and observed differential phases, mostly by manipulating the position angle of the system, but only at the expense of the visibilities. We were also not able to reproduce the broad base of the spectral line through any combination of parameters. \\
It is possible that a non-axisymmetric model, featuring a tilted magnetosphere, is better suited to resolve some of these discrepancies. The photocentre shift profile in particular, and its dependency on the inclination, is likely to be affected by the formation of preferential funnel flows per hemisphere, as is predicted in such a case.  However, even such a modification to the model is able to reconcile the observed large-scale emission in some objects with the compact scale of typical magnetospheres, nor would it account for the broad base of the emission lines or explain \bg emission coming from beyond the co-rotation radius. \\
With this in mind, based on the observational results and the work done on the model cube, we reaffirm our original classification of our sample objects: sources for which magnetospheric accretion is the main driver of \bg emission (TW Hya and DoAr 44), those which show spatially extended emission, likely predominantly originating in a disk or stellar wind (DG Tau and AS 353), and those that exhibit features not consistent with the model, but with characteristic sizes still of the order of the 4-7 R$_*$ range associated with typical magnetospheres. For the latter group, a mixed contribution to the total \bg flux coming from multiple emission components is likely  responsible for the detected size and photocentre profiles.

\subsection{Extended \bg emission: Disk and stellar winds}
While hydrogen line emission in T~Tauri stars is strongly associated with the magnetic heating mechanisms in the accretion flows, magnetospheric accretion alone can only account for \bg radiation produced in a compact region close to the stellar surface. This implies that spatially extended emission must arise from one or multiple other components in the system. 
In the context of Herbig Ae/Be type YSOs, the hot accretion disk itself is readily proposed as a possible source of \bg emission (e.g. \citet{Kraus2008}, \citet{Tambovtseva2016}), given the hot temperature environments found in the inner disk of intermediate mass objects. This idea has been extended to the case of the T~Tauri stars in the early studies published by \citet{Eisner2010} and \cite{Eisner2014}, who chose a Keplerian disk model and an infall/outflow model when attempting to reproduce their interferometric data and found no significant difference between the two in terms of the quality of their fit.
However, computations of the radial temperature profile in the inner disk of T~Tauri stars \citep{Bertout1988} suggest disk temperatures, even at distances of 0.02 au, do not exceed 4000 K and thus fall short of the 8000 - 10000 K range we would expect from a strong \bg emitter. By contrast, it has been shown that in magnetically driven winds hydrogen gas can be heated to temperatures of 10$^4$ K via ambipolar diffusion very quickly after being launched \citep{Garcia2001}. This makes a wind a suitable candidate for an extended \bg emitting component in the inner disk system.
Such a wind might come in the form of an actual disk wind, launched from the surface of the inner disk, or originate from the polar region of the star as a stellar wind (e.g. \citealt{Ferreira2006}, \citealt{MattPudritz2005}). Interactions between stellar and disk winds, which can occur simultaneously, can also lead to material being ejected from the interface region between disk and stellar wind \citep{Zanni2013}. 
Ultimately it is unclear whether it would be possible to distinguish between different types of outflow purely based on interferometric observations, as wide opening angles may lead to certain overlap of the possible scales of wind-driven emission. For the purpose and in the context of this discussion, we only refer to a wind as a \bg emission component that is spatially extended beyond the magnetosphere and not differentiate any further. In any case, given that the rotation of the accretion disk is likely to transfer to and kinematically dominate the launch region of a disk wind, such a scenario could potentially reconcile the low temperatures of the accretion disk with the quasi-rotational photocentre profiles that we found for some of the our objects with extended emission regions. \\
Disk and stellar winds as \bg emission sources have in the past also been discussed for Herbig Ae/Be stars: \citet{Kurosawa2016} examined spectro-interferometric data taken with VLTI AMBER, the predecessor to GRAVITY, and modelled the \bg feature and associated interferometric observables with both a magnetosphere and a disk wind. \citet{GarciaLopez2015} similarly explored AMBER data on the Herbig star VV Ser and found that a disk wind model was able to best reproduce the interferometric data, but did not match the line shapes, which were better reproduced by a bipolar outflow model. \citet{Weigelt2011} examined the scenario of a wind launched from a rotating disk for a Herbig Be star and produced model images for a \bg disk wind with which they were able to reproduce their interferometric observables. The images they present show changes in the intensity distribution across the line that appears very similar to a rotating magnetosphere observed at low inclination, and would likely lead to a similarly rotational photocentre profile. 
\citet{Wilson2022} presented a spectroscopic study on hydrogen emission lines in a number of T~Tauri stars, including DG Tau, which they compared against radiative transfer models of accretion with an additional stellar wind. Their model was able to reproduce the characteristics of the H$\alpha$ line, but showed significant divergence from the observed infrared emission lines, including the \bg feature.  \\
Most objects in the GRAVITY T~Tauri sample show emission lines which are either comparable in width to those derived from the radiative transfer model for low to moderate inclinations, or are asymmetric to the point that they are best described by a superposition of two Gaussians. Of the two Gaussians, one is typically broader and shifted to a higher blueshifted velocity, whereas the second one is comparatively narrow and centred. The narrow component is across the sample consistently similar in width to the magnetospheric model emission line, extending between $\pm$ 200 km/s, whereas the broad component appears at quite different widths and positions for the different targets, and can change significantly even for the same object between two epochs (see Fig. \ref{fig:RULup2021}).\\
It is possible that the asymmetry is a manifestation of a multi-component \bg emission region, consisting of a magnetospheric accretion component and additional outflow components in the form of disk or stellar winds. Such an interpretation would help to explain the prevalence of the asymmetries and broad bases of the line shapes throughout the sample, given that the presence of outflows is generally expected in accreting objects due to their role in managing angular momentum in the system. This is especially true for objects with known associated large-scale jets, such as AS 353, DG Tau, and RU Lup. We note that AS 353 and RU Lup feature emission beyond the co-rotation radius as well as broad high velocity Gaussian components centred between -150 km/s to -50 km/s, which is similar to what has been reported by \citet{Garcia2001b} for the high velocity component of an atomic wind close to the star. \\
In AS 353, \bg radiation was detected up to a HWHM of more than 13 R$_*$, which is about twice as far out as the co-rotation radius of the system. The photocentre profile is quasi-rotational, which would match the idea of a wind base in rotation. However, it is not clear if the scale of the photocentre profile is consistent with such an extended wind, as even the most displaced photocentre exhibits a shift magnitude of less than 2 R$_*$.  \\
In RU Lup we equally observe \bg emission from outside the co-rotation radius, but at 6.3 R$_*$ in 2018 and 5 R$_*$ in 2021 the emission region HWHM is still consistent with typical magnetospheric scales. If there is an additional wind present in the system, its contribution to the total \bg output would likely be comparatively weaker than in the case of AS 353. The strong U-shape in the size profile at both epochs could also be explained in the context of a wind. In the case of a magneto-centrifugally driven wind, observed at low inclination, an accelerated flow of hydrogen ejected at a wide opening angle would reach higher velocities at more extended spatial scales. The complex photocentre profile of RU Lup is also not well explained by either a rotating wind or a rotating axisymmetric magnetosphere alone, as both of these scenarios on their own should produce a profile with a singular axis alignment. Whether a combination of a rotating magnetosphere and a disk wind could lead to an RU Lup-like profile is a question that ultimately requires more advanced modelling to answer. \\
The line shape observed in DG Tau is peculiar insofar as that it is one of only two objects in the sample for which the blueshifted excess is not at all detected, despite the fact that it has one of the largest emission region HWHMs of 12.9 R$_*$ and is also the only object which shows \bg radiation originating beyond the NIR continuum HWHM. 
The line extends almost to a range of $\pm$ 250 - 300 km/s and is as such broader than predicted by the magnetospheric accretion model. It is possible that inclination effects play a part here, but since DG Tau is observed at a similar inclination as AS 353, this alone would not account for the discrepancy in line shape between the two objects. Either way, the case of DG Tau casts some doubt on whether the Gaussian decomposition into broad and narrow components can be straightforwardly tied to physical emission components. 
Within the uncertainties, the photocentre profile can be interpreted as effectively rotational and well aligned with the disk profile, which also supports the idea of a wind launched from a rotating base, especially since in this case we know the photocentre profile mimics the one reported for the large-scale rotating CO disk. \\
Overall five of the seven objects show signs of extended emission. For two of them (DG Tau and AS 353) the \bg feature is likely driven predominantly by a rotating wind, whereas two (RU Lup and S CrA N) could see a more even ratio between the \bg flux originating from the magnetosphere and the flux originating in an extended component, based on the region size compared to the co-rotation radius and how well rotation alone describes the photocentre profile.  The case of VV CrA is not conclusive due to the large uncertainties on the photocentre positions and half-flux radii, although overall the observables also indicate a mixed origin of \bg emission.\\
In the broader context of the analysis of the \bg emission size as a function of mass accretion (see Fig. \ref{sizelum1}), we remind that the accretion rates were derived using the empirical relationship of \citet{Alcala2014}, established from a population with significantly weaker accretion properties than most of the objects of our sample show. However, the validity of this relationship for our study is well supported by the fact that it remains unchanged when including results for stronger accretors from \citet{Muzerolle1998} and essentially \citet{Calvet2004} (see panel C.7 for \bg in \citet{Alcala2014}). Still, the strong dispersion of the experimental points translates into large uncertainties on the coefficients $a$ and $b$ in Eq.~\ref{eq:alcala}, itself resulting in large error bars reported on the mass-accretion rates. 
In the light of our results, we propose that improved constraints on the connection between line luminosity and accretion rate could potentially be obtained through further quantitative estimates of the balance between 
mass inflow and outflow mechanisms (e.g. magnetospheric accretion vs. disk winds) in the inner disk.

\section{Summary}
We presented spectro-interferometric data for a sample of seven T~Tauri stars, observed in the K-band with VLTI GRAVITY between 2017 and 2021, focussing on the \bg hydrogen emission line at 2.16 $\mathrm{\mu}$m. For the first time, we were able to spatially resolve the innermost star-disk interaction regions for these objects, which are home to complex and time-variable phenomena, with an unprecedented accuracy of down to $\sim$ 0.001 au scales, improving upon previous campaigns, such as presented in \citet{Eisner2014}, by an order of magnitude.  \\
Based on the \bg spectra, spectrally dispersed interferometric visibilities, and differential phases, we determined the HWHM of a geometrical Gaussian disk model in a number of spectral channels of the line, probing the change of the emission region size at different radial velocities. We also extracted the relative photocentre shift of the pure \bg emission region with respect to the continuum photocentre from the differential phases and showed the distribution of \bg emission photocentres in the different wavelength channels and corresponding velocities. Based on these results, we commented on generalised trends within the sample and attempted to identify distinct groups of objects based on their characteristics. These results were interpreted with respect to the likely origin of \bg radiation in the respective systems, especially in the context of magnetospheric accretion, which is thought to be a major driver of \bg emission in T~Tauri stars. To assist in the interpretation and to help distinguish \bg emission produced in accretion flows from other potential mechanisms, we used a radiative transfer model of an axisymmetric accreting magnetosphere to establish characteristic interferometric signatures which could serve as baselines against which to compare our observational results. To summarise the most important findings and conclusions:
\begin{itemize}
 \item Out of a sample of seven objects in total, five exhibit \bg emission lines with asymmetric line shapes. These lines feature a broad base with significant excess in the blue wing and can be decomposed into a broad high-velocity Gaussian and a narrow low-velocity Gaussian component. The lines of the remaining two sources (DG Tau and DoAr 44) are well approximated by a single Gaussian function.
\item Across the sample, two objects (TW Hya and DoAr 44) show emission coming predominantly from within the co-rotation radius and Gaussian disk HWHMs that are comparable with the typical magnetospheric radii of around 5 R$_*$. This is in agreement with previous findings in \citet{GarciaLopez2020} and \citet{Bouvier2020a}.
 \item For another two objects (DG Tau and AS 353), \bg half-flux radii extend to scales outside the co-rotation radius and beyond 10 R$_*$. We expect that other, spatially extended, sources of \bg emission are present in these systems and are major, possibly dominant, contributors to the total line-flux output. Winds are likely candidates for such components. DG Tau is also the only source in the sample for which the \bg HWHM is larger than the NIR continuum HWHM. Strong jets were previously reported for both of these objects.
 \item \bg emission region sizes for the remaining objects are of the order of 4 to 7 R$_*$, but the data indicate that \bg emission is nonetheless extending to outside the co-rotation radius or beyond the upper limit of estimated truncation radii for plausible magnetic field strengths. Such behaviour can indicate simultaneous emission from both a magnetospheric and a more extended component, possibly featuring a wind launched from close to the disk-magnetosphere interface region.
 \item The photocentre profiles can be roughly classified based on whether they show a quasi-rotational alignment of photocentres along a single axis, similar to what would be expected from a disk in Keplerian rotation, or if they exhibit more complex patterns. Large uncertainties on the position of the photocentres complicate this classification. Out of the seven objects, one (AS 353) shows a distribution of photocentres across the line that most resembles a rotational profile. Two other objects (DG Tau and DoAr 44) can be interpreted as quasi rotational when their uncertainties are taken into account. One (TW Hya) shows an indistinct clustering of points within a single stellar radius that would be consistent with a rotating system observed at a close to pole-on configuration.
 \item The synthetic data derived from the radiative transfer models allowed us to identify trends across the line that match the observations of TW Hya and DoAr 44 in terms of their size and photocentre profiles. At intermediate inclinations, the photocentre profile of the rotating axisymmetric magnetosphere shows a rotational profile that is misaligned with the disk semi-major axis by about 30$^{\circ}$. At higher inclinations, we found that methodological effects when determining the region size can have a large impact on the derived profile.
 \end{itemize}
 
 In conclusion, we find that the observational data fit the case of magnetospheric accretion best for the two weakest accretors in the sample (TW Hya and DoAr 44), while objects with higher mass-accretion rates also appear to feature significant influences from more spatially extended \bg emission origins, which may come in the form of disk or stellar winds. Observations at very high angular resolution, such as those that only long baseline interferometry is able to provide in the foreseeable future, will be vital for any attempt to disentangle the different contributions in order to improve our understanding of the underlying processes in the star-disk interaction region. In particular, we are looking forward to the possibilities that will be unlocked by the implementation of the GRAVITY+ project, as the upgrade in sensitivity will make new reservoirs of potential sources available for future investigations. Additionally, multi-technique campaigns which can, for example, include polarimetric measurements of the magnetic field topologies, as well as repeated observations of individual sources to probe the time variability of the innermost regions, will offer great opportunities to gain new insights into the connection between multiple phenomena in the inner disk and strengthen the depth of the analysis.

\section*{Acknowledgements}

This work was supported by the “Programme National de Physique Stellaire” (PNPS) of CNRS/INSU co-funded by CEA and CNES. This project has received funding from the European Research Council (ERC) under the European Union’s Horizon 2020 research and innovation programme (grant agreement no. 742095; SPIDI: StarPlanets-Inner Disk-
Interactions, \hyperlink{http://www.spidi-eu.org}{http://www.spidi-eu.org}, and grant agreement no. 740651
NewWorlds). 
This work has made use of data from the European Space Agency
(ESA) mission Gaia (\hyperlink{https://www.cosmos.esa.int/gaia}{https://www.cosmos.esa.int/gaia}), processed by
the Gaia Data Processing and Analysis Consortium (DPAC, \hyperlink{https://www.
cosmos.esa.int/web/gaia/dpac/consortium}{https://www.
cosmos.esa.int/web/gaia/dpac/consortium}). Funding for the DPAC has
been provided by national institutions, in particular the institutions participating in the Gaia Multilateral Agreement. 
This research has made use of the Jean-Marie Mariotti Center \texttt{Aspro} service \footnote{Available at http://www.jmmc.fr/aspro}
V.G was supported for this research through a stipend from the International Max Planck Research School (IMPRS) for Astronomy and Astrophysics at the Universities of Bonn and Cologne, and from the Bonn-Cologne Graduate School of Physics and Astronomy (BCGS). 
A.C.G.has been supported by PRIN-INAF-MAIN-STREAM 2017 “Protoplanetary disks seen through the eyes of new-generation
instruments” and by PRIN-INAF 2019 “Spectroscopically tracing the disk dispersal evolution (STRADE)”
A.A.  and P.G. acknowledge supported by Fundação para a Ciência e a Tecnolo-
gia, with grants reference UIDB/00099/2020, SFRH/BSAB/142940/2018 and
PTDC/FIS-AST/7002/2020.
R.G.L. acknowledges support by Science Foundation Ireland under Grant
No. 18/SIRG/5597. 

\bibliographystyle{aa}
\bibliography{sample} 

\appendix
\section {Methodology}\label{sec:5}

Following previous treatments on line emission features in spectrally dispersed interferometric data for Herbig Ae/Be (e.g. \citealt{Kraus2008}) and T~Tauri (e.g. \citealt{GarciaLopez2020}) type YSOs, we consider the \bg emission region as a singular component in the construction of our total visibility function and use a simple Gaussian disk model to connect the interferometric quantities to its spatial properties.
As such the emission region would be characterised by a single set of geometric disk parameters in the form of size, inclination and position angle, effectively coarse graining out the internal structure arising from different \bg origin mechanisms which may contribute to the signals in our data.

\subsection{Visibilities}

\begin{figure}
\centering
\begin{minipage}{\linewidth}
\centering
\includegraphics[width=\linewidth]{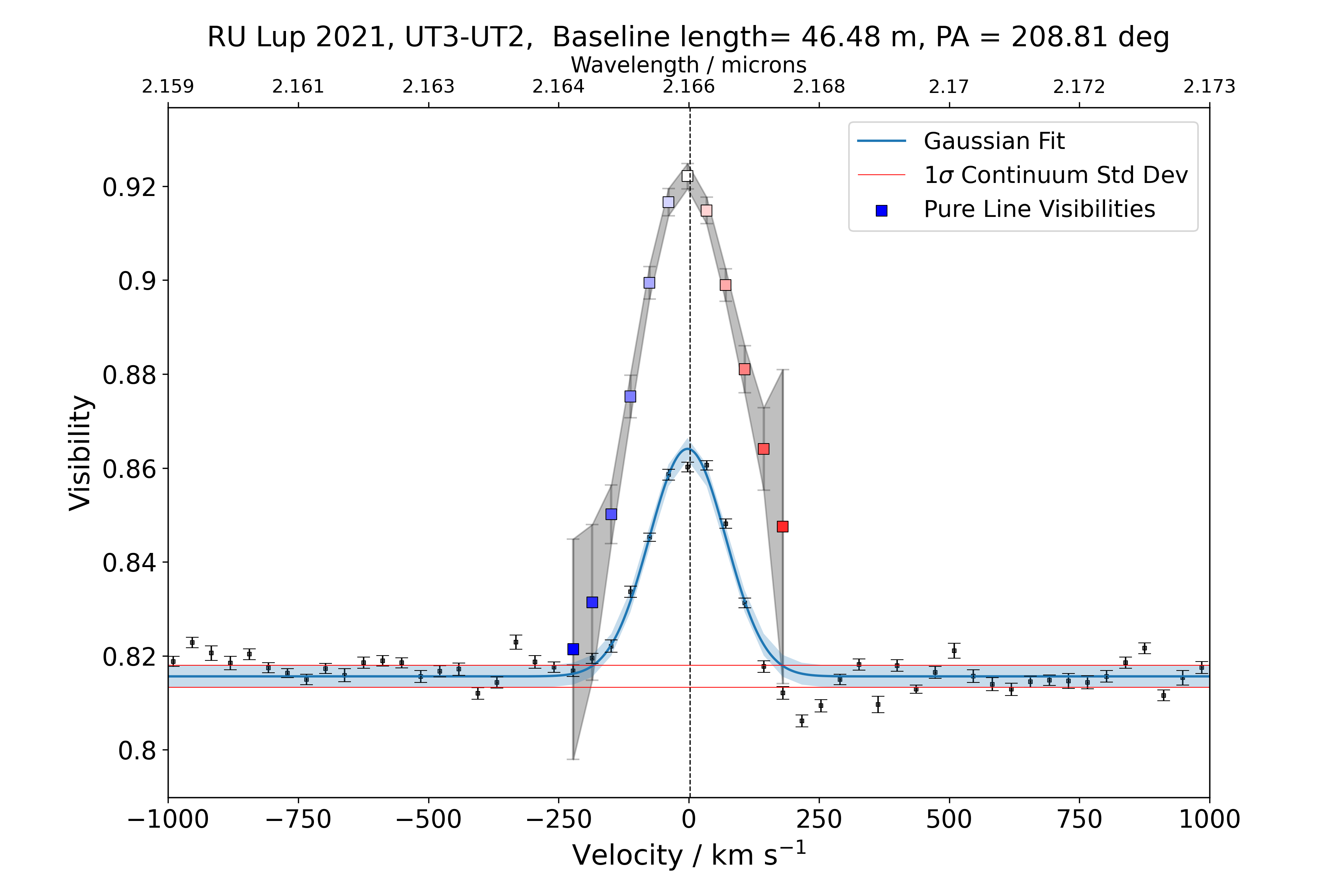}
\end{minipage}
\label{fig:visdata}
\begin{minipage}{\linewidth}
\centering
\includegraphics[width=0.9\linewidth]{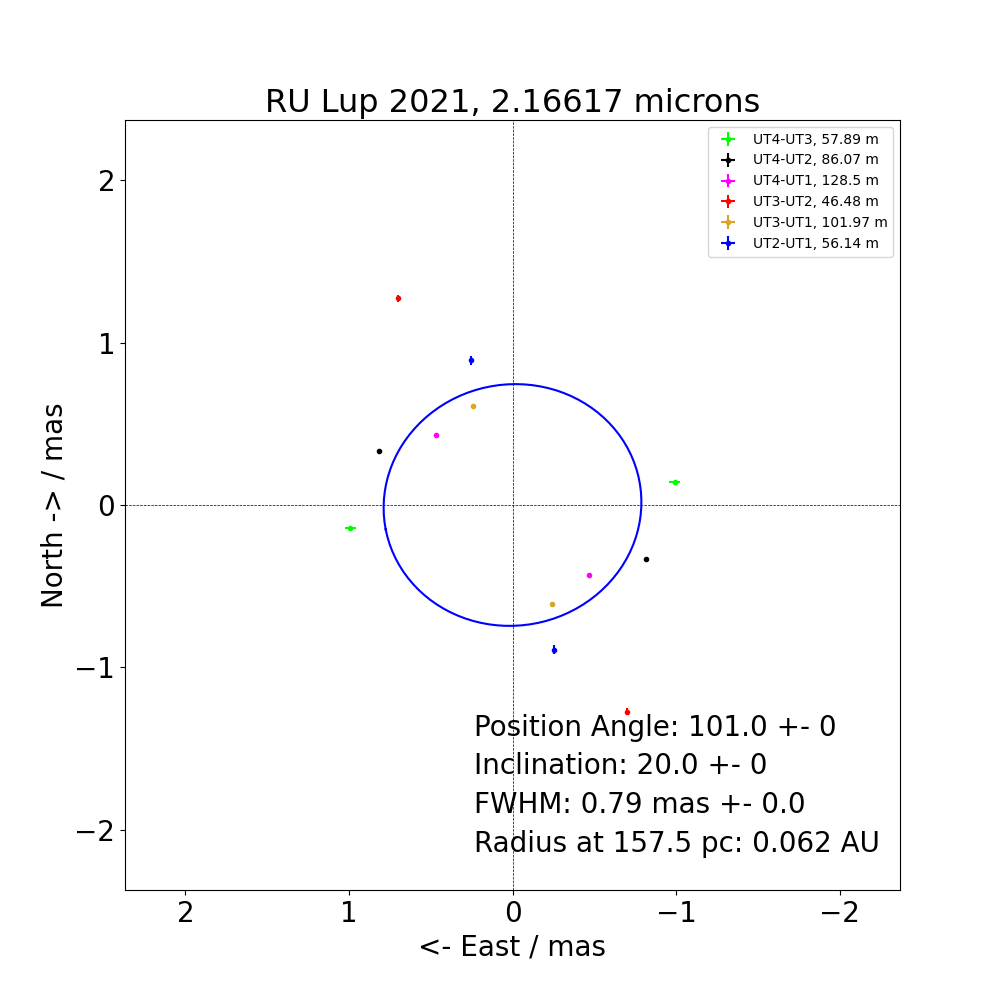}
\end{minipage}
      \caption{Fitting of the visibility amplitude data. \\
      \textbf{Top:}  Visibility data from the UT3-UT2 baseline of the RU Lup observation in 2021. The coloured squares indicate the pure line visibilities after the extraction of the continuum contribution from the \bg signal as laid out in Eq. \ref{eq:vline}. 
      \textbf{Bottom:} An ellipse representing the 2D emission region as projected onto the six baselines of GRAVITY in a single spectral channel. We used a Gaussian disk model to determine the projected half width at half maximum (HWHM) at each baseline and then fitted an ellipse to this set of points to obtain the HWHM of an inclined 2D Gaussian. The overall HWHM of the region is then given by the semi-major axis of the ellipse.} 
         \label{norm}
        
   \end{figure}

Interferometric complex visibility data contains information about the ratio of correlated, spatially unresolved flux to total observed flux coming from an object and thus encodes the size of the object as measured along a certain axis given by the orientation of a projected two-telescope baseline on the sky plane. \\
The total visibility amplitude is one of the principal interferometric observables and is the product of multiple contributing components of the object brightness distribution, such as the continuum contribution from the star, the continuum contribution from the inner dusty disk and the actual line emission region. \\
 For the purpose of treating the \bg emission origin as a singular geometric region, it is sufficient to distinguish between a single line emission component with visibility amplitude $\VL$ and flux contribution $\FL$ and a single continuum emission component $\VC$ with flux $\FC$, as further consideration of the internal distinction between the unresolved star and the marginally resolved dusty disk does not affect the extraction of $\VL$ from the total system visibility $\VT$ (see Appendix \ref{AppendixI}):
 
\begin{align}
\VL &=\frac{ \FLC\VT - \VC}{\FLC-1} .
\label{eq:vline}
\end{align}

 $\FLC$ denotes the total line-to-continuum flux ratio as taken from the normalised spectrum observed by GRAVITY, see Fig. \ref{norm}.
\begin{align}
  \FLC&=\frac{\FC+\FL}{\FC} \label{flc}.
\end{align}

The distinction between pure line and total visibility amplitude is visualised in Fig. \ref{norm} for an exemplary set of baselines. For the purpose of calculating $\VL$, the continuum visibility was determined from the average of the four channels of the low resolution fringe tracker data which are closest to the $\bg$ line. \\
From the expression Eq. \ref{eq:vline} a projected size for the pure emission region along a baseline \textbf{B} can be computed under the assumption of a certain geometric morphology of the underlying brightness distribution. In accordance with similar work done by  \citet{Ganci2021} and \citet{GarciaLopez2020}, for instance, we chose to model the intensity distribution as a Gaussian disk so that $\VL$ can be related to the angular size parameter $\Theta$, representing the FWHM of the distribution \citep{Berger2007}: 

\begin{align}
 \VL =  \text{exp}\left(-\frac{(\pi\Theta B)^2}{4ln(2)\lambda^2}\right). \label{gaussian}
\end{align}

From Eq. \ref{gaussian} we determined a set of 6 projected sizes as measured along the orientation of the GRAVITY baselines. \\
As we assumed the morphology of the emission region to be in first order geometrically close to an inclined disk, the projected sizes provided us with a set of 12 points on the sky plane, 2 per baseline due to centrosymmetry. We then obtained the position angle, inclination and half width half maximum of the overall distribution by fitting an inclined ellipse to the 12 points, with the half-flux radius being the semi-major axis of the ellipse.
 Initial attempts to determine all three parameters from the fit were shown to be flawed, as incomplete coverage by the VLTI UTs left a gap in the distribution of points which lead to fitting solutions biased towards an elongated ellipse across the gap. Subsequently the position angle and inclination of the inner dusty disk, as determined from the K-band continuum data \citep{Perraut2021}, were used to constrain those parameters under the assumption that the gas emission region and the continuum emitting inner rim were broadly aligned, leaving only the size of the region as a free parameter of the fit. \\ The ellipse was fitted for each of the considered spectral channels separately to determine the change of the emission region size as we go through the different velocity components of the spectral line. 

\subsection{Differential phases}
\begin{figure}
\centering
\begin{minipage}{\linewidth}
\centering
\includegraphics[width=\linewidth]{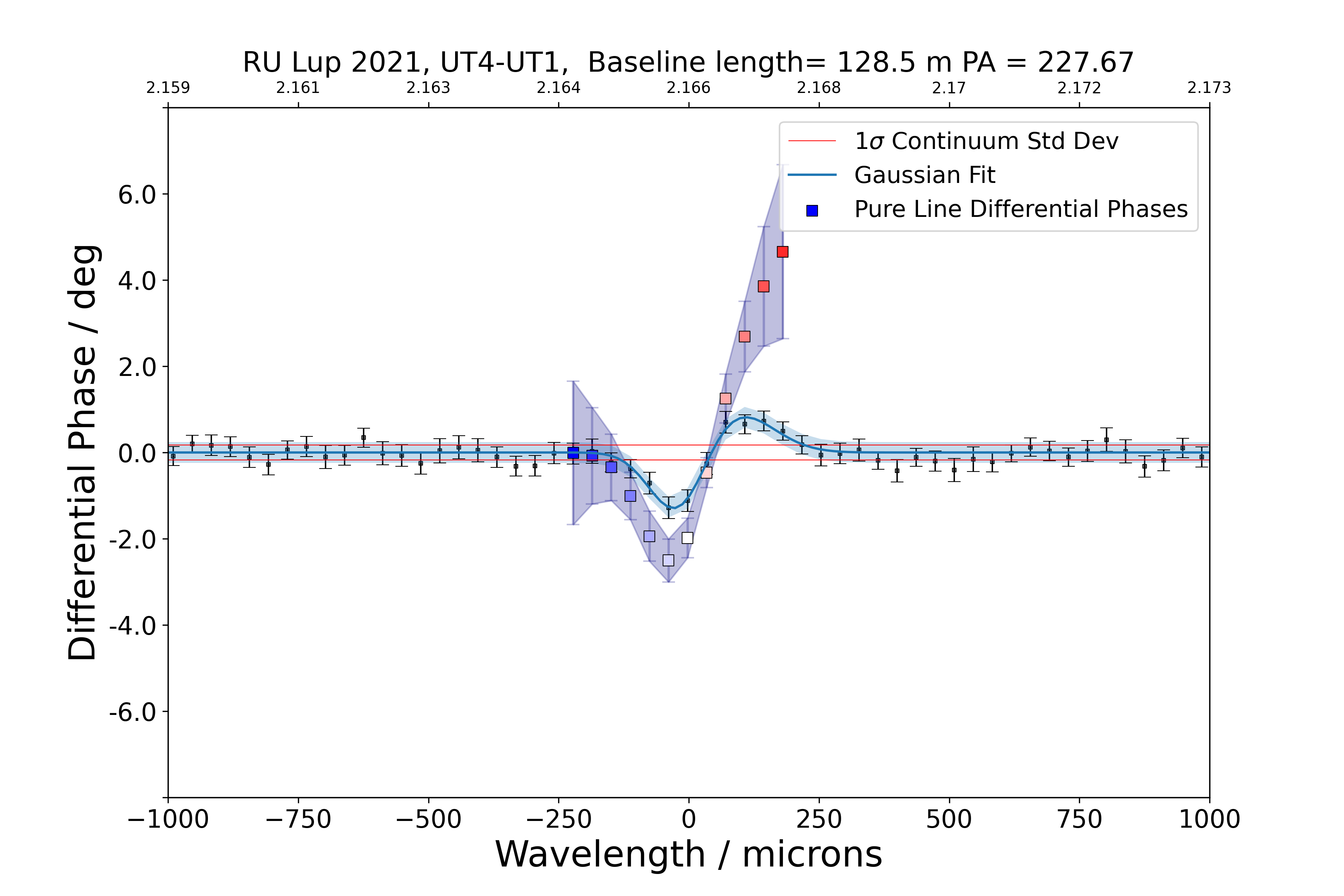}
\end{minipage}

\begin{minipage}{\linewidth}
\centering
\includegraphics[width=\linewidth]{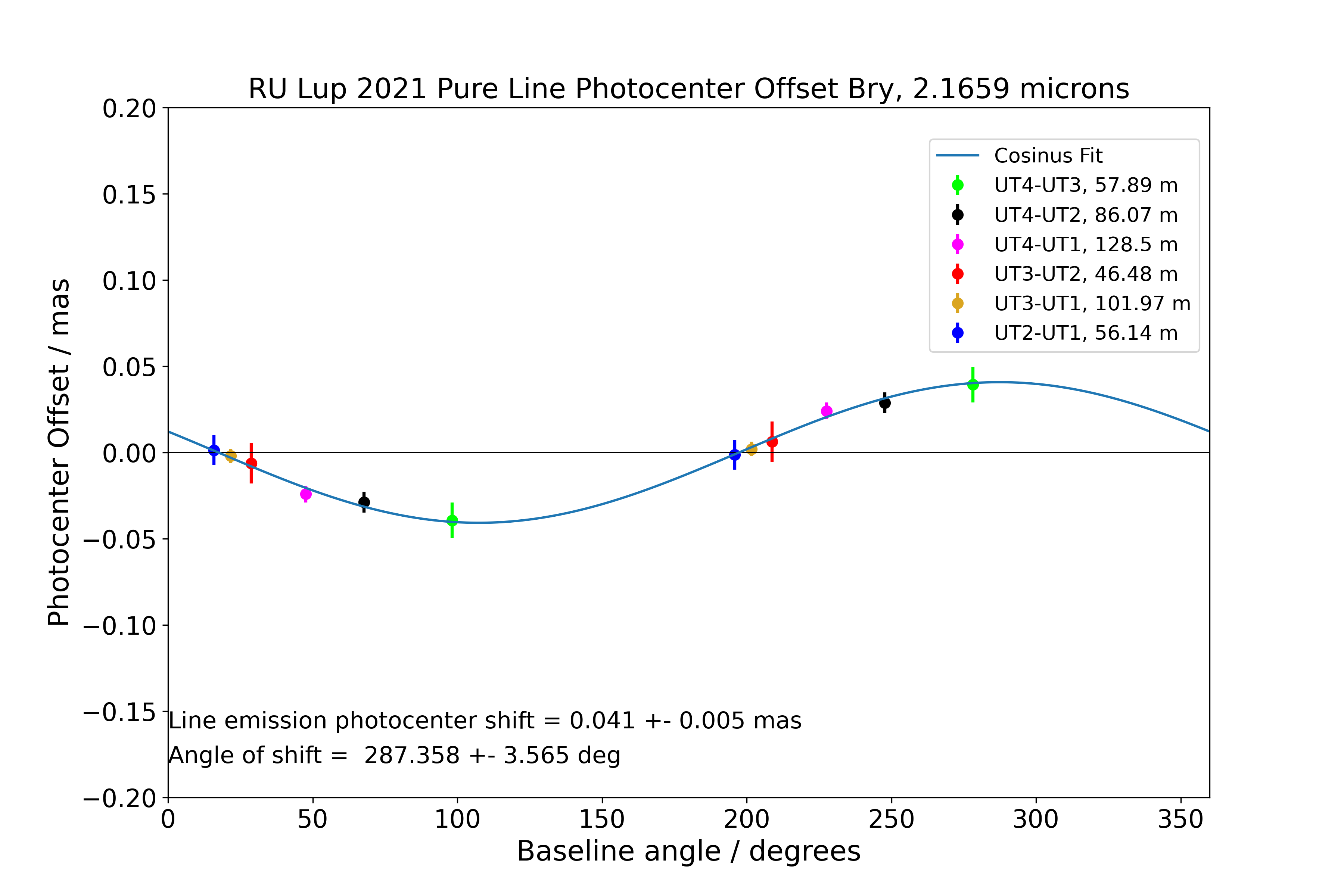}
\end{minipage}

\caption{Fitting of the differential phase data. \\
\textbf{Top:} Differential phase data from the UT4-UT1 baseline of the RU Lup observation in 2021. Here the coloured squares indicate again the pure line differential phases, excluding the influence of the continuum via Eq. \ref{pdh}. \protect\\
\textbf{Bottom:} From the pure line differential phases we obtained the photocentre shifts as projected onto the six baselines. We deprojected the overall shift vector by fitting a cosine function to the projected shift magnitudes at the respective baseline angles.}
\label{size1}
\end{figure}

The phase $\phi$ of the complex visibility function can be related to the position of the barycentre of the brightness distribution (e.g. \citealt{Waisberg2017}) if the source is marginally resolved. As the pure phase information is scrambled due to atmospheric turbulence, the absolute centroid offset cannot be recovered directly and differential quantities such as the closure phase or the differential phase have to be relied upon instead to extract such information. 
\begin{align}
\Phi_{tot}= \phi_{(Cont+Line)} - \phi_{Cont}.
\end{align}
The total differential phase $\Phi_{tot}$ is computed by subtracting the absolute average continuum phase $\phi_{Cont}$ from the total phase in a given line spectral channel $\phi_{Cont+Line}$, under the implicit assumption that the noise contribution from the atmosphere is constant in the immediate vicinity of the spectral line and thus removed by subtraction. The remaining difference can then be related to the relative shift of the total (i.e. continuum + line) photocentre with respect to the photocentre of only the continuum.  \\
The resulting total differential phase still contains contributions from the continuum that need to be removed in order to recover the pure line differential phase in a process that is largely analogous to the extraction of the pure line visibilities (see Appendix \ref{AppendixI} for a full treatment). \\
Doing so leads to an expression for the pure line differential phase $\Phi_{Line}$ that relies on $\Phi_{tot}$, the pure line and total visibilities $\VL$ and $\VT$, and also the flux ratio $\FLC$.
\begin{align}
\label{pdh}
\Phi_{Line} &= \text{arcsin}\left(\frac{\FLC}{\FLC-1} \frac{\VT}{\VL} \text{sin}(\Phi_{tot}). \right)
\end{align}

In the regime of marginally resolved sources, there is an approximate direct relationship between the vector of the total photocentre P and the differential phase $\Phi$. 
\begin{align}
\Phi_i \approx -2\pi \frac{\mathbf{B_i}}{\lambda} \mathbf{P}. \label{proj}
\end{align}

Here $\mathbf{B_i}$ is the vector of the baseline along which the measurement. \\
Since the measured differential phase is again a projection of the true shift vector \textbf{P} with $\mathbf{P} \cdot \mathbf{B_i}=\mathrm{B_i} \cdot P cos(\delta_i)$, $\mathrm{\delta_i}$ being the angle between the two vectors, we can rewrite Eq. \ref{proj} in terms of the projected shift per baseline i, $\mathrm{p_i=Pcos(\delta_i)}$, leading to the well known relationship between differential phase and shift along a baseline \citep{LeBouquin2009}:

\begin{align}
p_i=-\frac{\Phi_i \ \lambda}{2\pi B_i}.  \label{mendi}
\end{align}

From the resulting set of 6 projected shifts, the true shift vector \textbf{P} can be recovered by fitting a cosine function of the form $\mathrm{p_i=P \ cos(\Psi-\theta_i)}$ to the pairs of projected shifts and baseline angles $\mathrm{\theta_i}$, where P and $\mathrm{\Psi}$ are the magnitude and angle of \textbf{P}, respectively. The baseline-dependent shifts were deprojected in each the considered spectral channels in order to obtain the photocentre shifts of the emission region in every channel, see Fig. \ref{fig:AS353}, for example.

\subsection{Uncertainties}
For our analysis we relied on the instrumental error bars for flux, visibilities, differential and closure phases, as determined by the GRAVITY standard pipeline. In addition, we adopted a floor value for the uncertainties on the interferometric quantities in order not to underestimate them for data with high continuum dispersion. To this end we determined the floor value as the 1$\sigma$ standard deviation computed on the science channel continuum adjacent to the \bg feature.\\
Based on the relevant equations detailed in the previous sections, the instrumental error bars are translated into uncertainties on the inferred pure line quantities and then further on the projected sizes and offsets per channel and baseline via standard error propagation. This results in the error bars depicted in the bottom plots of Fig. \ref{size1} and Fig. \ref{proj}, respectively. These propagated uncertainties take into account the impact of the individual errors on the flux ratios, the total visibilities, the continuum visibilities and, for the photocentre shift, also on the differential phases. They are in a final step considered by the fitting routines for the emission region size and the photocentre offset. For the cosinus fit, depicted in Fig. \ref{proj}, we use a fitting routine provided by the Python library Scipy, while for the ellipse fit shown in Fig. \ref{size1} a custom fitting routine was employed. In both cases the error bars on the best fit parameters are determined from the covariance matrix.

\section{Visibility modelling}
\label{AppendixI}
\subsection{Pure line visibility amplitudes}
 Whilst it can be instructive to consider the visibility amplitude as a measure of the 'degree of resolvedness' of the source when interpreting the amplitude intuitively, it is mathematically more opportune to invoke the Van Cittert - Zernike theorem which relates the intensity distribution I($\alpha$,$\beta$) to the complex visibility function o(u,v) via Fourier transformation in order to recover the projected size of the source.
\begin{align}
o(u,v)= \int I(\alpha,\beta) exp(-2i\pi (u\alpha + v\beta) d\alpha d\beta . \label{vczII}
\end{align}

The linearity property of the Fourier transform allows for the convenient construction of total visibility functions by addition of different flux contributing components to the system. 

\begin{align}
\OT&=\frac{\FC \OC + \FL \OL}{\FC + \FL} \label{cv}  \nonumber \\  \nonumber \\
&=\frac{ \OC + \frac{\FL}{\FC} \OL}{1 + \frac{\FL}{\FC}} \nonumber \\\nonumber \\ 
 &=\frac{ \OC + (\FLC-1) \OL}{\FLC},
\end{align}

with a phase $\phi$, which contains information about the position of the object, and an amplitude V, which encodes the spatial geometry of the emission region. 

In Eq. (\ref{cv}) we introduce the \textbf{total line to total continuum ratio $\FLC$} with 
\begin{align}
\FLC=\frac{\FC+\FL}{\FC} = 1+\frac{\FL}{\FC},
\end{align}
which corresponds to the ratio we obtain from normalising the total flux of the image in each spectral channel to the continuum flux outside the line.
The \textbf{complex pure line visibility $\OL$} is subsequently

\begin{align}
\OL =\frac{\FLC \OT-\OC}{\FLC-1}, \label{cvf2}
\end{align}

where O is a complex quantity with a visibility amplitude V and a visibility phase $\phi$
\begin{align}
O=V e^{-i\phi}. \label{cvf3}
\end{align}

We need to further express Eq. (\ref{cvf2}) in terms of the \textbf{total differential phase $\Phi$} since the absolute phase information is lost due to atmospheric scrambling:

\begin{align}
&\OC \cdot \OL^*=\VC \VL \dphl \label{eq:pldphi1} \\
\iff & \VL \dphl = \frac{\FLC \VT \dpht \ -\VC}{\FLC-1}. \label{eq:pldphi2}
\end{align}

From this an expression for the \textbf{pure line visibility amplitude $\VL$}, from which the size of the line emission region is computed, can be derived:
\begin{align}
\VL=\lvert \VL \dphl \rvert = \left| \frac{\FLC  \VT \dpht \ -\VC}{\FLC-1} \right|.
\end{align}

For complex numbers z we can use that $\lvert z \rvert = \sqrt{zz^*}$ and arrive at

\begin{eqnarray}
\VL&=& \left( \left( \frac{\FLC}{\FLC-1} \VT\right)^2 + \left( \frac{1}{\FLC-1} \VS \right)^2 \right. \nonumber \\
 && \left. - 2 \frac{\FLC}{(\FLC-1)^2} \VT \VS cos(\Phi_{tot}) \right)^{1/2} . \label{trupl}
\end{eqnarray}

As we typically operate with differential phases small enough so that $cos(\Phi_{tot}) \approx 1$, we can further approximate Eq. (\ref{trupl}) as:

\begin{align}
\VL &\approx \left( \left( \frac{\FLC}{\FLC-1} \VT\right)^2 + \left( \frac{1}{\FLC-1} \VC \right)^2 \right. \nonumber \\
 & \left. - 2 \frac{\FLC}{(\FLC-1)^2} \VT \VC \right)^{1/2} \\ \nonumber \\
&=\left( \left( \frac{\FLC}{\FLC-1} \VT - \frac{1}{\FLC-1} \VC \right)^2 \right)^{1/2} \\\nonumber \\ 
&=\frac{\FLC \VT - \VC}{\FLC-1} .\label{result1}
\end{align}

\subsection{Pure line differential phase}
The \textbf{pure line differential phase $\Phi_{Line}$} can be extracted  by rearranging Eq. (\ref{eq:pldphi2})

\begin{align}
 \dphl &=\frac{\FLC}{\FLC-1} \frac{\VT}{\VL} \dpht - \frac{1}{\FLC-1} \frac{\VC}{\VL}.
\end{align}

Via use of Euler's formula $ e^{i \Phi}=cos(\Phi)+isin(\Phi)$, this can be written as 

\begin{align}
cos(\Phi_{Line})+isin(\Phi_{Line}) &=\left(\frac{\FLC}{\FLC-1} \frac{\VT}{\VL} cos(\Phi_{tot})   - \frac{1}{\FLC-1} \frac{\VC}{\VL} \right) \nonumber \\ 
&+ i \left(\frac{\FLC}{\FLC-1} \frac{\VT}{\VL} sin(\Phi_{tot})\right),
\end{align}

and, by comparison of the imaginary coefficients, we arrive at

\begin{align}
sin(\Phi_{Line}) &= \frac{\FLC}{\FLC-1} \frac{\VT}{\VL} sin(\Phi_{tot})  \\\nonumber \\ 
\iff \Phi_{Line} &= arcsin\left(\frac{\FLC}{\FLC-1} \frac{\VT}{\VL} sin(\Phi_{tot}) \right). \label{result2}
\end{align}


\section{Model cube observables}
\label{AppendixII}
\subsection{Model visibilities}

The data cubes of the radiative transfer models contain 2D images of the brightness distribution in the vicinity of the star at wavelengths close to the \bg peak, featuring the magnetosphere and a stellar (and magnetospheric) continuum component including the photosphere and the shock region. \\ 
To extract interferometric observables from such an image, it is Fourier transformed as per Eq. (\ref{vczII}). By our previous considerations, the visibility amplitude and the phase can then be recovered from the obtained complex visibilities $O(\mathbf{B},\lambda)$ per spectral channel via
\begin{align}
V(\mathbf{B},\lambda)=\mathrm{abs}(O(\mathbf{B},\lambda)) \\
\phi((\mathbf{B},\lambda)=\mathrm{arg}(O(\mathbf{B},\lambda)).
\end{align}

These quantities are specific to the baseline vector $\mathbf{B}$ as they encode the projections of the geometric features along a specific baseline alignment at a resolution defined by the baseline length. \\
Such observables, directly computed from the image, are not suitable for comparison with observational data as the latter encompasses influences from spatially more extended regions which are not part of the model. \\
As the image only contains continuum emission from the shock region, the photosphere and adjacent magnetospheric region and does not feature an extended dusty disk, the computed observables need to be modified to reflect this contribution and ensure comparability with the GRAVITY data. \\

\subsection{Modified flux ŕatio of the model}

Consider the original \textbf{ total line to total continuum flux ratio $\FLCP$} of the model, which only features a purely stellar continuum 

\begin{align}
     \FLCP &= \frac{\FS + \FL}{\FS} \\  \nonumber \\
     &= 1 + \frac{\FL}{\FS}.
\end{align}
 
We now define the modified flux ratio $\FLC$ to include both a disk and a halo contribution to the total flux:

\begin{align}
\FLC&=\frac{\FS+\FD+\FH+\FL}{\FS+\FD+\FH} \\ \nonumber \\
&=\frac{1+\frac{\FL}{\FS}+\frac{\FD+\FH}{\FS}}{1+\frac{\FD+\FH}{\FS}} \\ \nonumber \\
&=\frac{\FLCP+\frac{\FD+\FH}{\FS}}{1+\frac{\FD+\FH}{\FS}}. \label{oldnew1}
\end{align}

In Eq. (\ref{oldnew1}) we expressed the new ratio in terms of the original flux ratio of the model cube and the infrared excess $\frac{\FD+\FH}{\FS}$, which in this context is either externally defined by the user or determined from fitting the synthetic data to experimental observations. \\
The principal difference between disk and halo flux lies in the fact that, while both contribute equally to the total flux, the halo is overresolved by the baselines and does not contribute to the total visibilities.

\subsection{Modified visibility amplitude of the model}

Applying the result from Appendix \ref{AppendixI} to the cube images, we can state the total visibility function for the magnetospheric system:

\begin{align}
\VTP &= \frac{\FS \VS + \FL \VL}{\FS + \FL} \\ \nonumber \\
&= \frac{\VS + \frac{\FL}{\FS} \VL}{1+\frac{\FL}{\FS}} \\ \nonumber \\
&= \frac{\VS + (\FLCP -1) \VL}{\FLCP} \label{VTP}.\\ \nonumber 
\end{align}

A total visibility function which includes both disk and halo component discussed in the previous section can be written as:

\begin{align}
\VT &= \frac{\FL \VL + \FS \VS + \FD \VD + \FH \VH}{\FL + \FS + \FD + \FH} \\ \nonumber \\
&= \frac{\VS + \frac{\FL}{\FS} \VL +\frac{\FD}{\FS} \VD }{1+\frac{\FL}{\FS}+\frac{\FD+\FH}{\FS}} \label{tot_vis_halo1}\\ \nonumber \\
&= \frac{\FLCP \VTP +\frac{\FD}{\FS} \VD }{\FLCP +\frac{\FD+\FH}{\FS}}. \\ \nonumber 
\end{align}

Here we made use of the fact that, by definition, $\VH=0$, and used the original magnetospheric visibility function Eq. (\ref{VTP}) to express the disk and halo contribution as a modification of those original visibilities derived from the image cube. \\

Alternatively, using $\frac{\FL}{\FS}=(\FLC -1 )(\IREX +1)$ and $\VC=\frac{\frac{\FD}{\FS}\VD + \VS}{\FD +\FH + 1}$, Eq. (\ref{tot_vis_halo1}) can be written in an analogous form to the magnetospheric model total visibility function:
\begin{align}
    \VT&=\frac{\VC+  (\FLC-1)\VL}{\FLC} \label{tot_vis_halo2}.
\end{align}

\subsection{Modified differential phase of the model}

We refer again to Appendix \ref{AppendixI} to state the pure line differential phase function for the purely magnetospheric system of the model cube:

\begin{align}
\Phi^{\prime}_{Line}=arcsin\left(\frac{\FLCP}{\FLCP-1} \frac{\VTP}{\VL} sin(\Phi^{\prime}_{tot}) \right),
\end{align}

and for the modified model including disk and halo contribution:

\begin{align}
\Phi_{Line}=arcsin\left(\frac{\FLC}{\FLC-1} \frac{\VT}{\VL} sin(\Phi_{tot}) \right),
\end{align}

which self-evidently follows from Eq. (\ref{tot_vis_halo2}) in an analogous manner. \\
Since, by necessity, $\Phi_{Line}=\Phi^{\prime}_{Line}$, we can relate the total differential phase of the modified model to the total differential phase of the original magnetospheric system via

\begin{align}
 &\frac{\FLCP}{\FLCP-1} \frac{\VTP}{\VL} sin(\Phi^{\prime}_{tot}) = \frac{\FLC}{\FLC-1} \frac{\VT}{\VL} sin(\Phi_{tot}) \\ \nonumber \\
\iff &sin(\Phi_{tot})= \frac{\FLCP}{\FLC}\frac{\FLC-1}{\FLCP-1} \frac{\VTP}{\VT}  sin(\Phi^{\prime}_{tot})  \\ \nonumber \\
&= \frac{\FLCP \VTP}{\IREX \VD + \FLCP \VTP}  sin(\Phi^{\prime}_{tot}).
\end{align}

The total differential phases for a system including a disk continuum can thus be computed from the infrared excess, the disk visibility at the respective baseline and the datacube observables as

\begin{align}
\Phi_{tot}=arcsin\left(\frac{\FLCP \VTP}{\IREX \VD + \FLCP \VTP}  sin(\Phi^{\prime}_{tot}) \right).
\end{align}

\section{Observational data}
\label{AppendixIII}

\begin{figure*}[h!]
\begin{minipage}{\linewidth}
    \centering
    \includegraphics[width=0.9\linewidth]{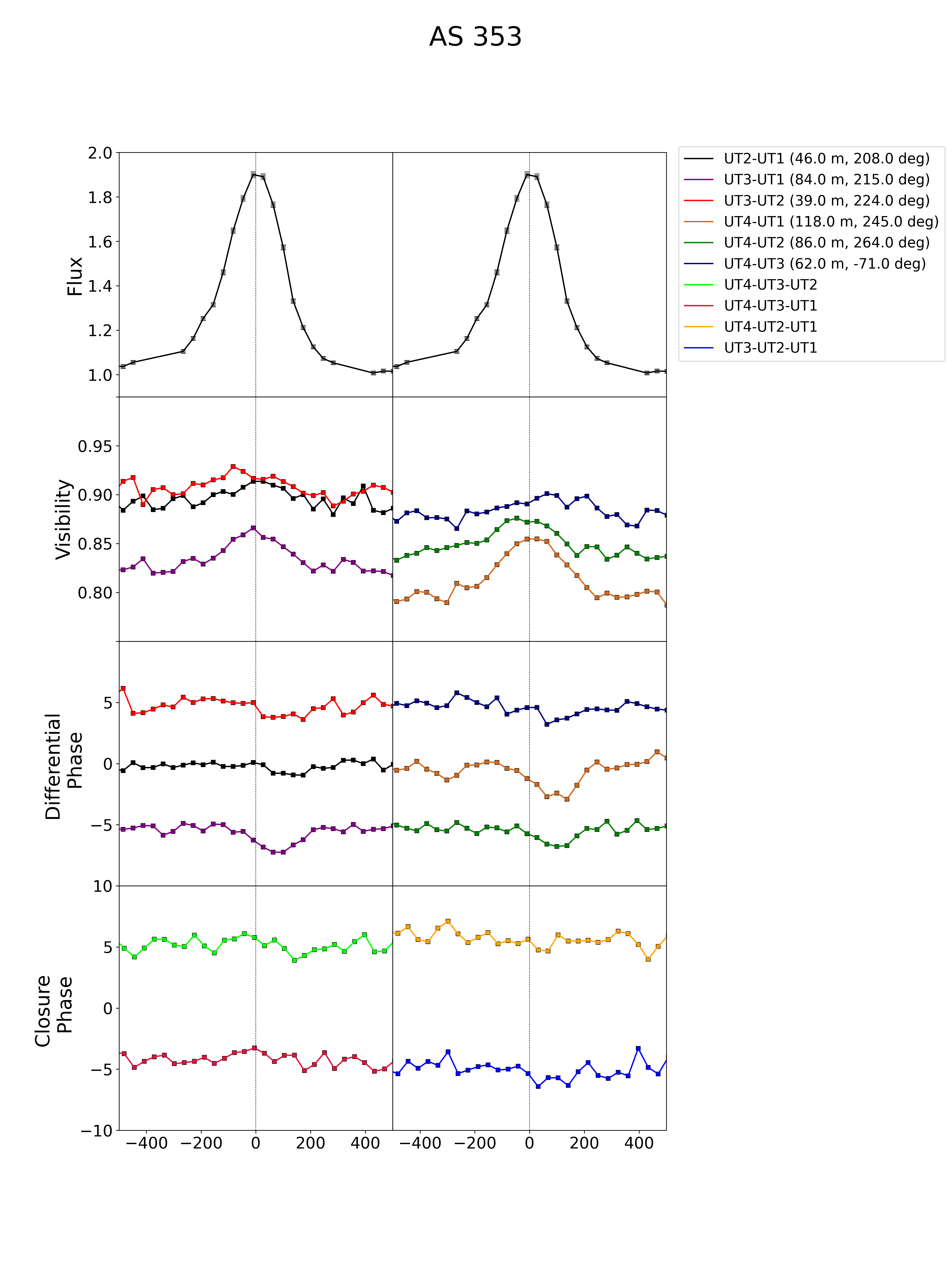}
    \caption{Observational data for AS 353. Differential phases and closure phases have been shifted by $\pm$5 deg at certain baselines.}
    \label{fig:AS353Data}
    \end{minipage}
\end{figure*}
\clearpage

\begin{figure*}[!htbp]
\begin{minipage}{\linewidth}
    \centering
    \includegraphics[width=0.9\linewidth]{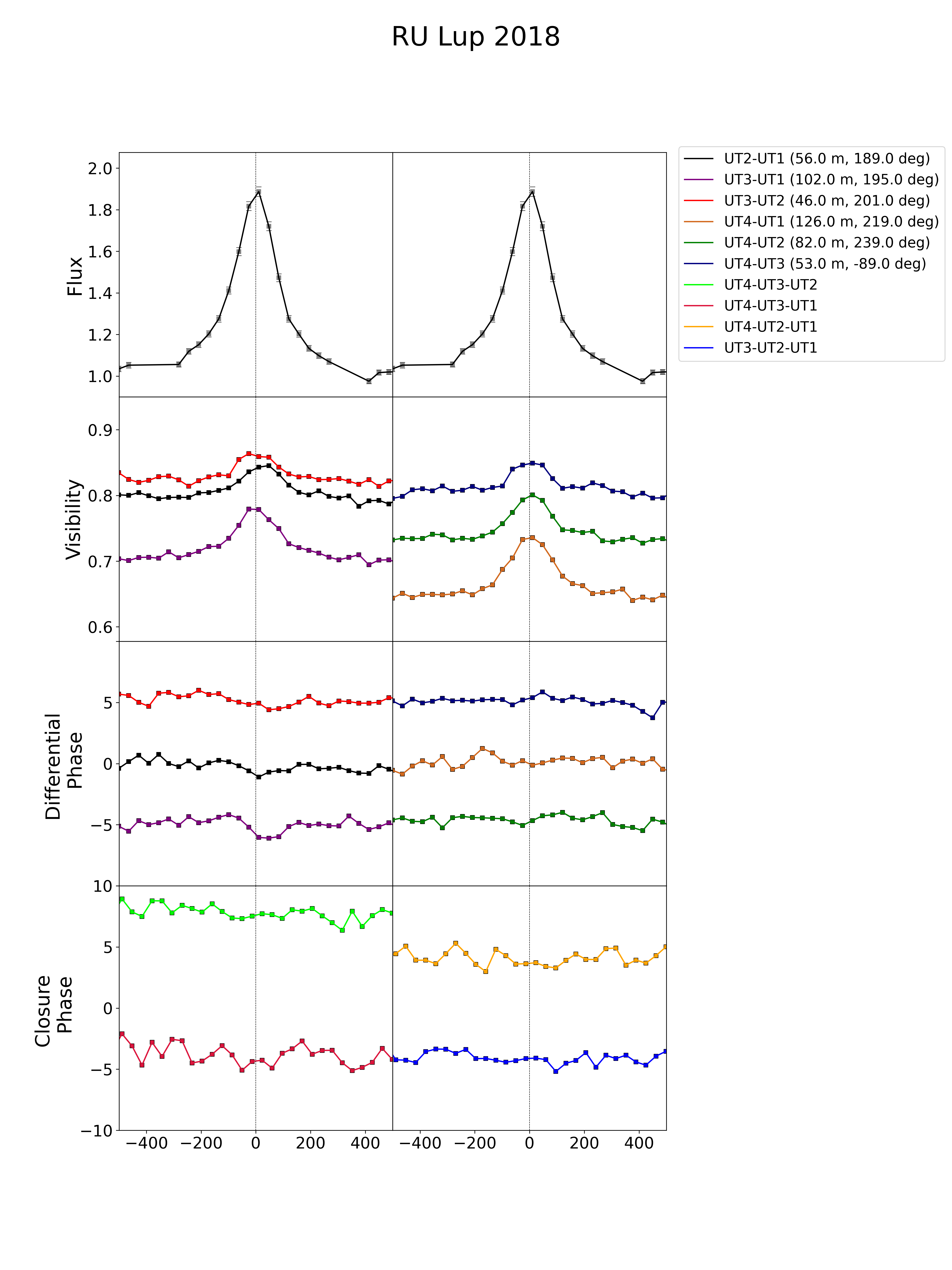}
    \caption{Observational data for AS 353. Differential phases and closure phases have been shifted by $\pm$5 deg at certain baselines.}
    \label{fig:RULup18Data}
    \end{minipage}
\end{figure*}
\clearpage

\begin{figure*}[!htbp]
\begin{minipage}{\linewidth}
    \centering
    \includegraphics[width=0.9\linewidth]{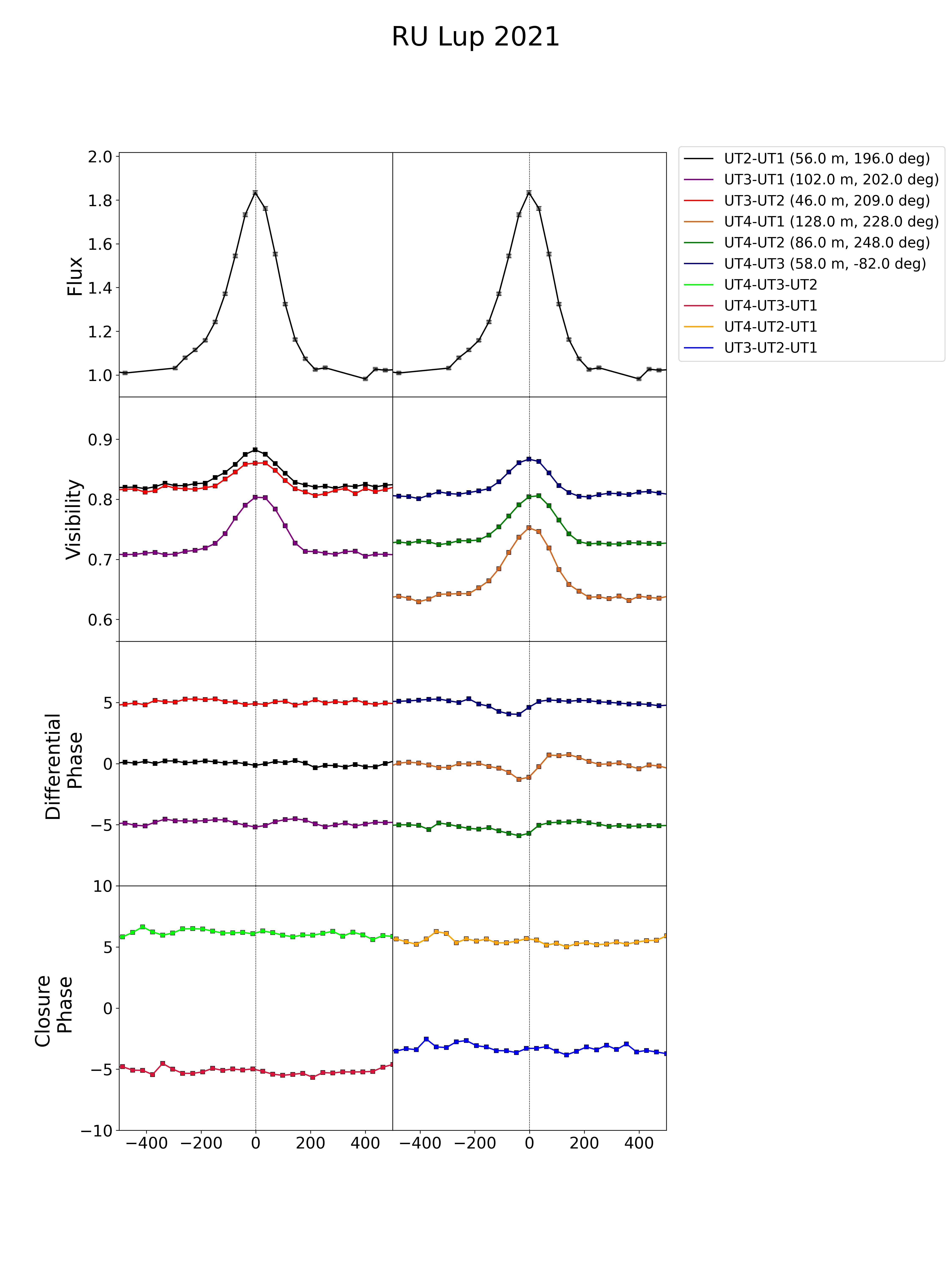}
    \caption{Observational data for AS 353. Differential phases and closure phases have been shifted by $\pm$5 deg at certain baselines.}
    \label{fig:RULup21Data}
    \end{minipage}
\end{figure*}
\clearpage

\begin{figure*}[!htbp]
\begin{minipage}{\linewidth}
    \centering
    \includegraphics[width=0.5\linewidth]{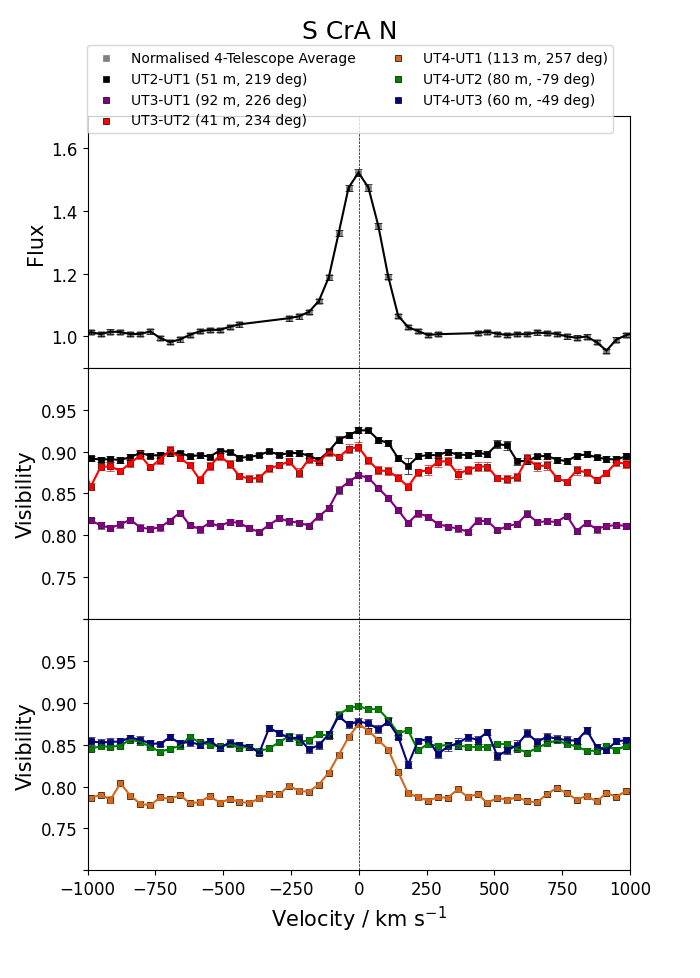}
    \caption{Observational data for AS 353. Differential phases and closure phases were omitted due to poor data quality.}
    \label{fig:SCrAData}
    \end{minipage}
\end{figure*}
\clearpage

\begin{figure*}[!htbp]
\begin{minipage}{\linewidth}
    \centering
    \includegraphics[width=0.9\linewidth]{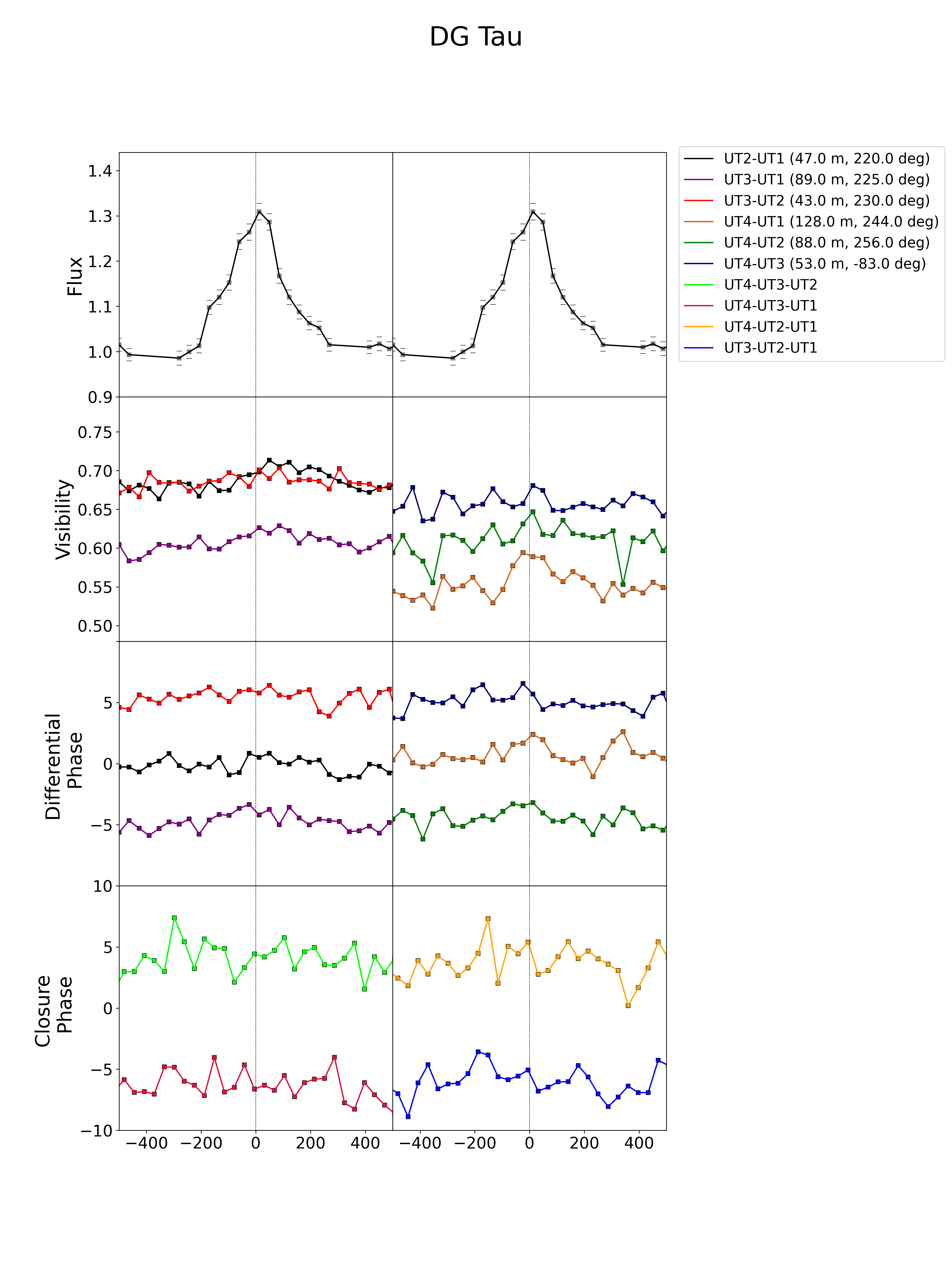}
    \caption{Observational data for AS 353. Differential phases and closure phases have been shifted by $\pm$5 deg at certain baselines.}
    \label{fig:DGTauData}
    \end{minipage}
\end{figure*}
\clearpage

\begin{figure*}[!htbp]
\begin{minipage}{\linewidth}
    \centering
    \includegraphics[width=0.9\linewidth]{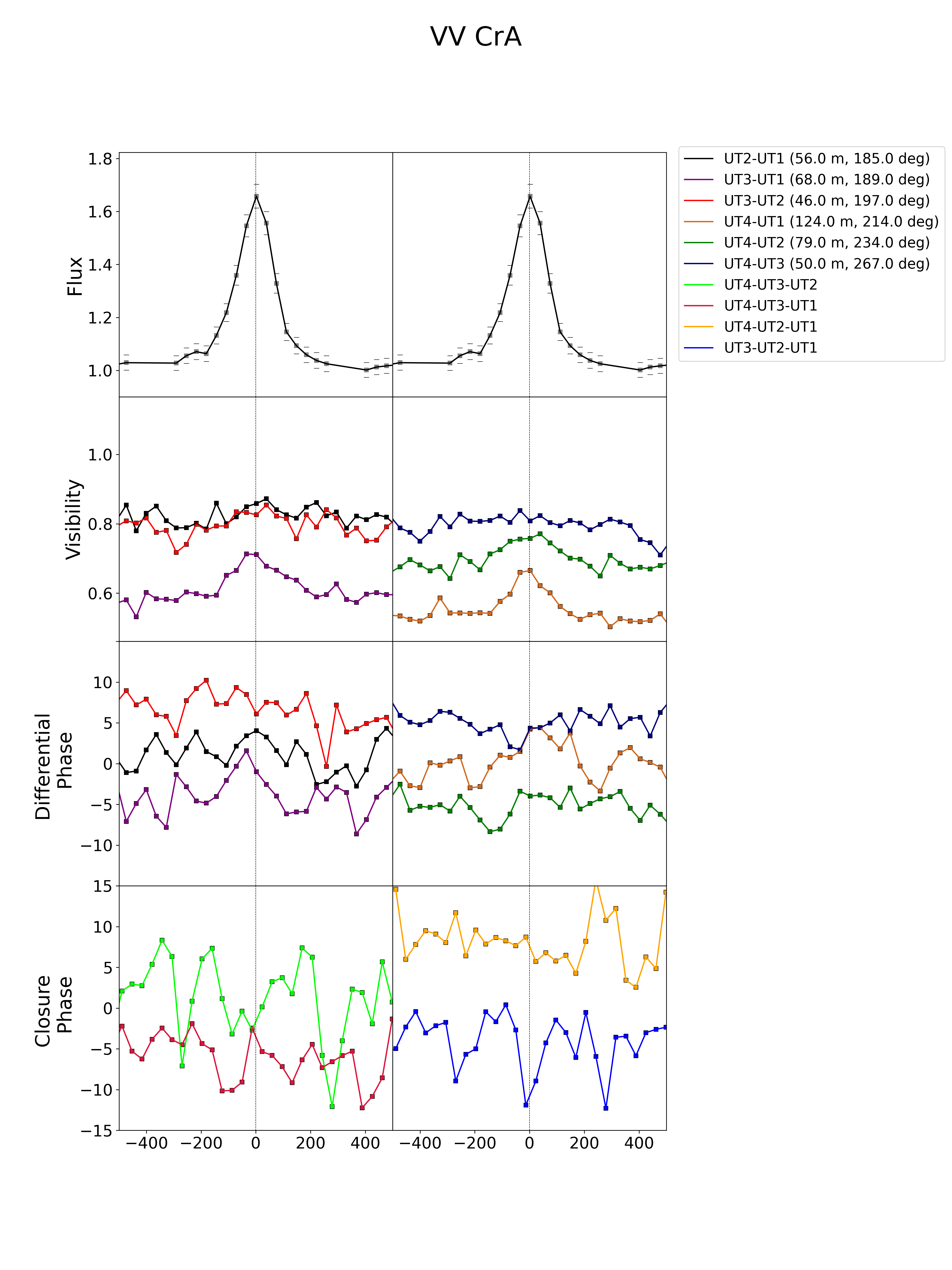}
    \caption{Observational data for AS 353. Differential phases and closure phases have been shifted by $\pm$5 deg at certain baselines.}
    \label{fig:VVCrAData}
    \end{minipage}
\end{figure*}
\clearpage

\begin{figure*}[!htbp]
\begin{minipage}{\linewidth}
    \centering
    \includegraphics[width=0.9\linewidth]{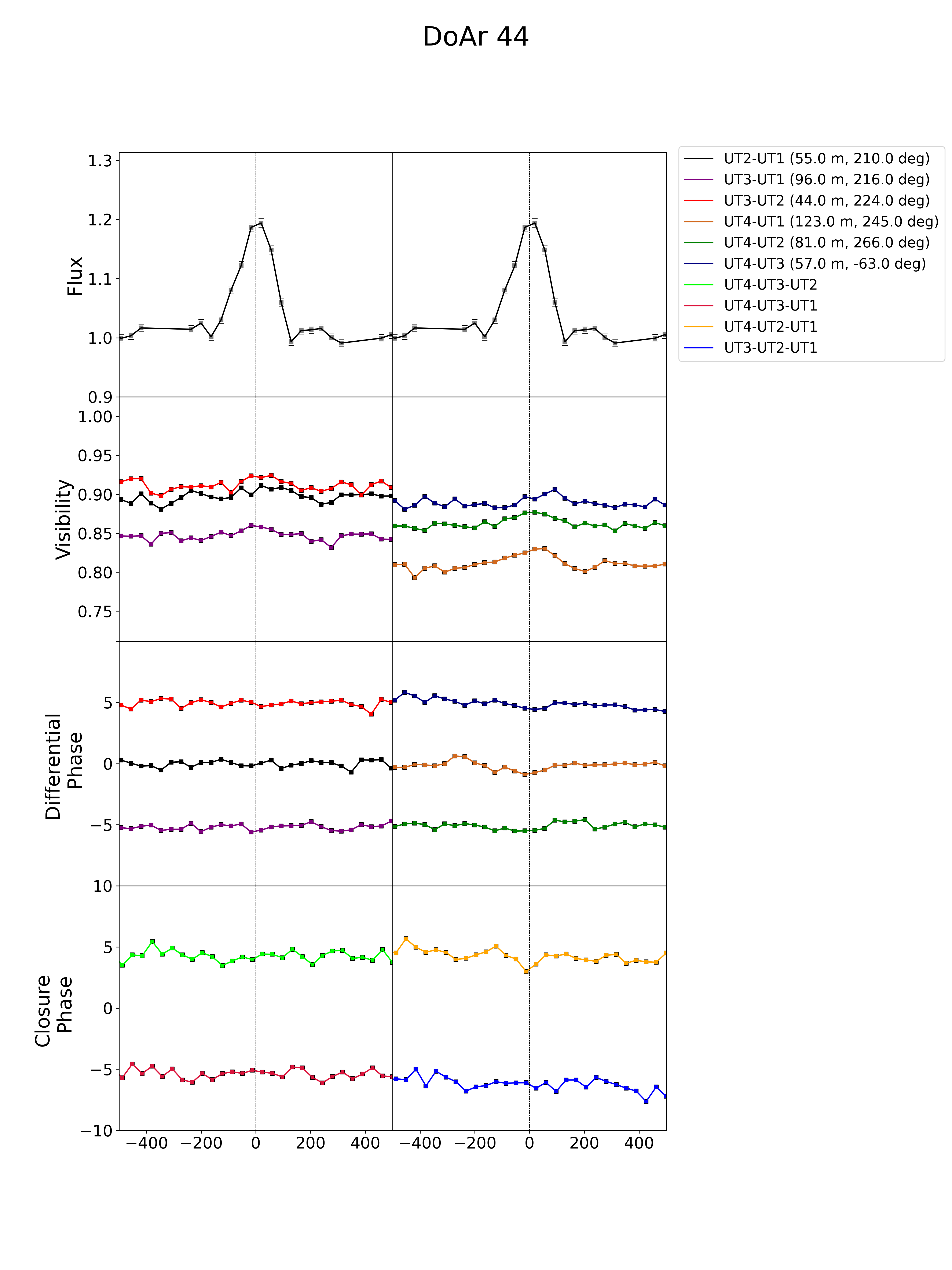}
    \caption{Observational data for AS 353. Differential phases and closure phases have been shifted by $\pm$5 deg at certain baselines.}
    \label{fig:DoAr44Data}
    \end{minipage}
\end{figure*}
\clearpage

\begin{figure*}[!htbp]
\begin{minipage}{\linewidth}
    \centering
    \includegraphics[width=0.9\linewidth]{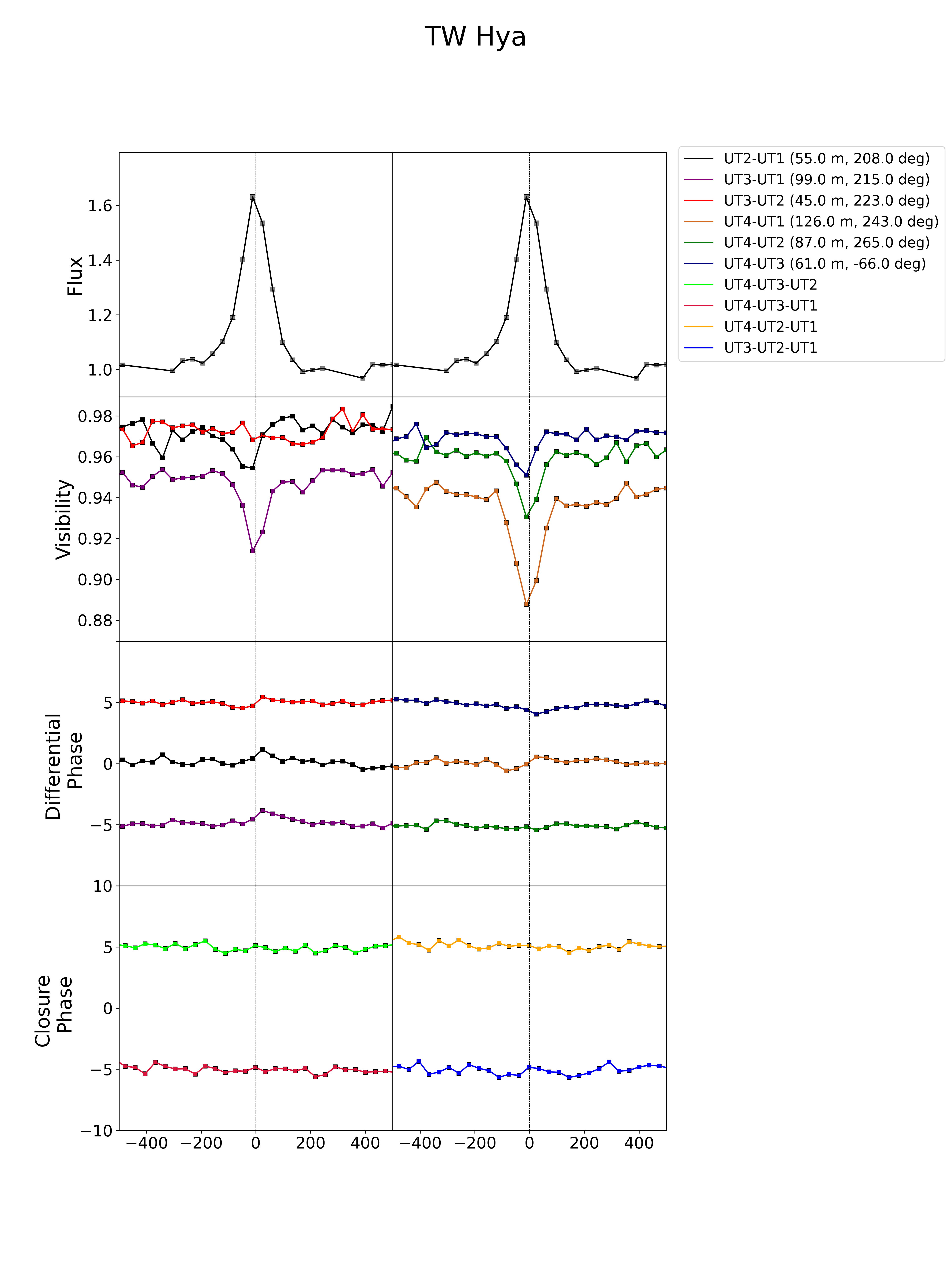}
    \caption{Observational data for AS 353. Differential phases and closure phases have been shifted by $\pm$5 deg at certain baselines.}
    \label{fig:TWHyaData}
\end{minipage}
\end{figure*}

\end{document}